\documentclass[preprint,11pt]{aastex}
\usepackage{emulateapj5}

\newcommand{\hdf}{HDF--N}
\newcommand{\hst}{\textit{HST}}
\newcommand{\wfu}{\hbox{$U_{300}$}}
\newcommand{\wfb}{\hbox{$B_{450}$}}
\newcommand{\wfv}{\hbox{$V_{606}$}}
\newcommand{\wfi}{\hbox{$I_{814}$}}
\newcommand{\nicj}{\hbox{$J_{110}$}}
\newcommand{\nich}{\hbox{$H_{160}$}}
\newcommand{\ks}{\hbox{$K_s$}}
\newcommand{\AAA}{\hbox{\AA}}
\newcommand{\lya}{Lyman~$\alpha$}
\newcommand{\hb}{\hbox{H$\beta$}}
\newcommand{\ha}{\hbox{H$\alpha$}}
\newcommand{\lsim}{\lesssim}
\newcommand{\gsim}{\gtrsim}
\newcommand{\lstar}{\hbox{$L^\ast$}}
\newcommand{\extinctA}{\hbox{$A_{1700}$}}

\newcommand{\mathM}{\hbox{$\mathcal{M}$}}
\newcommand{\NLBG}{33}
\newcommand{\dchi}{\hbox{$\Delta \chi^2$}}
\newcommand{\ebv}{\hbox{$E(B-V)$}}
\newcommand{\royalsociety}{Phil.\ Trans.\ R.\ Soc.\ Lond.\ A}
\newcommand{\etal}{et al.}
\newcommand{\eg}{e.g.}
\newcommand{\ie}{i.e.}
\newcommand{\Msol}{\hbox{$\mathcal{M}_\odot$}}
\newcommand{\Zsol}{\hbox{$Z_\odot$}}
\newcommand{\msol}{\hbox{$\mathcal{M}_\odot$}}
\newcommand{\zsol}{\hbox{$Z_\odot$}}
\newcommand{\kms}{\hbox{km~s$^{-1}$}}

\newcommand{\fion}[2]{\hbox{[\ion#1#2]}}

\newcommand{\efolding}{$e$--folding}

\shorttitle{THE STELLAR POPULATIONS AND EVOLUTION OF LBGS}
\shortauthors{PAPOVICH, DICKINSON, \& FERGUSON}

\submitted{Accepted for publication in the Astrophysical Journal}

\begin{document}

\title{THE STELLAR POPULATIONS AND EVOLUTION OF LYMAN BREAK
GALAXIES\altaffilmark{1}}

\author{Casey Papovich\altaffilmark{2}} \affil{Department of
Physics and Astronomy, The Johns Hopkins University, \\ 3400
N. Charles St., Baltimore, MD 21218}
\email{papovich@stsci.edu}

\author{\sc Mark Dickinson\altaffilmark{3}, and Henry C. Ferguson}
\affil{Space Telescope Science Institute, 3700 San Martin Dr.,
Baltimore, MD 21218}
\email{{med@stsci.edu},{ferguson@stsci.edu}}

\altaffiltext{1}{Based on observations taken with the NASA/ESA Hubble
Space Telescope, which is operated by the Association of Universities
for Research in Astronomy, Inc.\ (AURA) under NASA contract
NAS5--26555}   \altaffiltext{2}{Also Space Telescope Science
Institute} \altaffiltext{3}{Visiting Astronomer, Kitt Peak National
Observatory,  National Optical Astronomy Observatories, which is
operated by the  Association of Universities for Research in
Astronomy, Inc.  (AURA)  under cooperative agreement with the National
Science Foundation.}


\begin{abstract}

Using deep near--infrared and optical observations of the Hubble Deep
Field North from the {\it Hubble Space Telescope}\ NICMOS and WFPC2
instruments and from the ground, we examine the spectral energy
distributions of Lyman break galaxies (LBGs) at $2.0 \lsim z \lsim
3.5$  in order to investigate their stellar population properties.
The ultraviolet--to--optical rest--frame spectral energy distributions
(SEDs) of the galaxies are much bluer than those of present--day
spiral and elliptical galaxies, and are generally similar to those  of
local starburst galaxies with modest amounts of reddening.  We use
stellar population synthesis models to study  the properties  of the
stars that dominate the light from LBGs.  Under the assumption that
the star--formation rate is continuous or decreasing with time, the
best--fitting models provide a lower bound on the LBG mass estimates.
LBGs with ``\lstar'' UV luminosities  are estimated to have minimum
stellar masses $\sim 10^{10}$~\msol, or roughly $1/10$th that of a
present--day \lstar\ galaxy, similar  to the mass of the Milky Way
bulge.  By considering  the photometric effects of a second
stellar--population component of maximally--old stars, we set an upper
bound on the stellar masses that is $\sim 3-8\times$ the minimum mass
estimate.  The stellar masses  derived for bright LBGs are similar to
published estimates of their  dynamical masses based on
nebular--emission--line widths, suggesting that  such kinematic
measurements may substantially underestimate the total  masses of the
dark matter halos.  We find only loose constraints  on the individual
galaxy ages, extinction, metallicities, initial mass functions, and
prior star--formation histories.   Most LBGs are well fit by models
with population ages that range from 30 Myr to $\sim 1$~Gyr, although
for models with sub--solar  metallicities a significant minority of
galaxies are well fit by very young ($\lsim 10$~Myr), very dusty
stellar populations, $A$(1700~\AA)$ > 2.5$~mag.  We find no galaxies
whose SEDs are consistent with young ($\lsim 10^8$~yr), dust--free
objects, which suggests that LBGs are not dominated by ``first
generation'' stars, and that such objects are rare at these redshifts.
We also find that the typical ages for the observed star--formation
events are significantly younger than the time interval covered by
this redshift range ($\sim 1.5$~Gyr).   From this, and from the
relative absence of candidates for quiescent, non--star--forming
galaxies at these redshifts in the NICMOS data that might correspond
to the fading remnants of galaxies formed at higher redshift, we
suggest that star formation in LBGs may be recurrent, with short duty
cycles and a timescale between star--formation events of $\lsim 1$~Gyr.

\end{abstract}
 


\section{Introduction}

The last few years have seen rapid advances in the study of galaxies
at very high redshifts, especially at $z > 2$.  In a large part, this
has been due to the development of simple photometric techniques  to
select high redshift objects by characteristic color signatures.
Neutral hydrogen absorption within galaxies and from the intergalactic
medium strongly attenuates flux shortward of Lyman~$\alpha$ (1216~\AA)
and the 912~\AA\ Lyman limit, producing a spectral ``break'' that
provides  a color signature to identify objects at high redshift.
Such color  selection methods have long been used to identify high
redshift QSOs  (\eg, Warren \etal\ 1987), and were latter applied to
set limits on  the number of star--forming, faint galaxies at
$z\approx 3$  (Guhathakurta, Tyson, \& Majewski 1990; Songalia, Cowie,
\& Lilly 1990).  Using this method with a custom suite of broad--band
filters, \citet{ste92}  and \citet{ste95} established the existence of
a significant number  of high redshift galaxy candidates, now commonly
described as Lyman break galaxies (LBGs), which were then
spectroscopically confirmed to have   $z \approx 3$
\citep{ste96a,ste96b}.  By now, nearly 1000 LBGs have  measured
redshifts, providing a rich sample for studying the  properties of
star--forming galaxies at high redshift.

Understanding the epoch at which galaxies assembled the bulk of their
stellar mass remains one of the key cosmological questions.  Although
new surveys are producing a growing inventory of photometric and
spectroscopic data on LBGs, it is not yet clear how these objects  fit
into the ancestral history of the present--day galaxy population.  The
process of galaxy assembly depends on the particular parameters of the
cosmological model, \eg, the biasing of baryonic mass relative to the
underlying dark  matter distribution, and the physics and feedback of
star formation.  In hierarchical models, galaxy formation is a
continuous process driven by mergers of lower mass subcomponents over
a wide range of redshifts.  In other scenarios, star formation and
galaxy assembly takes place on rapid timescales at high redshift, with
the galaxies evolving \textit{in situ} thereafter, more or less
passively, to the present day.  These different models predict
substantially different histories and timescales for the formation and
assembly of early stellar populations, especially at $z \gsim 1$.
Observational constraints of the mass assembly history are
inconclusive.  The presence of old stars ($\gsim 10$~Gyr) in the
Galactic bulge  and the spheroidal components of M31 (e.g., Renzini
1999,  Rich \& McWilliam 2000), the existence of a population of old
ellipticals  at $z \gsim 1$ \citep{dic95,dun96,dad00}, and the small
scatter in the  color--magnitude relation of cluster ellipticals to
$z\sim 1$ \citep{sta98}  all support the picture of rapid collapse and
enrichment at high redshifts,  and imply that these systems formed a
significant fraction of their stellar  mass by $z \gsim 2$.
Conversely, the observed clustering properties  of galaxies, the steep
faint--end slope of the local luminosity  function (\eg, Folkes \etal\
1999, and references therein),  and evidence for the relative paucity
of massive galaxies  at $z > 1$ \citep{cow96,kau98} have been cited to
support  the hierarchical picture.

One important and unresolved issue is the stellar population content
of LBGs, and their stellar and dark matter masses.  Steidel and
collaborators (\eg, Steidel \etal\ 1996b; Pettini \etal\ 2000) have
found that the spectra of LBGs are broadly consistent with ongoing
star formation, and are similar to those of local starburst galaxies.
\citet{gia96} and \citet{low97} have emphasized the compact sizes  of
LBGs, with half--light radii of only a few kpc.  The strong clustering
observed for LBGs favors relatively large dark matter  halo masses
\citep{gia98,ade98}.  These observations are consistent with
simulations (\eg, Governato \etal\ 1998; Baugh \etal\ 1998) that
predict that LBGs evolve to become the bulges  and spheroidal
components of present--day, high--mass ($\sim \lstar$) galaxies,
preferentially in dense environments.  In contrast, \citet{low97}
argued that if $z \sim 3$ LBGs were to fade without continued star
formation or subsequent mergers, then their sizes and luminosities
would ultimately resemble those of present--day, low--mass dwarf
elliptical/spheroidal galaxies.  They suggest that the LBGs are
possibly low--mass objects, undergoing intense, brief periods  of
intense star formation, which place them above the magnitude  limit
and within the color--selection criteria.  Observations of
nebular--emission--line widths for LBGs find typical values in the
range $\sigma \simeq 60-120$~\kms\ \citep{pet98,pet01,tep00a,moo00}.
Combined with the sizes for the galaxies, these line widths suggest
virial masses $\sim 10^{10}\Msol$, although it is unclear whether such
measurements really trace the total mass of the dark matter halos of
these galaxies, which may extend well beyond the limits of the  star
forming, UV--bright, line emitting region.

Optical photometry has shown that the rest--frame ultraviolet (UV)
continua from LBGs are typically redder than is expected from ongoing,
unreddened star formation, suggesting the presence of dust (\eg,
Meurer \etal\ 1997; Dickinson 1998; Sawicki \& Yee 1998; Meurer,
Heckman, \& Calzetti 1999; Steidel \etal\ 1999; Adelberger \& Steidel
2000).  Nebular--line measurements for LBGs, mentioned above, have
sometimes yielded estimates of star--formation rates (SFRs) that are
higher than those derived from the UV continuum measurements  (but see
Pettini \etal\ 2001), again suggesting some amount of  dust extinction
($A_{1700} = 1-2$~mag).

Most photometric studies of LBGs have focused  on their rest--frame UV
light, which is dominated by short--lived, massive  stars.  Limited to
this portion of the spectral energy distribution (SED), it is not
possible to disentangle effects of metallicity, age,  extinction, and
past star--formation history.  By extending the photometric  baseline
to rest--frame optical wavelengths using near--infrared (NIR) data,
we may measure light from both short-- and longer--lived stars, and
may hope  to constrain the stellar population mix in these objects.
For all but the youngest galaxies, the optical rest--frame spectrum
has a strong contribution from the light of later--type,
main--sequence stars  (A--type and later) and evolved red giants, and
we may therefore expect  that NIR measurements will be more sensitive
to the total stellar mass,  which is dominated by lower--mass stars.
Also, the longer wavelengths  are less affected by dust extinction
that may strongly attenuate ultraviolet light.

\citet{saw98} have studied the SEDs  of 17 LBGs from the Hubble Deep
Field North (\hdf, Williams \etal\ 1996)  using optical WFPC2
photometry augmented by ground--based infrared $JHK_s$  measurements
from data obtained by Dickinson and collaborators  (cf.\ Dickinson
1998).  They concluded that these objects are generally  quite young
(median age $\sim$ 25~Myr) and highly reddened [median  extinction
$\ebv \simeq 0.3$, $\extinctA \simeq 3.2$~mag].  Based on  the derived
SFRs and timescales, they concluded that  the light from LBGs is
dominated by short bursts of star formation  and will likely only
produce $\sim 5$\% of the total stellar mass  of a present--day
\lstar\ galaxy.  However, the \hdf\ LBGs are  typically quite faint,
and the available ground--based infrared  data provide measurements
with rather low signal--to--noise  ratios ($S/N$).  Additionally,
there are special challenges to matching  the ground--based and WFPC2
photometry given the different resolutions and $S/N$ of the images.

In this paper, we investigate the properties of a sample of \NLBG\
LBGs with spectroscopically confirmed redshifts $2.0 \leq z \leq 3.5$
drawn from the \hdf, using new data from the Near--Infrared Camera and
Multiobject Spectrograph (NICMOS, Thompson \etal\ 1998) on board the
\textit{Hubble Space Telescope} (\hst). In \S2, we present the
broad--band  photometry used for the ensemble of objects.  In \S3, we
compare the LBG  SEDs to empirical spectral templates for various
galaxy types.  In \S4,  we constrain the stellar population mix in
these objects by fitting the observed SEDs with stellar population
synthesis models.  In \S5, we discuss the implications for the
galactic stellar masses and galaxy formation/evolution scenarios from
the best--fit models. Finally, in \S6, we present the conclusions.
Throughout this paper, we use a cosmology with $\Omega_M = 0.3$,
$\Lambda = 0.7$, and  $h = (H_0 / 100$~\kms\ Mpc$^{-1}) = 0.7$.  At
the typical redshift $z \approx 2.7$ for the galaxies studied here,
masses and SFRs derived from our model fitting procedure would be
$\sim50$\% smaller in an Einstein--de~Sitter universe with the same
Hubble constant,  and $\sim 10$\% smaller for an $\Omega_M = 0.3$
universe with  $\Lambda = 0$.  We present all magnitudes in the AB
system,  $m_\mathrm{AB} = 31.4 - 2.5\log(f_\nu/1\,\mathrm{nJy})$.   We
will denote galaxy magnitudes from the WFPC2 and NICMOS bandpasses,
F300W, F450W, F606W, F814W, F110W and F160W as \wfu, \wfb, \wfv, \wfi,
\nicj, and \nich, respectively.


\section{The Data and the Lyman Break Galaxy Sample
\label{section:data}}

The original \hdf\ WFPC2 observations provide high--quality photometry
of faint galaxies in four bandpasses from 0.3--0.8~\micron.  At $z
\gsim 2$,  the Lyman limit is shifted into the WFPC2 F300W filter,
permitting color selection of UV--bright star--forming galaxies with
$2 \lsim z \lsim 3.5$.  The WFPC2 images sample rest--frame UV
wavelengths at these redshifts  ($\lambda_0 \sim 1000-2000$~\AA), and
are thus primarily sensitive  to light from massive young stars
modulated by dust extinction. To study the photometric properties of
the LBGs at rest--frame optical  wavelengths, where longer--lived
stars may contribute to the luminosities, where dust extinction is
less severe, and where the most is known about  the low--redshift
galaxy population, we must observe in the near infrared.  This was one
of the primary justifications for our NICMOS survey of the  \hdf, in
which we mapped the complete WFPC2 field of view at 1.1~\micron\ and
1.6~\micron.

The NICMOS data reach rest--frame optical wavelengths $\lambda \geq
0.4$~\micron\ only for galaxies with $z \leq 3$.  Observations at
still  longer wavelengths are desirable for many reasons.  For
galaxies at $2 \lsim z \lsim 3.5$, the \ks--band  (2.16~\micron)
samples the light from  $\lambda_0 \sim 5000-7500$~\AA, i.e., well
into the optical rest frame, providing a better ``lever arm'' to
measure the properties of lower mass, longer--lived stars.  Deep
observations with  \hst\ longward of $\sim 2$~\micron\ are impractical
due to the warm telescope  assembly.  Therefore, we extended our data
to the \ks--band using observations with the Infrared Imager (IRIM) at
the KPNO 4m Mayall telescope \citep{dic98,dic01}.   Using a technique
developed by \citet{fer99}, we have analyzed the \ks\ image to
optimally  extract photometry matched to the WFPC2  and NICMOS data.
The details of  this method will be presented elsewhere (Papovich \&
Dickinson 2001),  but in summary, we use the \nich\ data to create
two--dimensional image templates for each object, convolve these to
match the point spread function (PSF) of the ground--based data, and
finally scale and fit the convolved templates to the \ks--band image
to extract fluxes.  This method  eliminates concerns about PSF and
aperture matching effects on the relative  photometry between the
\hst\ and ground--based images, and permits deblending  of images for
objects partially merged by the  ground--based seeing.   We have
performed detailed simulations to test the reliability of this
technique and to understand the distribution of  flux uncertainties.
In both cases where the objects are well detected  or are below the
flux limit of the ground--based image, the method measures robust
fluxes or upper limits.  However, most of the LBGs used  in this work
are well detected at \ks.  Although the \ks--band data are  not as
deep as the \hst\ data, they provide our only access to the
rest--frame optical wavelengths for $z \gsim 3$ galaxies.

A detailed discussion of the infrared observations and photometric
catalogs  will be presented in detail elsewhere (Dickinson \etal\
2001, see also  Dickinson 1999, 2000);  a summary of the data used
here is given in  Table~\ref{table:hdfdata}.  Briefly, we resampled
the NICMOS, WFPC2,  and IRIM images to the same plate scale
(0\farcs08~pixel$^{-1}$) and  convolved the WFPC2 and \nicj\ images to
match the PSF of the \nich\  image.  Our tests indicate that this PSF
convolution matches point  source photometry between bands to within
5\% for aperture radii  $> 0\farcs1$.  Object detection and photometry
were done with SExtractor  \citep{ber96} on a combined F110W + F160W
image.  Relative photometry from  the \hst\ WFPC2 and NICMOS data was
measured through matched isophotal  apertures defined on the NICMOS
images.  The isophotal apertures were used  for the relative colors
because our tests indicate that these most closely  match the \ks\
photometry derived using by the template fitting method  (see Papovich
\& Dickinson 2001).  In order to correct for galaxy light outside  the
isophotal apertures, we then scaled the isophotal photometry for  each
object by the ratio of the flux within an elliptical aperture  defined
by radial moments of each galaxies' F160W light profile  (i.e., the
SExtractor ``MAG\_AUTO'' measurements) to the isophotal flux.  The
flux correction was $< 20$\% for nearly all of the galaxies, with  a
median value of 4.4\%.

In order to ensure that the NICMOS and \ks\ photometry reach  optical
rest frame wavelengths, we limit our galaxy sample to  a redshift
interval similar to that selected by the ``\wfu--dropout''  Lyman
break criteria.  Extensive spectroscopy from the Keck telescope  has
measured redshifts for a subset of \hdf\ LBGs  (mostly from Steidel
\etal\ 1996, Lowenthal \etal\ 1997, Dickinson 1998,  and Cohen \etal\
2000).  In order to remove the redshift as a variable  when fitting
and analyzing the spectral energy distributions of the  galaxies, we
restrict our primary analysis to those galaxies with  spectroscopic
redshifts in the range $2.0 \lsim z \lsim 3.5$.  These objects are
listed in Table~\ref{table:lbgdata}.   In several cases, our catalogs
have split objects into separate entries that might be regarded as a
single object with complex, multiple structure.  For one object (HDF
2--239, using the catalog number from Williams \etal\ 1996) we have
merged two very faint components with the brighter, main body of the
galaxy.  In general, however, we have left the pieces separate.  In
several cases, the pieces have notably different  colors, and thus
independent analysis may be instructive.   In \S4.3 we note that two
galaxies have SEDs that seem  inconsistent with their reported
spectroscopic redshifts, and thus disregard them from the fitting
analysis.  We also exclude one other object, HDF 4-852.12 (NIC 824),
which is apparently a broad--line AGN at $z = 3.479$ \citep{coh01}.

\begin{figure*}[th]
\epsscale{1.5}
\plotone{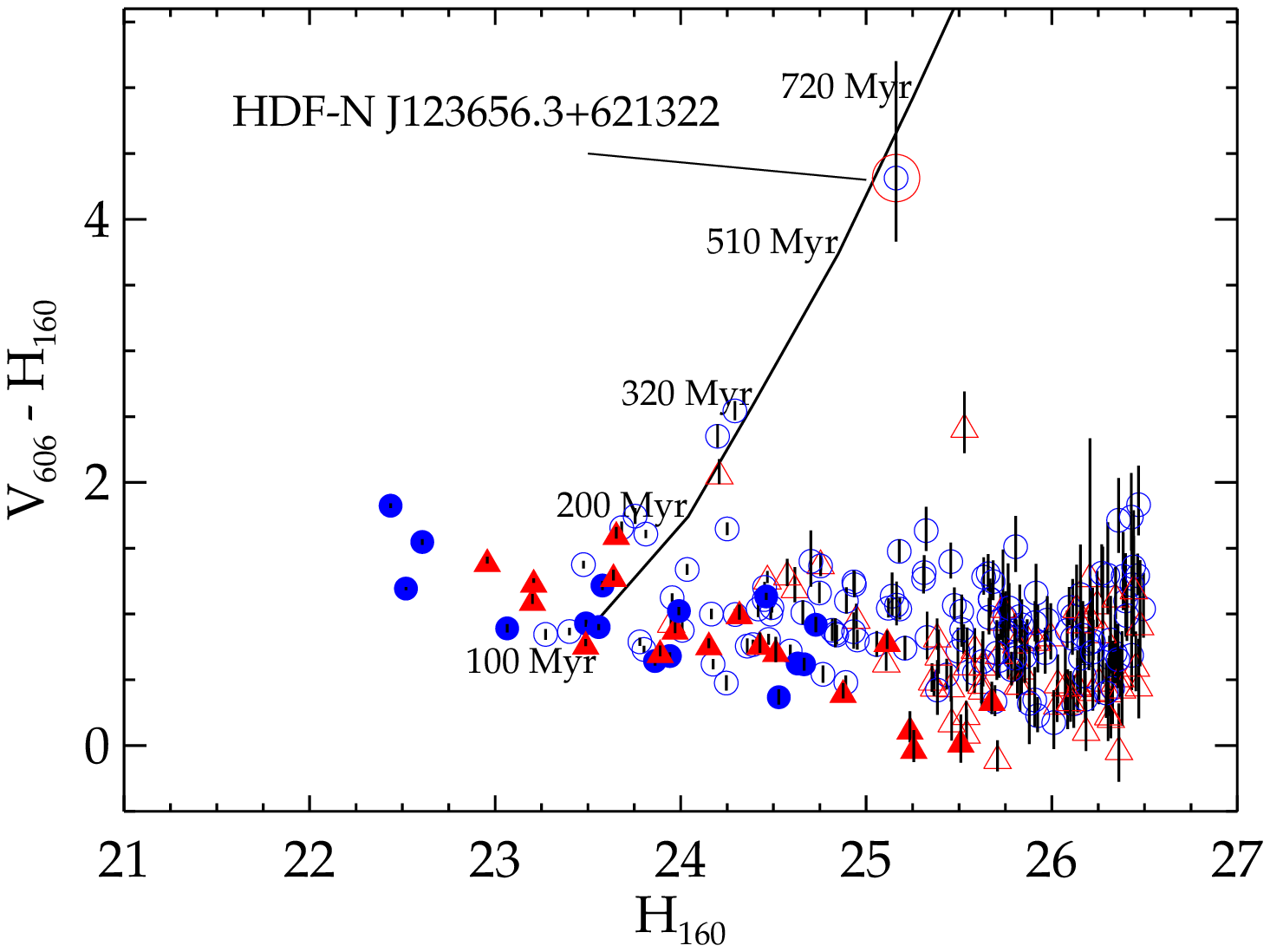}
\caption{Color--magnitude diagram for galaxies at $2 < z < 3.5$  in
the \hdf.  The redshift distribution has been divided into two
subsamples: $1.9 < z < 2.7$ (blue circles), and $2.7 < z < 3.5$
(red triangles).  Filled symbols show the spectroscopically confirmed LBGs
in our sample, while open symbols are \hdf\ galaxies with photometric
redshifts $1.95 < z_\mathrm{phot} < 3.50$ \citep{bud00,dic01}.    The
line shows the color--magnitude evolution as a function of age
(labeled) for a $\mathM = 10^{10}$~\msol\ stellar population  with a
$\delta$--function star--formation history, and observed  at $z = 2.7$.
There are few photometric candidates for red,  non--star--forming HDF
galaxies at these redshifts.  One possible candidate, the
``$J$--dropout'' object \hdf~J123656.3+621322  is indicated.  The
$S/N$ for the $\wfv$ measurement on this object is $< 2$, and thus the
$\wfv - \nich$ color is better considered  as a lower limit.  Although
its nature and redshift are unknown, this object might plausibly be an 
example of a non--star--forming or dust--obscured galaxy in this redshift 
range \citep{dic00a}.}
\label{fig:cmd-vmh}
\end{figure*}

Because all the objects in our subsample are bright enough for
spectroscopy, they represent only the bright end of the luminosity
function.  They therefore may be a biased subsample, either by color
(i.e., the most UV--bright and actively star--forming galaxies) or
possibly by mass.  However, it is important to remember that the \hdf\
LBGs are on average less luminous than typical LBGs in  the
ground--based samples of Steidel and collaborators.  In
Figure~\ref{fig:cmd-vmh}, we show a color--magnitude diagram for  all
\hdf\ galaxies selected from the NICMOS F110W+F160W data with  $\nich
< 26.5$ and with either spectroscopic  or photometric redshifts (the
latter from Budav\'ari \etal\ 2000, using the NICMOS and \ks\ infrared
photometry) in the range $1.95 < z < 3.5$.  In principle, the
infrared--selected photometric redshift sample could identify red
galaxies with little or no UV flux that might be missed in optically
selected Lyman break samples.  However, as noted by \citet{dic00b},
there are actually very few photometric candidates for galaxies in
this redshift range that would {\it not} be identified in an optically
selected sample.  The color--magnitude distribution shows most \hdf\
galaxies in this redshift range have broadly similar $\wfv-\nich$
colors, and that the spectroscopically confirmed LBGs span the range
of $\wfv-\nich$ colors observed for the large majority of objects in
the photometric redshift sample.  Figure~\ref{fig:cmd-vmh} includes a
line indicating the colors and magnitudes expected as a function of
age at $z = 2.7$  for a stellar population formed in a single burst
with a stellar  mass $10^{10}~\msol$.  There are a few galaxies
$\wfv-\nich$ colors slightly redder than the majority of the LBGs (but
which are still well detected in the WFPC2  data), including
interesting objects such as the radio and X--ray  source
J123651.7+621221 (Richards \etal\ 1998, Dickinson \etal\ 2000,
Hornschemeier \etal\ 2000, Brandt \etal\ 2001), with photometric
redshift $z_{phot} \approx 2.6$.  One very red object, the so--called
``J--dropout'' J123656.3+621322, is only marginally detected  in the
WFPC2 images and might plausibly be a reddened or  non--star--forming
galaxy at $z > 2$, or more speculatively an object at $z > 10$
(Dickinson \etal\ 2000).  However, with this possible exception, of
the $\approx 200$~galaxies with (photometric or spectroscopic)
redshifts, $2.0 \lsim z \lsim 3.5$, there are  no other good
candidates for relatively massive, non--star--forming HDF galaxies at
$z > 2$ and with ages $\gsim 0.3$~Gyr.

\begin{figure*}[th]
\epsscale{1.4}
\plotone{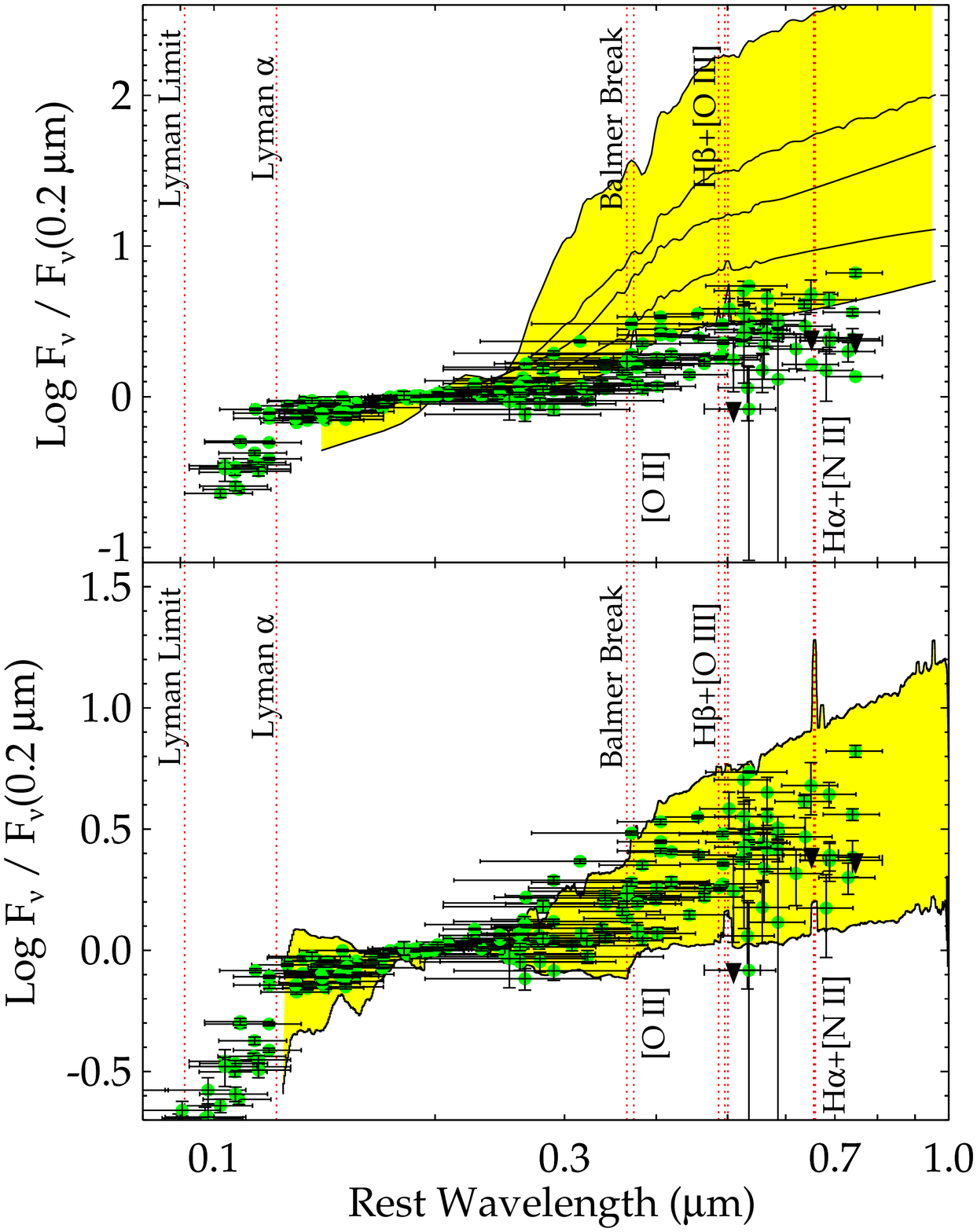}
\caption{Ensemble photometry of the \NLBG\ LBGs compared to empirical
galaxy templates.   The green data points are the measured LBG fluxes
from the WFPC2 (\wfu,\wfb,\wfv,\wfi), NICMOS (\nicj,\nich), and IRIM
(\ks) images, normalized to common flux density at
$F_{\nu}(2000\;\AAA)$ (black triangles indicate upper limits).  Top
panel:  The yellow shaded region is defined by spectral templates for
ordinary galaxies from \citet{col80}, spanning the range from the
earliest types (elliptical), through Sab, Sbc, and Scd spirals, to the
bluest, Magellanic Irregular (Im) templates.  While the Im template is
similar to the average LBG, the earlier--type, Hubble sequence
galaxies are all much redder than any of the LBGs.  Bottom Panel:  The
yellow shaded region spans the range of empirical  starburst galaxy
SEDs from \cite{kin96}, from the bluest  [NGC 1705, $\ebv \simeq 0.0$]
to reddest [$0.6 < \ebv < 0.7$]   templates.  Common spectral features
are indicated by dotted red lines.  The LBG SEDs fit comfortably
within the envelope defined by the starburst templates.
\label{fig:ensemble}}
\end{figure*}

We note a color--magnitude trend in Figure~\ref{fig:cmd-vmh}, in the
sense that the brightest galaxies at \nich\ tend to have  redder
optical--infrared colors.  Because this is an infrared--selected,
complete sample, there is no reason to think that this is due to any
selection effect.  To test the robustness of this correlation,  we
performed a Monte Carlo boot--strap test by randomly  reassigning
colors  from the $\wfv-\nich$ distribution (with substitution) to the
$\nich$ magnitudes.  We find that the observed trend is randomly
reproduced with a linear slope as steep or steeper in only $\simeq
0.03$\%\ of the Monte Carlo simulations, which supports a strong
likelihood that this correlation is not obtained from random sampling.
This trend might suggest  that the most luminous galaxies in this
redshift range are comparatively ``older'' (i.e., have lower ongoing
SFRs for their total mass), or that they are dustier.  A similar
color--magnitude trend has been noted for the UV light from HDF LBGs
(Meurer \etal\ 1999),  and interpreted as a measure of varying dust
content and/or metallicity.


\section{Comparisons to Empirical Galaxy Spectra
\label{section:empirical}}

The \hdf\ Lyman break galaxies have significantly bluer colors than
those of local spiral and elliptical galaxies.  In
Figure~\ref{fig:ensemble},  we compare the SEDs of the \hdf\ LBGs with
empirical UV--to--optical spectral templates for local galaxies from
Coleman, Weedman, \& Wu 1980  (henceforth CWW).  The LBG SEDs are
uniformly bluer than even the actively star--forming Scd spiral galaxy
template from CWW, and most are bluer than  even the CWW Magellanic
Irregular SED.  The CWW Scd model would predict  $\wfv-\nich \gsim 2$
for $2 \lsim z \lsim 3.5$, redder than all but a very few  of the
NICMOS--selected galaxies in either the spectroscopic or photometric
redshift based samples shown in Figure~\ref{fig:cmd-vmh}.   Even for a
sample selected from deep infrared images, it appears that nearly all
\hdf\ galaxies at $2 \lsim z \lsim 3.5$ have specific SFRs (i.e., the
instantaneous, ongoing rate of star formation relative to the
integrated past star formation or total stellar mass) that are higher
than those of present--day Hubble Sequence galaxies.  Such rapid star
formation is more characteristic of that in local starburst galaxies,
which are not represented among the CWW models.

In Figure~\ref{fig:ensemble} we also compare the LBG photometry to six
empirical starburst galaxy templates \citep{kin96} with varying dust
extinction ranging from $\ebv \simeq 0.0-0.7$.  The LBGs all fall
within  the envelope defined by these templates, which demonstrates
that  their UV--optical SEDs are broadly similar to those of local
starburst  analogs.  The UV spectra are generally redder than those of
the unreddened  starbursts, and there is also a spectral inflection
around the Balmer  break region that indicates a strong contribution
to the SED from  longer--lived stars (A--type and later).

\begin{figure*}[th]
\epsscale{1.0} 
\plotone{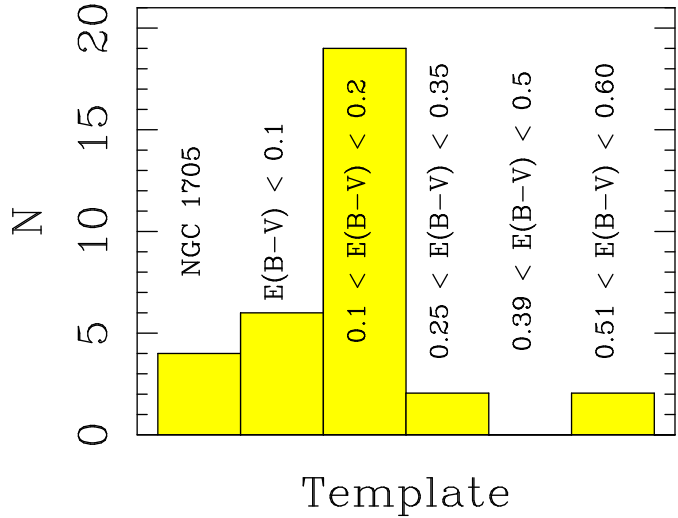}
\caption{Distribution of the starburst galaxy template types
\citet{kin96} that most nearly match galaxies in the LBG sample.  The
columns correspond to each starburst template, as labeled.   The first
column refers to the very blue starburst galaxy,  NGC 1705 [$\ebv
\simeq 0.0$].  The next five columns are  labeled by the range of
reddening values that correspond to  the averaged starburst templates.
\label{fig:kinney_dist}}
\end{figure*}

We also fit all the Kinney \etal\ galaxy templates (five quiescent
galaxy types -- Elliptical, S0, Sa, Sb, Sc -- six reddened starbursts,
and the  spectrum of the very blue, essentially unreddened starburst
NGC 1705)  to the SED of each LBG.  The distribution of the best
fitting templates is shown in Figure~\ref{fig:kinney_dist}.  The
photometry for most of the LBGs, 19/\NLBG\ ($\simeq 58$\%), are best
fit by starburst templates  with a moderate level of dust extinction,
$\ebv \sim 0.1 - 0.2$.  Of the other LBGs in the sample, 10/\NLBG\ of
the galaxies are well fit by essentially unreddened starburst models
or the extreme blue template for NGC 1705.  Only 4/\NLBG\ cases have
best fit templates with large color excesses, $\ebv \gsim 0.25$.  None
of the LBGs were well fit by the quiescent galaxy templates.  Based on
these results, therefore, the typical LBG SED is broadly  similar to
local starburst galaxies with modest amounts of the dust  extinction
--- typical UV continuum suppressions at 1700~\AA\ on the  order of,
$\extinctA \sim 0.5-1.0$~mag.\footnote{Kinney \etal\ derive the
reddening in their starburst sample from the Balmer--line decrement,
the ratio of \hb--to--\ha\ fluxes, which is a measure of the color
excess in the nebular--emission regions, \ebv$_g$.  Calzetti, Kinney,
\& Storchi--Bergmann (1994) and Calzetti \etal\ (2000)  have shown
that there is a scaling relationship between this and the effective
color excess of the stellar continuum [denoted as \ebv$_s$].
Therefore, the color excesses of the Kinney \etal\ starburst templates
correspond  to a suppression of the UV continuum at 1700~\AA\ by
$\extinctA \simeq 4.6 E(B-V)_g$~mag.}  These values are somewhat lower
than those derived from other methods \citep[see also
\S5.1]{saw98,meu99,ste99}.

\begin{figure*}[bh]
\epsscale{1.0}
\plotone{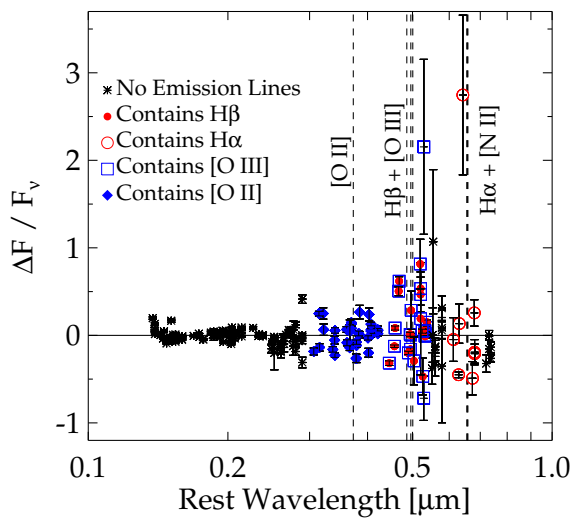}
\caption{Residuals between the observed LBG photometry and the
best--fitting starburst galaxy templates.  The data points show the
fractional difference between the data and model versus the
rest--frame wavelength, and error bars indicate the observed flux
uncertainty.  Each data point is coded to indicate whether the
bandpass spans a prominent emission line, as indicated in the figure.
\label{fig:lines_scatter}}
\end{figure*}

One possible concern with the infrared LBG photometry might be the
presence of a significant flux contribution from nebular emission
lines, arising from heated and shocked gas associated with regions of
star formation  (the most common being \fion{O}{2}, \fion{O}{3},
\ha+\fion{N}{2}, and \hb,  which are labeled in
Figure~\ref{fig:ensemble}).  Very strong emission lines can enhance
the flux in a given bandpass and alter the galaxy colors.  In several
instances, strong nebular emission lines have been known to
substantially affect broad--band infrared magnitudes and colors for
high redshift radio galaxies (e.g., Eisenhardt \& Dickinson 1992,
Eales \etal\ 1993), although generally radio galaxies have much
stronger line emission than do ordinary, star--forming galaxies.  To
investigate their contribution to the integrated fluxes, we consider
the effect of an emission line with equivalent width $W \equiv F_{\rm
line} / F_\lambda^c$, where $F_{\rm line}$ is the integrated flux in
the emission line and $F_\lambda^c$ is the flux density of the
continuum.  Assuming a (roughly) flat continuum, the average flux
density measured through a bandpass is approximately, $F_\lambda
\simeq F_\lambda^c + F_{\rm line}\, \Delta\lambda^{-1}$, where
$\Delta\lambda$ is the width of of the bandpass.  This can be
reexpressed as
\begin{equation}
\frac{F_\lambda}{F_\lambda^c} \simeq 1 + \frac{W_0
(1+z)}{\Delta\lambda},
\label{eqn:EW}
\end{equation}
where we substitute the rest frame equivalent width $W_0 =
W(1+z)^{-1}$.   Therefore, an emission line with rest frame equivalent
width $W_0$ introduces a magnitude increase of
\begin{equation}
\Delta m \simeq -2.5 \log\left[ 1+\frac{W_0 (1+z)}{\Delta\lambda}
\right].
\end{equation}
For galaxies in the redshift range of interest here, the optical
nebular lines pass through the NICMOS and \ks\ filters.  The range of
nebular emission lines observed in both the Kinney \etal\ local
starburst templates and the \citet{pet01} spectroscopic observations
of $z\sim 3$  galaxies show that the equivalent width of \fion{O}{3}
can be as high as  $W_0 \sim 200$~\AA.  For the redshifts considered
here, this corresponds to an increase in the observed broad--band
fluxes of $\Delta m \simeq 0.2$~mag to the \nich-- or \ks--bands.

We investigated the emission--line contamination by examining the
residuals in the $\Delta F_i/ f_{\lambda_i}$ distribution, where
$\Delta F_i = f(\lambda_i) - s_i T_j(\lambda_i)$ is the difference
between the observed flux through a bandpass, $i$, and the
``best--fitting''  Kinney \etal\ template, $j$.  In
Figure~\ref{fig:lines_scatter}, we show this distribution for the LBG
sample.   Each data point is coded to indicate if the bandpass
contains one of  the strong emission lines, \fion{O}{2}, \fion{O}{3},
\fion{N}{2}, \ha, or \hb.  Properly accounting for the photometric
measurement uncertainties  (which are generally larger for the
infrared data), the scatter for  bandpasses spanning one or more
strong emission lines is equivalent  to that for bandpasses devoid of
emission lines.   To test this further, we refit the observed LBG
photometry to the Kinney \etal\ templates with the  major emission
lines removed by interpolating across those wavelength intervals (\ie,
\fion{O}{2}, \fion{O}{3}, \hb, \ha, and \fion{N}{2}).  The strongest
emission lines in the local starbursts have similar equivalent widths
to the largest values measured for LBGs by \citet{pet01}.  The
resulting distribution of best-fitting spectra is essentially
unchanged with respect to the previous results.  Unless the
nebular--line strengths for the HDF LBGs are significantly larger than
those measured for other LBGs by Pettini \etal, it seems unlikely that
the emission lines have a significant effect on the spectral template
fitting.


\section{Comparisons with Stellar Population Synthesis Models
\label{section:models}}

Although we have seen that the spectral energy distributions of LBGs
are broadly similar to those of present--day starburst galaxies, we
must  refer to models in order to characterize their stellar content
and star  formation histories.  Here we compare the LBG
spectrophotometry to stellar  population synthesis models to
investigate constraints on the galaxy  ages, star--formation
histories, dust extinction, and stellar content.   Because our
photometry spans a broad range of rest--frame wavelengths
(0.08--0.6~\micron\ at the mean redshift of the sample),  we are
sensitive to stellar population components with a wide range of ages
and mass, and to different aspects of the galaxies' star  formation
histories.  \citet{meu99} have demonstrated that the UV spectral slope
(1000-4000~\AA) in nearby starburst galaxies correlates with the ratio
of their far--IR to bolometric luminosities, and hence the degree of
extinction for star--forming regions.  The UV--to--optical flux ratio
(\eg, the inflection around the Balmer/4000~\AA\ break) gauges the
ratio of early--type (OB) to late--type (A and later) stars, which is
a diagnostic of the past star--formation history.

To generate model galaxy spectra, we used the newest version  of the
Bruzual \& Charlot (1993) stellar population synthesis code (Bruzual
\& Charlot, in preparation, BC2000).  The BC2000 models generate
spectra of integrated stellar populations for a specified metallicity,
initial mass function (IMF), and star--formation history as a function
of the population age.  This technique represents the galaxies'
stellar content as the sum of a series of isochrones as a function of
time for the specified star--formation history,  using theoretical
evolutionary tracks to predict effective  temperatures and
luminosities for stars as a function of mass and  age.  To compute the
emergent spectrum, the BC2000 code uses stellar--spectral libraries,
summing over the distribution of stars present at each time step.  For
a further description of the BC2000 models and references, see
\citet[and references therein]{bru00}.

We will initially consider stellar populations formed by a single,
continuous episode of star formation whose rate decays exponentially
with a characteristic timescale, which we vary as a free parameter.
The ``age'' of the galaxy is therefore defined as the time since the
onset of the formation of the stellar population that dominates  the
observed SED.  The LBG star--formation histories might, in principle,
be significantly more complex than these monotonically evolving
parameterizations, due to discrete events such as mergers, tidal
interactions, shocks, or gas accretion.  However, to model such
stochastic processes would require one to invoke an infinite number of
discrete, possible, star--formation histories.  The choice of more
complex star--formation histories would not greatly affect many of the
conclusions here, which pertain to the nature of the stars that
dominate the observed light in the ultraviolet and optical rest frame.
Our definition of galaxy age depends to some degree on the form of the
star--formation history, because stars from earlier (hence older)
star--formation episodes may remain undetected (see
\S\ref{section:oldstars}).  The exponential, star--formation histories
provide a lower bound on the  stellar masses of the LBGs, because they
include possible models with very young ages, very active star
formation,  and hence the smallest values of $\mathM/L$.  Thus, we
should consider  the stellar--mass estimates from these simple models
strictly as \textit{lower} limits.  Below, in
\S\ref{section:oldstars}, we will  consider the effects of a
two--component star--formation history,  in which the first component
represents the simple, monotonic, exponentially decreasing
star--formation histories as above, and the second results from a
stellar population formed in an instantaneous  burst at $z = 1000$.
Such a stellar component has a maximal  $\mathM/L$ by design, and as
such, we argue that the total stellar--mass estimates from these
models represent a conservative \textit{upper} bound on the true
stellar mass of the LBGs.  Therefore, using these two sets of
star--formation histories, we are able to bound the possible range of
the total stellar masses of the LBG stellar populations in this sample.

\subsection{Model Parameters}

We generated a suite of population synthesis models spanning a wide
range  of the physical parameter space of age, star--formation
timescale,  metallicity, IMF, and dust extinction.    We allowed for
several forms of the IMF \citep{sal55,mil79,sca86}, all of which adopt
lower and upper mass cutoffs at 0.1 and 100~\Msol.  We parameterized
the star--formation history by adopting an instantaneous SFR of the
form, $\Psi(t)\sim \exp (-t/\tau)$, where $t$ is the time since the
onset of star formation, and $\tau$ is the characteristic
star--formation,\efolding\ time.  We considered values in the range
$\tau = 0.01 - 10$~Gyr, in quasi--logarithmic intervals, as well as a
constant star--formation model (in effect, $\tau \approx \infty$.)  We
tracked the time evolution of the model spectra in logarithmic
intervals for ages from $0.126$~Myr to $20$~Gyr.

The BC2000 models offer the choice of two sets of stellar spectral
libraries.  One set, primarily based on Kurucz model atmospheres (see
Lejeune, Cuisinier, \& Buser 1997), covers a wide range of
metallicities,  $Z\simeq 1/200-5$~\Zsol.   Another set uses empirical
stellar spectra from the compilation of \citet{pic98}, but is
available only for solar  metallicity.  We compared results from the
solar metallicity model  atmosphere libraries to those from the
empirical spectral libraries,  and found essentially no change in the
results when fitting the  broad--band LBG photometry.

The metallicities of LBGs are only loosely constrained from current
data.  Optical and infrared spectra of the  gravitationally lensed
$z=2.7$ galaxy cB--58 \citep{pet00a,tep00a} suggest values $\sim 1/4 -
1/3$~\Zsol, and NIR spectra of several other LBGs suggest values
consistent with this \citep{tep00b,pet01}.  This is higher than
measurements for damped Lyman~$\alpha$ (DLA) systems at  similar
redshift, which generally have metallicities  $Z \lsim 0.1$~\Zsol\
\citep{pet97,pet00b}.  One might reasonably expect  higher
metallicities for LBGs than for DLA systems at the same redshift
because the former are sites of active star formation, which may
rapidly enrich the surrounding medium.  Therefore, we have used the
stellar  atmosphere models so that we were able to investigate the
effects  of varying metallicity on the model fitting results.
However,  it is important to remember that the theoretical stellar
atmospheres  at sub--solar metallicities are relatively untested by
empirical  data.  \citet{lei01} have recently created a library of
\hst\  ultraviolet spectra of LMC and SMC O--type stars, and find a
fairly  weak metallicity dependence for some UV line indices, but
there  is not yet much UV spectral information for later--type stars,
nor data on the overall UV--to--optical SEDs  of composite stellar
populations at various metallicities.   In \S\ref{section:discussion}
below we will note various ways  in which the results of our parameter
fitting depend on the model metallicities.    These should be regarded
with some caution, due to the current dearth of empirical calibration
for the low--metallicity models.

We included dust extinction using two different prescriptions:  the
\citet{cal00} attenuation curve inferred from local starburst
galaxies, and the \citet{pei92} parameterization of the extinction
curve for the Small Magellanic Cloud (SMC).   The empirical Calzetti
\etal\ starburst attenuation curve has a much ``greyer''  wavelength
dependence in the near ultraviolet than does the SMC law.  In part,
this is believed to be due to geometric effects resulting from  a
mixed distribution of stars and dust in extended galaxies.    The
classical stellar reddening curve for the SMC represents  a
combination of absorption and scattering terms.  For a galaxy, photons
that are scattered but not absorbed will eventually emerge.
\citet{pei92} provides separate parameterizations for scattering and
absorption in the SMC extinction curve, and it might be appropriate to
use only his absorption terms.  Instead, we have retained the full SMC
extinction curve with both scattering and absorption  terms, precisely
because it differs most substantially from the  Calzetti relation.
Our motivation is to diagnose the degree to  which our results for LBG
fitting depend on the extinction model.   In this case, the steeper UV
extinction of the classical SMC law provides a useful test.  In
reality, however, we might expect dust extinction in LBGs to behave
more nearly like the local starburst examples.  Often, therefore, we
will focus  of our attention on results derived using the
\citet{cal00} starburst extinction law.\footnote{Note that
\citet{cal00} slightly updates the attenuation parameters previously
presented in \citet{cal94} and \citet{cal97}.  In particular,
\citet{cal00} revises the ratio of total to selective extinction,
$R_V$, to $4.05\pm0.80$, from 4.88 in \citet{cal97}.} In the
discussion that follows,  we parameterize the effects of dust as the
attenuation, $A_\lambda = \ebv k(\lambda)$, in units of magnitudes,
at some wavelength, $\lambda$.   At our UV reference wavelength,
1700~\AA,  $k(1700$\AA$) = 9.65$.

\subsection{Fitting the Synthesis Models to the LBG Spectrophotometry}

To fit the population synthesis models to the LBG photometric data, we
converted the synthetic spectra to bandpass--averaged fluxes.  The
BC2000 code generates synthetic spectra as specific luminosity
density, $\ell_\lambda$, in units of solar luminosity per Angstrom
per unit solar mass.  We transformed these to luminosity per unit
frequency and applied attenuation for dust extinction,
\begin{equation}
L_\nu(\lambda_0,t,\tau,A_{\lambda_0},\mathM ) = 
  \frac{\lambda_0^2\, \mathM }{c\, m^\ast(t)}\;\ell_\lambda(\lambda_0,t,\tau)
   \, 10^{-0.4\, A_{\lambda_0}},
\end{equation}
where $\lambda = \lambda_0 (1+z)$  relates observed-- and  rest--frame
wavelengths, and $\mathM$ is the total galaxy mass in stars.   The
BC2000 model spectra are normalized so that their total mass (gas +
stars) is 1~\msol, and the stellar--mass fraction $m^\ast(t)$ is
provided for each  time step.  From this we obtain the flux density,
\begin{eqnarray}
F_\nu(z,\lambda,t,\tau,A_{\lambda_0},\mathM) =&(1+z)\,
  \frac{L_{\nu_0}(\lambda_0,t,\tau,A_{\lambda_0},\mathM)}{4\pi D_L^2(z)}
	\nonumber \\
  &\times \;e^{-\tau_\mathrm{IGM}(z,\lambda)},
\end{eqnarray}
where $D_L(z)$ is the luminosity distance (cosmology dependent), and
$\tau_\mathrm{IGM}(z,\lambda)$ is the wavelength dependent flux
suppression due to the \ion{H}{1} opacity of the IGM at wavelengths
shortward of \lya\ \citep{mad95}.   We multiplied each spectrum with
the (dimensionless) bandpass throughput function  $T_\nu$, which is
the total throughput in the sense that it includes the response from
the telescope, filter, and instrument optical assemblies.  Thus, we
obtain integrated synthetic, bandpass--averaged fluxes,
\begin{equation}
\langle F_{\nu} (z,t,\tau,\extinctA,\mathM) \rangle =  \frac{
 \int T_\nu  F_\nu(z,\lambda,t,\tau,A_{\lambda_0})\,
 d\nu/\nu}  {\int T_\nu \, d\nu/\nu},
\end{equation}
where we have parameterized the dust attenuation as magnitudes at
(rest--frame) 1700~\AA, $\extinctA$.

Once a model has been fit to the broad band galaxy photometry, its
star--formation rate $\Psi(t) = \Psi_0 \exp(-t/\tau)$ can be
determined  from the model normalization.   This is the
\textit{intrinsic} rate,  \ie,  the physical rate that the stellar
population model is forming  stars at the observed age $t$.  The dust
correction has already been  taken into account in the template
fitting procedure.

Due to the nature of the BC2000 models, only discrete values for the
IMF and metallicity are available, and so we hold these parameters
fixed at specific values.  Although we tested a wide range of values, 
$Z = 0.01 - 3.0$~\Zsol, and several forms of the IMF, we concentrate
our presentation here on models with $Z=0.2$, 1.0~\Zsol\ and  Salpeter
or Scalo IMFs.   These metallicity values are consistent with  the
limited observational data on LBGs (see \S4.1).  As we will see below,
the results of our analysis do not strongly favor any particular range
of metallicity for \hdf\ LBGs, although the choice of metallicity does
sometimes affect the parameters of the best--fitting models.   To
determine the best fitting models, we allow the variables $t$, $\tau$,
$\extinctA$, and $\mathM$ to vary as free parameters.  We generated a
suite  of $\sim 10^5$ synthetic models that span a wide range of the
physical  parameter space ($t\simeq 0.001-20$~Gyr, $\extinctA\simeq
0.0-6.7$~mag, $\tau=0.01-10$~Gyr and $\tau \rightarrow 1000$~Gyr).

We derived best--fit models for each LBG by computing a $\chi^2$
statistic for each  model.  We defined $\chi^2$ for each model set of
synthetic data points,
\begin{equation}
	\chi^2 = \sum_{i} 
	\frac{ \left[ f_\nu^{(i)} - 
	\langle F_\nu^{(i)}(z,t,\tau,\extinctA,\mathM) \rangle 
	\right]^2}
	{ \sigma^2(f_\nu^{(i)}) + \sigma_{\mathrm{sys},i}^2} , 
	\label{eqn:chi2}
\end{equation} 
where the $f_\nu$ are the observed photometric flux densities,  and
$\sigma(f_\nu)$ are the pure photometric uncertainties.  The
$\sigma_{\rm sys}$ terms represent systematic errors that encompasses
several uncertainties, which we discuss below.  Although photometry is
available through seven bandpasses for all objects
(\wfu\wfb\wfv\wfi\nicj\nich\ks), we always exclude the F300W
measurement because this point is well below the Lyman limit for  all
LBGs and suffers from severe IGM attenuation.  The relation  provided
by \citet{mad95} represents an average value of the  IGM attenuation
as a function of wavelength and redshift, but  cannot account for
variations in $\tau_\mathrm{IGM}(\lambda,z)$  along individual sight
lines due to the stochastic distribution  of absorbing clouds.  For
$z>2.5$, the F450W bandpass extends  significantly blueward of \lya,
where variations in the \lya\ forest density may also affect the
observed flux, and therefore we also  excluded that band when fitting
objects at those redshifts.  By excluding these bandpasses, we prevent
unknown statistical deviations in the IGM from influencing the fit.

In principle, the expected value of $\chi^2_0$ per degree of freedom
should be unity.  In practice, we found that the minimum $\chi^2$
values for the best--fitting models were generally larger than this.
This implied that either our models incompletely describe the true
SEDs of the LBGs, that the measurement uncertainties were larger than
the simple photometric errors, $\sigma(f_\nu)$, or that the parameter
error distribution is not Gaussian (or some combination of these
effects).  It is not surprising that the stellar population synthesis
models are incomplete, because we are forced to use discrete parameter
values for metallicity and IMF.   We are also restricting the range of
possible star--formation histories to simple exponential models.
Moreover, the population synthesis models might well  have systematic
inaccuracies (see \eg, Bruzual \& Charlot 1993,  Charlot, Worthey, \&
Bressan 1996, Bruzual 2000), and there are almost certainly
uncertainties or inherent variations in the extinction properties
(see Calzetti 1997, Calzetti \etal\ 2000).  Finally, it is not
unlikely  that we have underestimated the flux uncertainties for the
\hst\ photometry.   The values we have used here are based on the
measured pixel--to--pixel image noise, but do not account for
systematic errors in the measurements  as implemented by the
photometry software (e.g., issues of background  determination,
aperture positioning, etc.), as well as additional systematic terms
due to flat fielding errors, PSF mismatching, photometric zeropoint
uncertainties, etc.   Many of these sources of systematic error would
introduce flux deviations (either errors or differences relative to a
model) proportional  to the source flux itself.   Using the results of
image simulations that were done to analyze the errors and detection
efficiency in our HDF/NICMOS catalogs (see Dickinson \etal\ 2001), we
included an additional error term $\sigma_{\rm sys}/f_\nu \approx
3.5$\% for the \hst\ photometry.   The \ks--band photometry was
measured using a different method (\S\ref{section:data}) whose
uncertainties have been extensively tested and calibrated, and thus do
not include this extra term.   Even with this additional uncertainty,
the reduced $\chi^2$ values were still larger than one, and we
therefore included an additional  4\% systematic flux error for all
bands in order to force  the values of the best--fit $\chi^2$ per
degree of freedom to be of order unity.  In general, including these
additional uncertainty terms does not substantially change the
distribution of $\chi^2$ values in the model space or the results for
the best--fitting model, but it does affect the mapping of the
$\chi^2$ distribution to percentage confidence contours on  the fitted
parameters.  Overall, it has the net effect of  \textit{increasing}
the uncertainty on the model parameters.   Therefore, we expect that
the parameter constraints from our modeling are, if anything,
conservative.

\subsection{Constraints on the Stellar Population Parameters
\label{section:constraints}}

In Table~\ref{table:bestfits}, we list the parameters for the  stellar
population synthesis models that best fit the observed photometry for
each galaxy, \ie, the models with the minimum $\chi^2$.  Columns~1 and
2 list the NICMOS ID number and redshift, respectively.  For each LBG,
there are four best models (rows) that show the results for two
parameterizations of the IMF (column~4) and two metallicities
(column~3).  The best--fit parameters (population age, star-formation
\efolding\ time, \extinctA, and total stellar mass) are given in
columns~$5-8$.  All models listed in Table~\ref{table:bestfits} assume
the Calzetti \etal\ starburst extinction law, although we have also
fit SMC extinction models, and will refer to these models in the
following discussion.  The instantaneous SFRs associated with the
best--fit model is shown in column~9.   The minimum $\chi^2$
corresponding to the best fit is given in column~10.

\begin{figure*}[th]
\figurenum{5a}
\epsscale{1.26}
\plotone{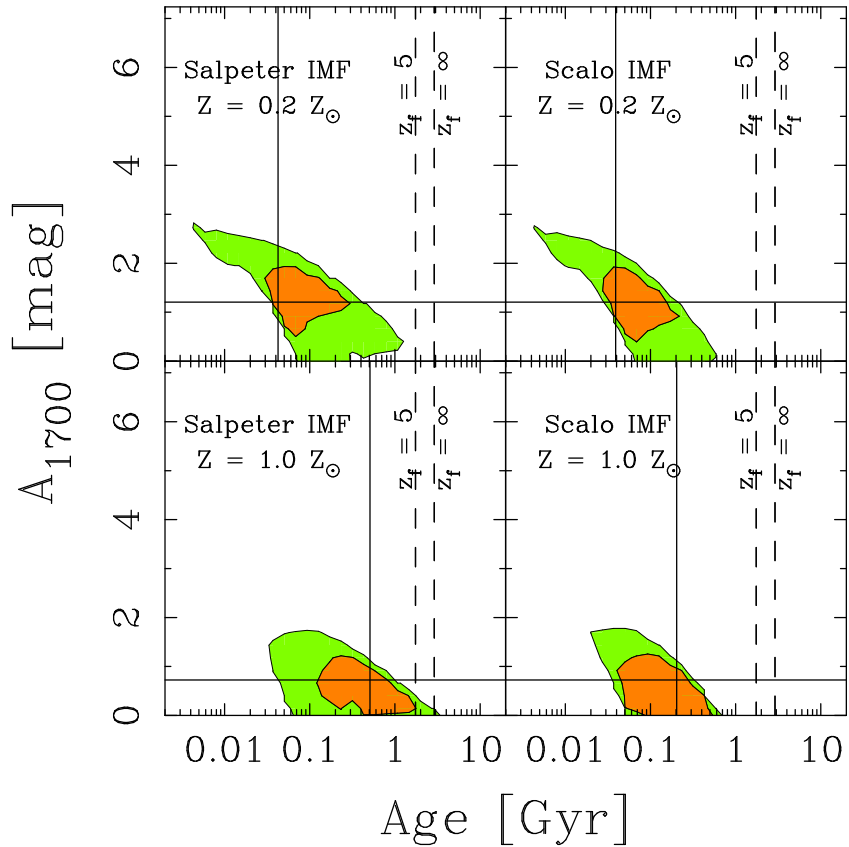}
\caption{Spectral synthesis model fitting results for one galaxy,  NIC
503 (HDF 2-903.0, $z=2.233$), which is fairly typical of the
spectroscopic LBG sample.  The star--formation history used here is
parameterized as decaying exponentially with $e$--folding time $\tau$.
As described in the text, these models provide a lower bound on the
total stellar mass of the galaxy.  In each panel, we show the
equivalent 68\%  and 95\% (orange and green shaded regions,
respectively) confidence intervals on various quantities plotted
against stellar population age, defined as the time since the onset of
star  formation: (a) 1700~\AA\ dust attenuation, (b) star--forming,
\efolding\ timescale, and (c) total stellar mass.   Each of the four
panels corresponds  to a different combination of the assumed
metallicity and functional  form of the IMF (as labeled).  The
best--fitting values  (see Table~\ref{table:bestfits}) are indicated
by the solid  cross--hairs.  The vertical dashed lines mark the
elapsed time  between the galaxy redshift and higher redshifts $z_f =
5$  and $z_f = 1000$.  In (c), a line indicates the characteristic
stellar mass of  present--day $L^\ast$ galaxies \citep{col01}.
\label{fig:lbg01}}
\end{figure*}

\begin{figure*}[hp]
\figurenum{5b} \epsscale{1.26} \plotone{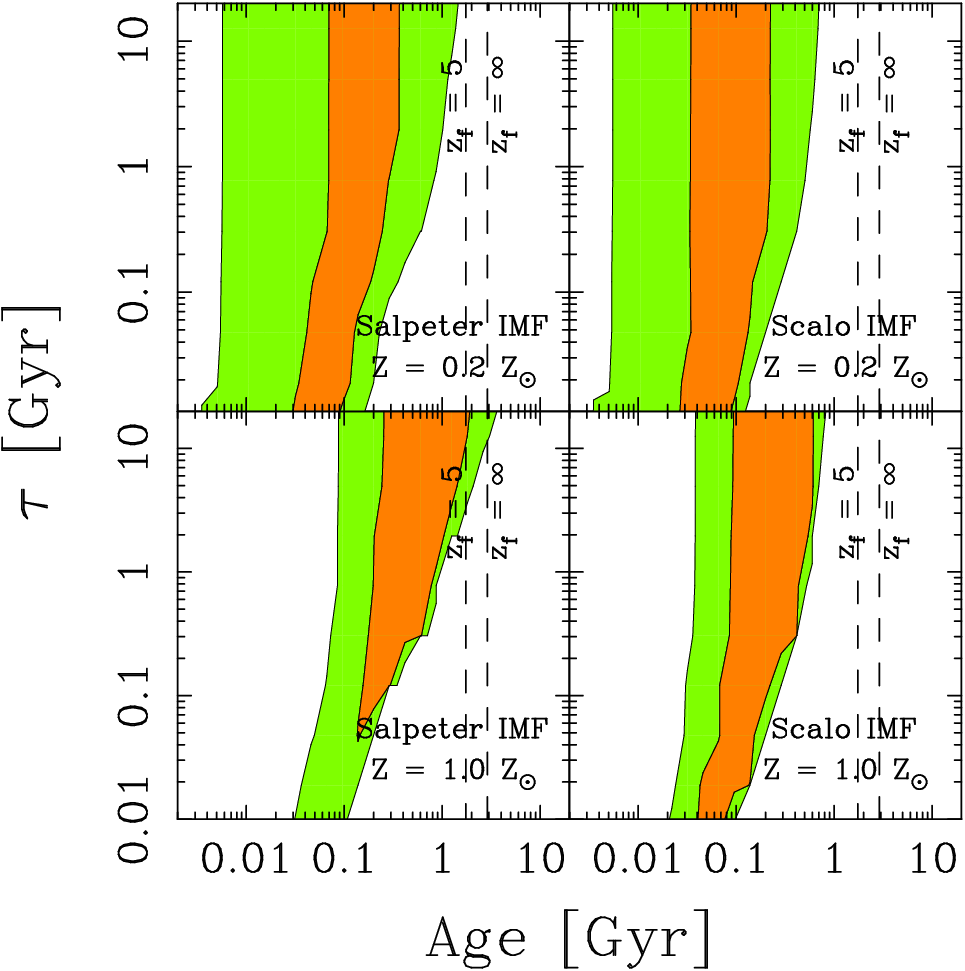}
\caption{\label{fig:lbg01b}}
\end{figure*}

\begin{figure*}[hp]
\figurenum{5c}
\epsscale{1.26}
\plotone{f5c.eps}
\caption{\label{fig:lbg01c}}
\end{figure*}

\addtocounter{figure}{1}


\begin{figure*}[hp]
\figurenum{6a}
\epsscale{1.26}
\plotone{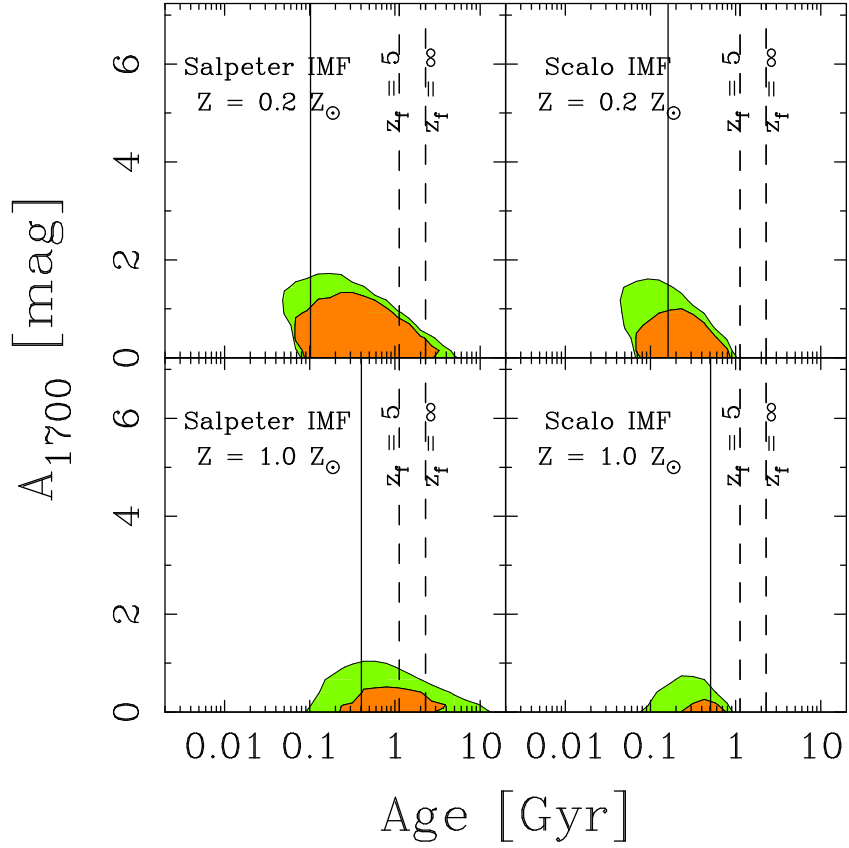}
\caption{Same as Figure~\ref{fig:lbg01} but for NIC 1352 (HDF 4-497.0, 
$z=2.800$),
which has one of the flattest (bluest) UV spectral  slopes among the
galaxies in the sample.
\label{fig:lbg26}}
\end{figure*}

\begin{figure*}[hp]
\figurenum{6b}
\epsscale{1.26}
\plotone{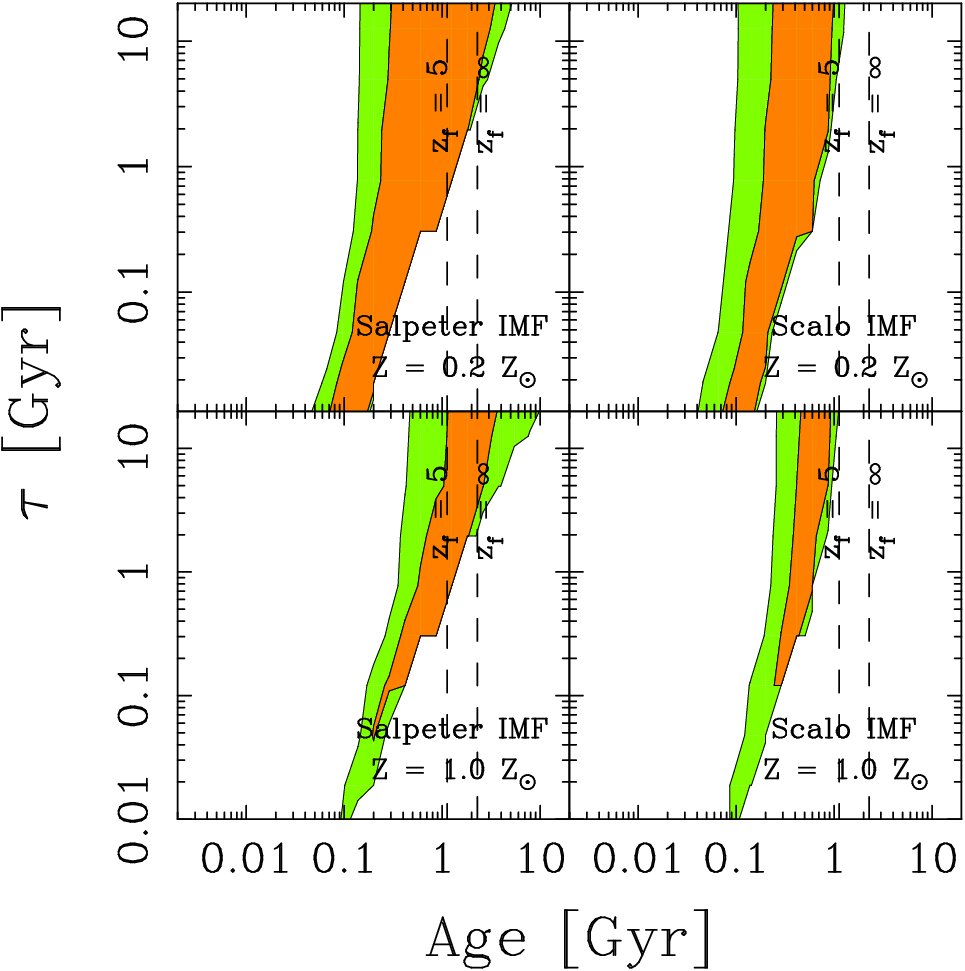}
\caption{\label{fig:lbg26b}}
\end{figure*}

\begin{figure*}[hp]
\figurenum{6c}
\epsscale{1.26}
\plotone{f6c.eps}
\caption{\label{fig:lbg26c}}
\end{figure*}

\addtocounter{figure}{1}


\begin{figure*}[hp]
\epsscale{1.26}
\figurenum{7a}
\plotone{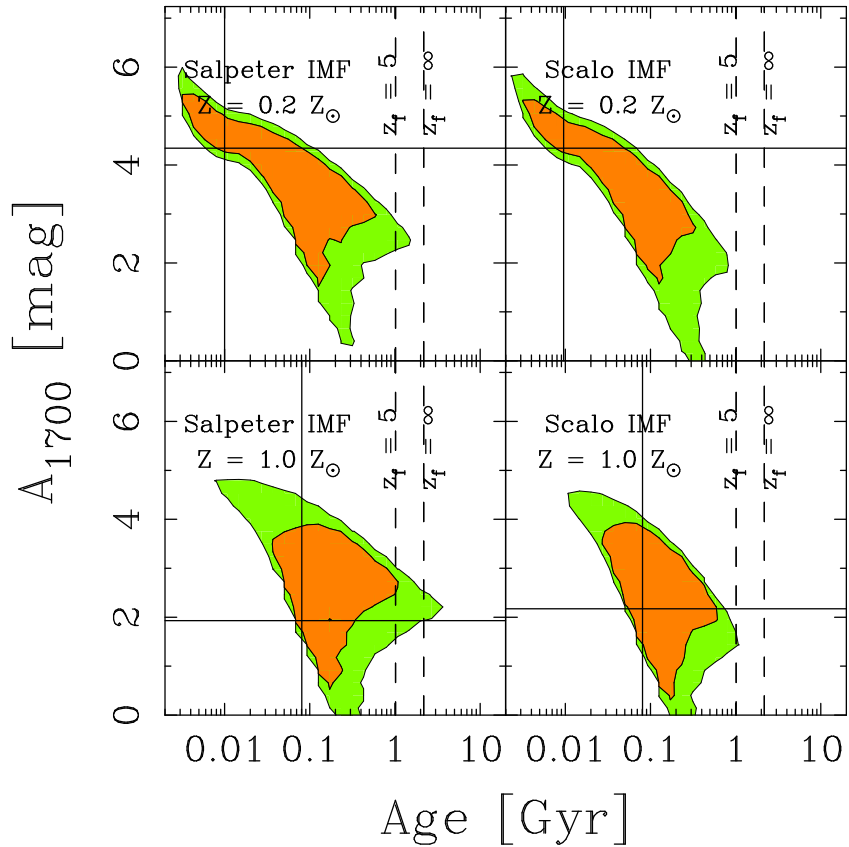}
\caption{Same as Figure~\ref{fig:lbg01} but for NIC 814 (HDF 3-93.0,
$z=2.931$), one of the reddest and apparently most massive  LBGs in
the sample.
\label{fig:lbg31}}
\end{figure*}

\begin{figure*}[hp]
\epsscale{1.26}
\figurenum{7b}
\plotone{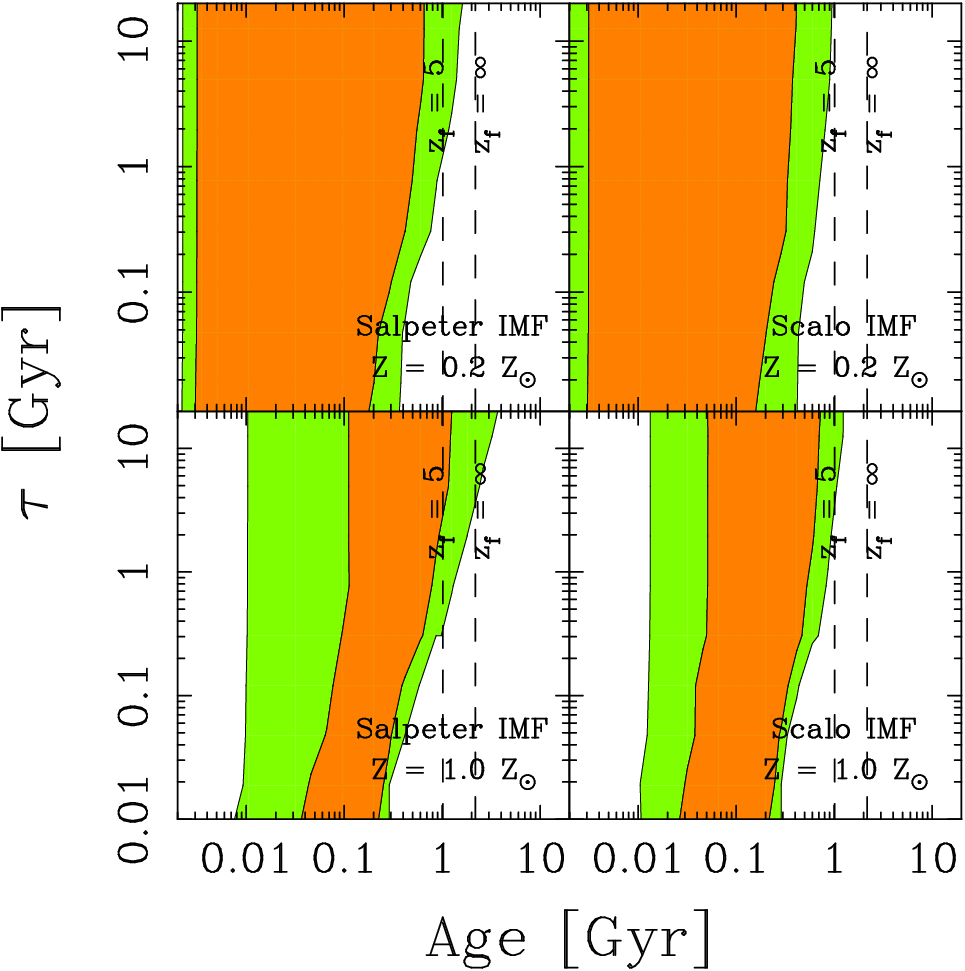}
\caption{\label{fig:lbg31b}}
\end{figure*}

\begin{figure*}[hp]
\epsscale{1.26}
\figurenum{7c}
\plotone{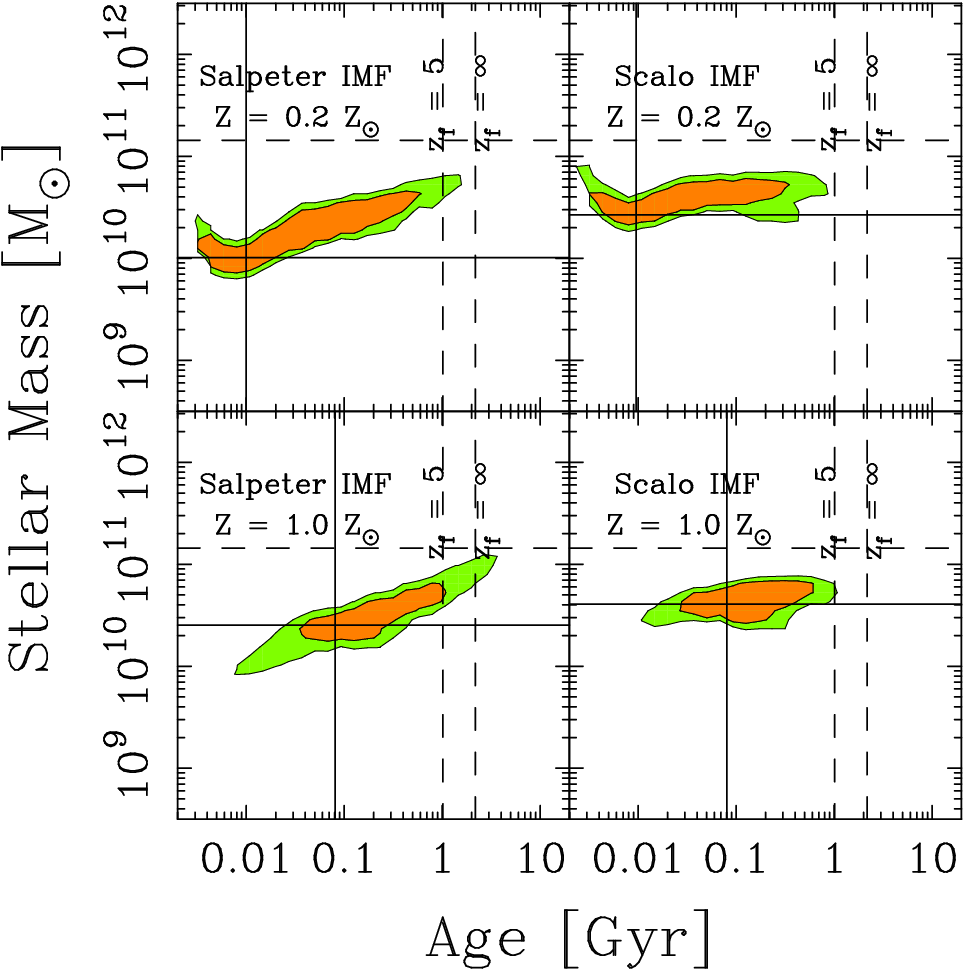}
\caption{\label{fig:lbg31c}}
\end{figure*}

\addtocounter{figure}{1}

The $\chi^2$ statistic can be used to construct confidence intervals
around the best--fit parameters.  With our suite of models, we
explored  the likelihood that other models produce equally good fits
to the data.  To derive this likelihood given a best--fit model with
$\chi^2_0$,  we computed the probability that the correct model has a
value $\chi^2$  that is greater than the observed $\chi^2_0$.  For
each model, we obtained  the difference in $\chi^2$ between the best
fit and every other model,  $\dchi = ( \chi^2 - \chi^2_0)$.  For each
LBG, curves of constant  \dchi\ in the multi--dimensional parameter
space translate directly to confidence distributions on the
parameters.  To make this transformation, we performed Monte Carlo
analyses using the LBGs' flux and uncertainty distributions (including
systematic uncertainties).  For each galaxy, we constructed 100
synthetic realizations of the data  by modulating the observed fluxes
with a random amount drawn from  the (normal) error distribution
(including both the photometric uncertainties and the systematic
errors).  We then refitted the models to the synthetic data sets and
derived new best--fitting models.   When compared to the original
best--fitting model, the curve of constant \dchi\ that encompasses
some fraction of the synthetic ``best--fits'' corresponds directly to
confidence percentages (equal to the contained fraction of synthetic
best--fitting models) on the parameters of best--fits to the true
data.  For 100 Monte Carlo realizations, the uncertainty on the
mapping of a curve of constant \dchi\ to a confidence percentage is
$\sigma(X)/X \approx 7.1$\% for parameter $X$.

Figures~5(a-c)--7(a-c) present examples of the types of confidence
intervals we derive on the fitted parameters, shown in
two--dimensional projections of the  parameter space.  Each figure
contains a series of panels showing the parameter dependences of
population age versus attenuation, age versus total stellar mass, and
age versus the characteristic star--formation time scale, for assumed
values for the metallicity and IMF.  In each figure we also mark the
lookback times to redshifts $z = 5$ and 1000.  In the mass--age
confidence interval panels, we indicate the characteristic
``$\mathM^\ast$'' stellar mass for present--day galaxies,
$1.44\times10^{11}\; h_{70}^{-2}$~\Msol, derived by \citet{col01} from
their measurement of the the local $K$--band luminosity function  and
the optical--IR color distribution of galaxies, assuming a Salpeter
IMF.

\begin{figure*}[th]
\epsscale{1.3}
\plotone{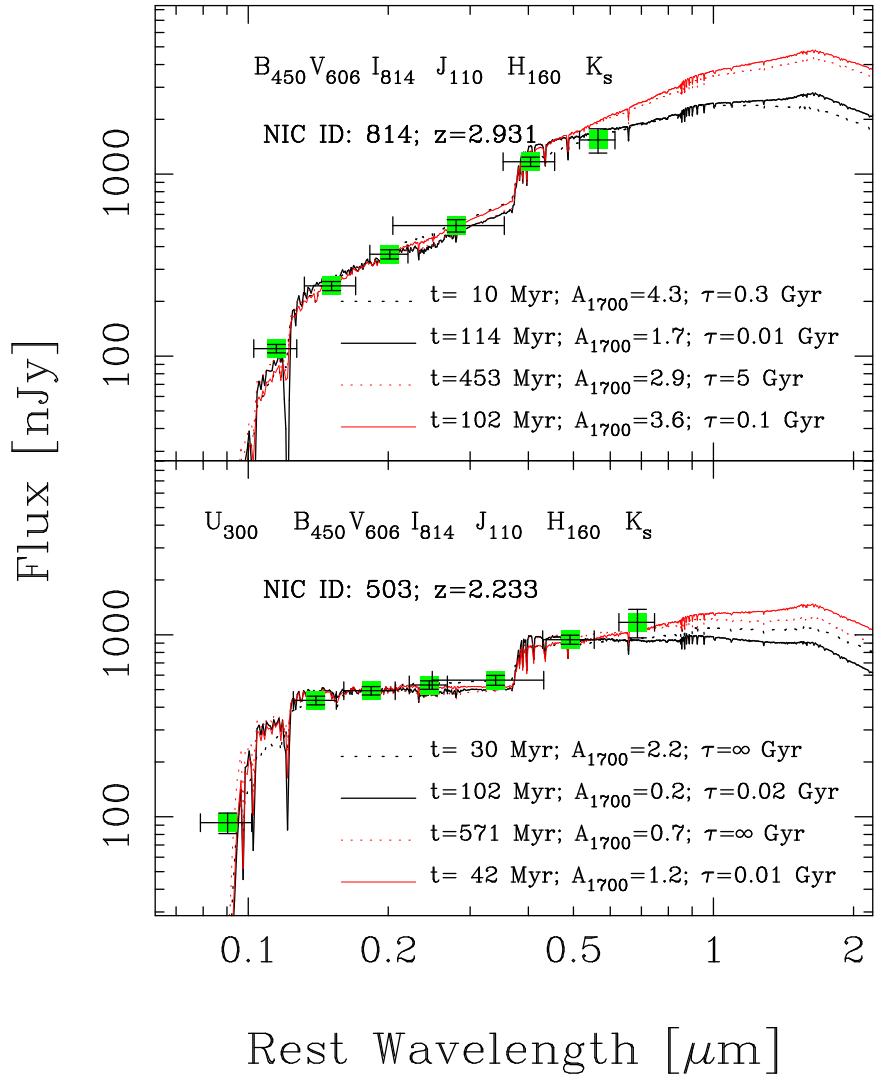}
\caption{Model spectra fit to the photometry for LBGs NIC 503 (HDF
2-903.0) and  NIC 814 (HDF 3-93.0).  Each panel shows several
different models drawn from  within the 68\% confidence region of
parameter space.  The model parameters are listed within each panel,
see also Figures~\ref{fig:lbg01} and \ref{fig:lbg31}.  The data points
(green shaded boxes) with error bars show the observed photometry.  The
horizontal error bars delineate the FWHM of the respective bandpasses.
In the case of the WFPC2 and NICMOS data,  most of the vertical
photometric error bars are smaller than the  symbol size.
\label{fig:lbgspec} }
\end{figure*}

In most cases there is a range of models that all fit the data
reasonably well.  In Figure~\ref{fig:lbgspec}, we show examples of
model spectra, each from within the 68\%\ confidence intervals, but
with substantially differing parameters.  With the available
photometry,  the models fit the observed SED very well in the
rest--frame UV blueward  of the 4000~\AA\ and Balmer breaks.  The
differences between the models  become larger around and redward of
the \ks-band [see especially the  top panel, NIC 814 (HDF 3-93.0),
where the redshift is such that the \nich--band  spans the Balmer
break].   Thus, these models are not well  differentiated.  Deeper
$K$--band data, and especially measurements  at longer rest--frame
wavelengths, \eg, from the {\it Space Infrared  Telescope Facility}
(SIRTF), should be quite effective in narrowing  the range of possible
stellar populations for galaxies at these  redshifts.

Some parameters are better constrained than others, and there are some
degeneracies between the fitting parameters for individual galaxies.
For example, both dust and age redden the colors of the galaxies.
While extinction strongly reddens the UV portion  of the spectrum, the
Balmer break amplitude is relatively unaffected, and thus serves as a
useful age indicator, constraining the relative number of early-- (OB)
to late--type stars (A and later).  However, even with precise NICMOS
photometry, we are able to derive only rather loose constraints on age
and reddening, and the confidence distributions  for those parameters
tend to be elongated along an axis of anticorrelation  (see, \eg,
Figures~\ref{fig:lbg01} and \ref{fig:lbg31}).  We note that this
age--extinction degeneracy is considerably weaker when fitting with
models using the SMC extinction relation.

The best parameter constraints are the derived LBG stellar--mass
estimates.  Based on fitting the simple star--formation histories to
the observed photometry for any given LBG, the 68\% range of allowable
stellar masses is constrained $\sigma(\log \mathM) \lsim 0.5$~dex,
with median~$\simeq 0.25$~dex.  In part, this is due  to the high
quality of the near--infrared data, which probes rest--frame optical
wavelengths that are less sensitive to mass--to--light ratio
($\mathM/L$) variations due  to population age or extinction.
However, even in the rest frame $B$--band,  $\mathM/L$ can vary
substantially for stellar populations with different ages.   For a
given galaxy, the range of possible, intrinsic values for $\mathM/L$
is restricted by the UV--to--optical SED fitting, which constrains the
age and reddening for each galaxy.  Although, we have seen, these
parameter constraints are somewhat degenerate, (\eg, for age versus
extinction or metallicity), these degeneracies tend to cancel out in
terms of their effect on the {\it extrinsic} $\mathM/L$ values, \ie,
the stellar mass over the emergent (after extinction) luminosity.
E.g., for spectral models that fit the photometry for a particular
galaxy, the younger models will have lower $\mathM/L$, but require
more extinction, which suppresses the light and raises $\mathM/L$.
The reverse situation applies to older, less reddened model fits.  In
this way, the allowable range of $\mathM/L$ is limited, resulting in
tighter constraints on the stellar mass for each galaxy (cf.\
Figures~\ref{fig:lbg01c}, \ref{fig:lbg26c}, \ref{fig:lbg31c}).   This
method for estimating LBG stellar masses is essentially the same as
that used by \citet{bri00} in their analysis of field  galaxies at $0
< z < 1$.

It is important to note that the stellar--masses derivations are based
on fitting the observed LBG photometry to simple, constant or
monotonically decreasing, star--formation histories.  The results,
therefore, are largely driven by the most luminous stellar populations
(namely, the most recently formed stars).  Thus, the models predict
best--fitting $\mathM/L$ ratios for the stars that dominate observed
light,  and because these models fit the light from the youngest,
highest--mass stars,  we must consider these $\mathM/L$ values as
lower bounds for the stellar component in these galaxies (neglecting
changes to the form of the assumed IMF, see also \S\ref{section:IMF}).
With these models, we have  relatively weak constraints on the
presence of any additional, older stellar components (e.g., from prior
episodes of star formation), which might have substantially higher
$\mathM/L$ ratios due to a predominance of lower--mass stars.
Therefore, we regard the stellar--mass estimates using these simple
star--formation histories strictly as minimum values for the galaxies'
total stellar mass.  In \S\ref{section:oldstars} below, we consider
the effects on the derived stellar parameters by adding a second,
maximally old, stellar component to the simple models above.

The stellar--mass estimates do systematically depend on the functional
form of the IMF and to the metallicity of the population synthesis
model.  Generally, increasing the model metallicities shifts the
stellar masses to higher values.  The size of this offset depends on
the  assumed form of the IMF.  For a Scalo IMF, increasing the
metallicity  from $Z=0.2$~\zsol\ to 1.0~\zsol\ shifts the derived
stellar masses by $\lsim 0.1 - 0.2$~dex.  However, for a Salpeter IMF,
changing the metallicity over the same range introduces a systematic
mass offset  in most cases of $\sim 0.3 - 0.5$~dex.  Furthermore,
changing from Salpeter IMF to Scalo IMF models (for fixed metallicity)
produces systematic stellar mass shifts of $\delta(\log \mathM) = 0.0$
to 0.6~dex (median~$\simeq 0.2$~dex) for $Z=0.2$~\zsol, and 0.0 to
0.3~dex (median~$\simeq 0.1$~dex) for $Z=1.0$~\zsol\ models.

Two of the galaxies, NIC 989 (HDF 4-445.0, $z=2.500$) and  NIC 782
(HDF 4-316.0, $z=2.801$), have very unusual distributions of parameter
values.  These galaxies have very narrow loci in the age--attenuation
plots,  favoring very young ages, extremely high extinction, and large
stellar  masses.  These models appear to be somewhat unphysical, with
instantaneous SFRs $\sim 30,\!000$~\Msol~yr$^{-1}$.  The best--fit
model spectra deviate from the observed \wfu-- and \wfb--band data by
many standard deviations.  This could suggest  that the published
spectroscopic redshifts for these objects  are erroneous.  For
example, NIC 782 is apparently well detected  at \wfu, which seems
unlikely for an object at $z = 2.8$.  The photometric redshift
analyses of \citet{bud00} favor lower redshifts  of $z\simeq 1.5$, and
$z\simeq 1.8$ for NIC 989 and NIC 782, respectively  (see also
Fern\'andez--Soto \etal\ 2001).  We cannot exclude the possibility
that these two objects have unusual SEDs that are not well fit by the
models we have used, but for lack of additional information we will
not consider  these objects further in the remaining discussion of the
LBG sample.


\begin{figure*}[th]
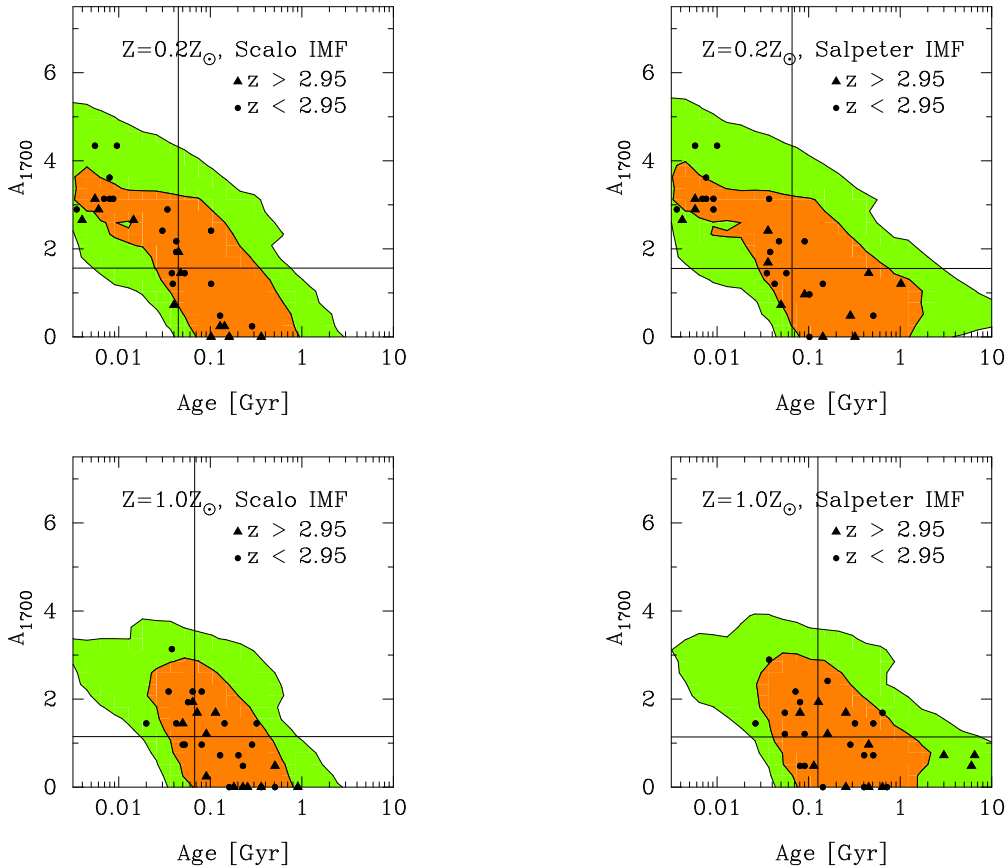

\epsscale{1.3333}
\vbox{\plottwo{f9a.eps}{f9b.eps}} 
\vspace{0.2in}
\vbox{\plottwo{f9c.eps}{f9d.eps}}
\caption{Composite probability distributions for the model parameters
of dust attenuation versus population age for the whole LBG sample.
The models used here assume the starburst extinction law.  The
contours  represent the 68, and 95\% confidence intervals averaged
over the  whole sample (orange and green shaded regions,
respectively), and thus trace the distribution of allowable model
parameters among the \hdf\ LBGs.  The points show the best--fit
parameter values for each individual LBG, and are coded to indicate
galaxies with $z < 2.95$ and $z> 2.95$, \ie, below and above the
redshift  at which the Balmer break redshifts into the \nich\ filter.
The cross--hairs show the (geometric) mean value for the distribution.
Each panel shows results for a different  combination of metallicity
and IMF.
\label{fig:comp_age_extinct}}
\end{figure*}

\begin{figure*}[ph]
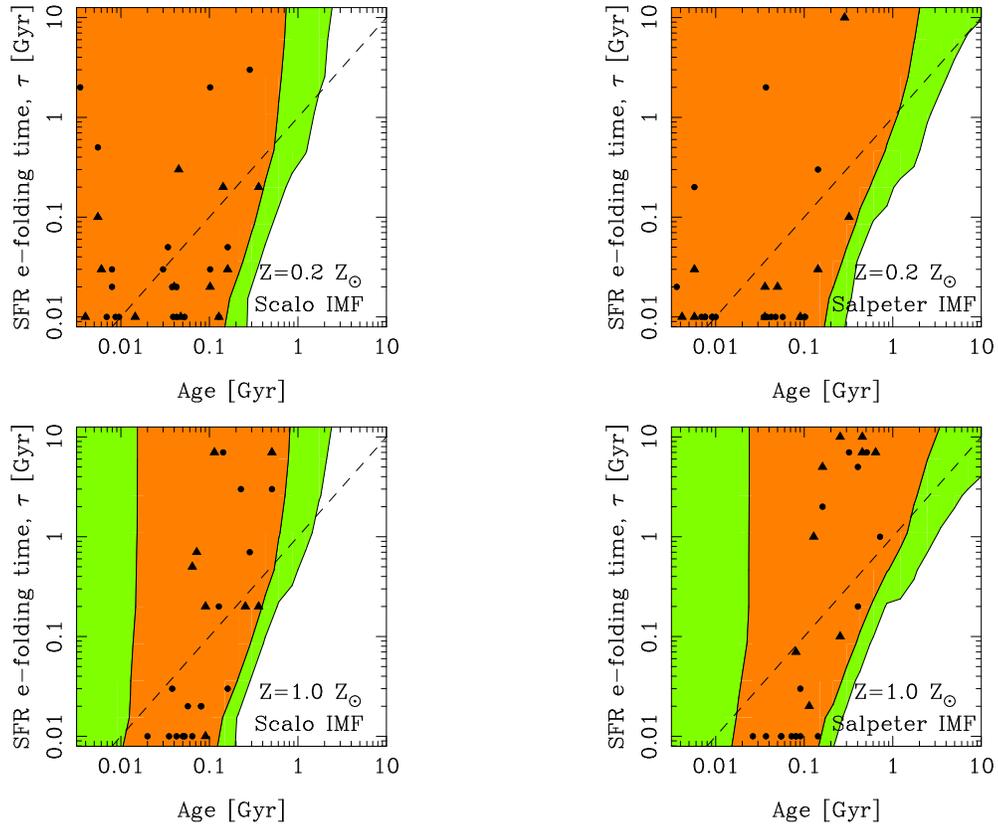

\epsscale{1.3} \vbox{\plottwo{f10a.eps}{f10b.eps}}
\vspace{0.1in}
\vbox{\plottwo{f10c.eps}{f10d.eps}}
\caption{Composite probability distributions for the model parameters
of star--forming, \efolding\ time versus population for the LBG
sample, assuming the  starburst extinction law.  The contours and
symbols are the same as in Figure~\ref{fig:comp_age_extinct}.  The
dashed line indicates the one--to--one correspondence.
\label{fig:comp_age_tau}}
\end{figure*}

\begin{figure*}[ph]
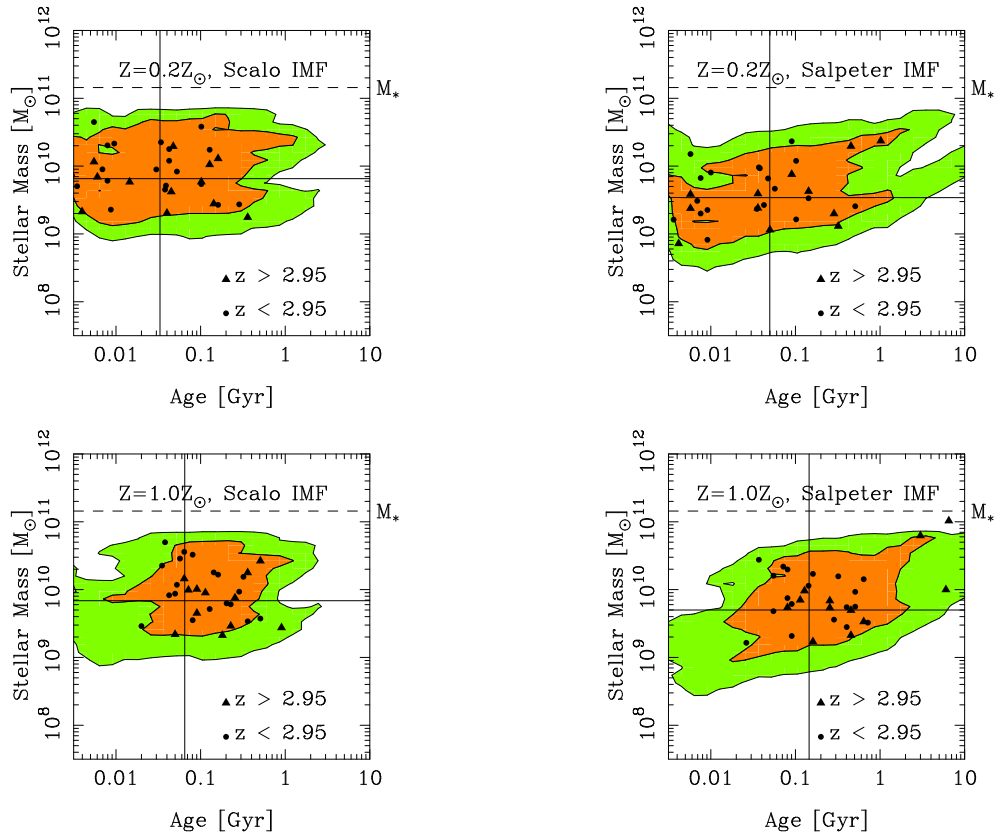

\epsscale{1.3}
\vbox{\plottwo{f11a.eps}{f11b.eps}}
\vspace{0.1in}
\vbox{\plottwo{f11c.eps}{f11d.eps}}
\caption{Composite probability distributions for the model parameters
of stellar mass versus population for the LBG sample, assuming the
starburst extinction law.  The contours and symbols are the same as in
Figure~\ref{fig:comp_age_extinct}.  The cross hairs show the
(geometric) mean  value for the distribution.
combination.\label{fig:comp_age_mass}}
\end{figure*}

\begin{figure*}[ph]
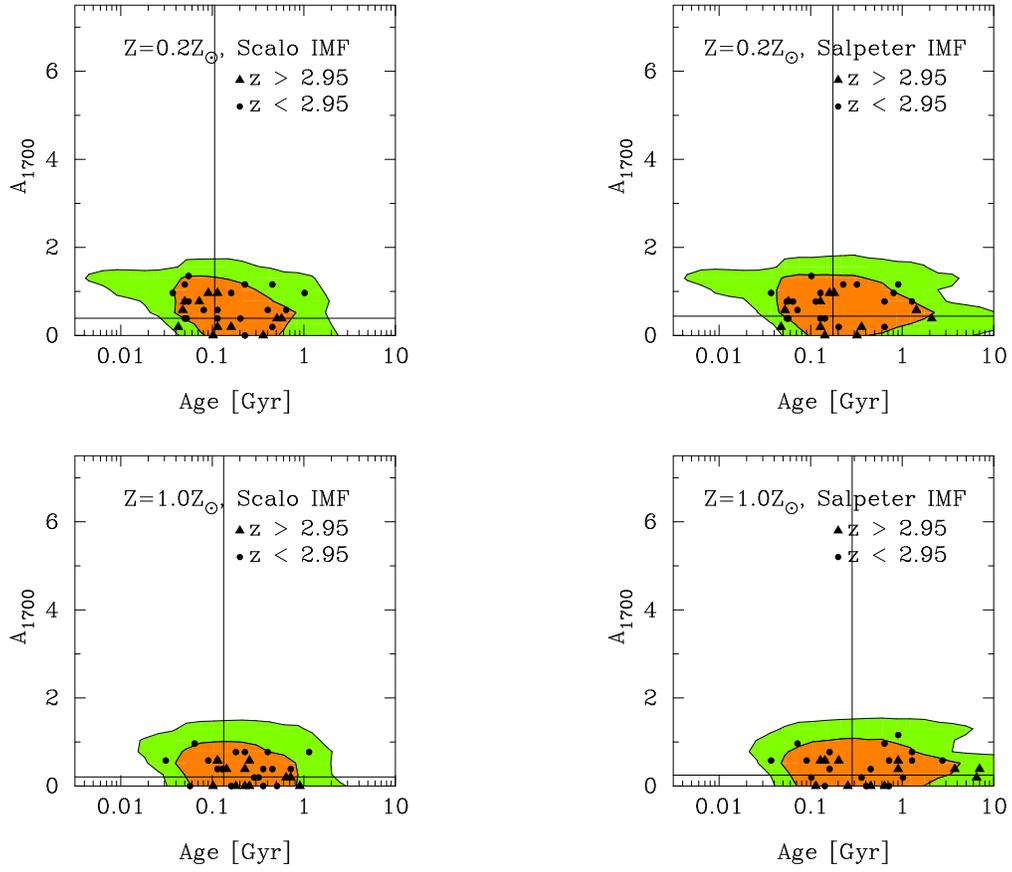

\epsscale{1.3333}
\vbox{\plottwo{f12a.eps}{f12b.eps}}
\vspace{0.2in}
\vbox{\plottwo{f12c.eps}{f12d.eps}}
\caption{Same as Figure~\ref{fig:comp_age_extinct}, but using the
SMC--like dust extinction law. \label{fig:comp_age_extinct_smc}}
\end{figure*}

\begin{figure*}[ph]
\epsscale{1.3333}
\vbox{\plottwo{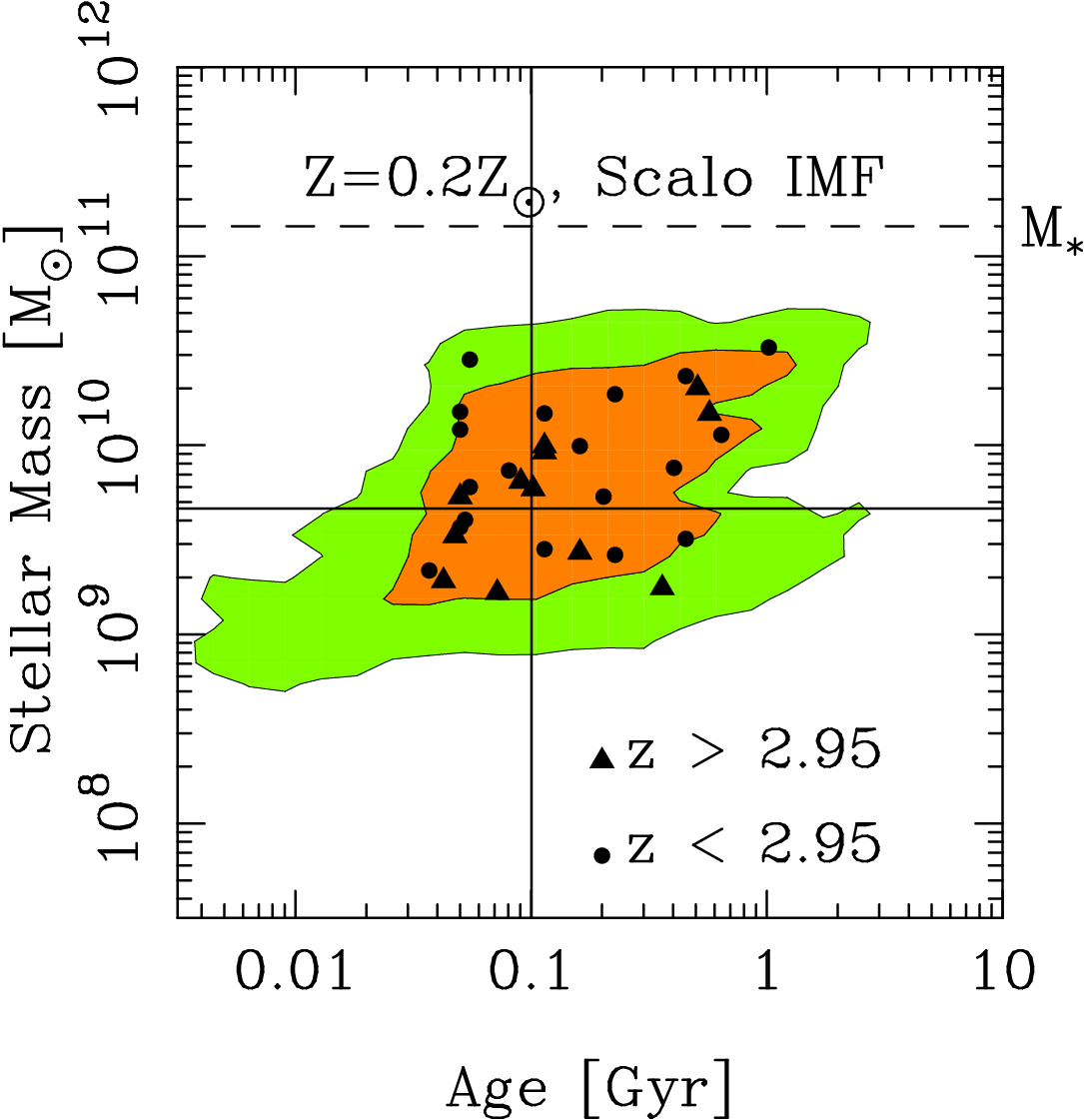}{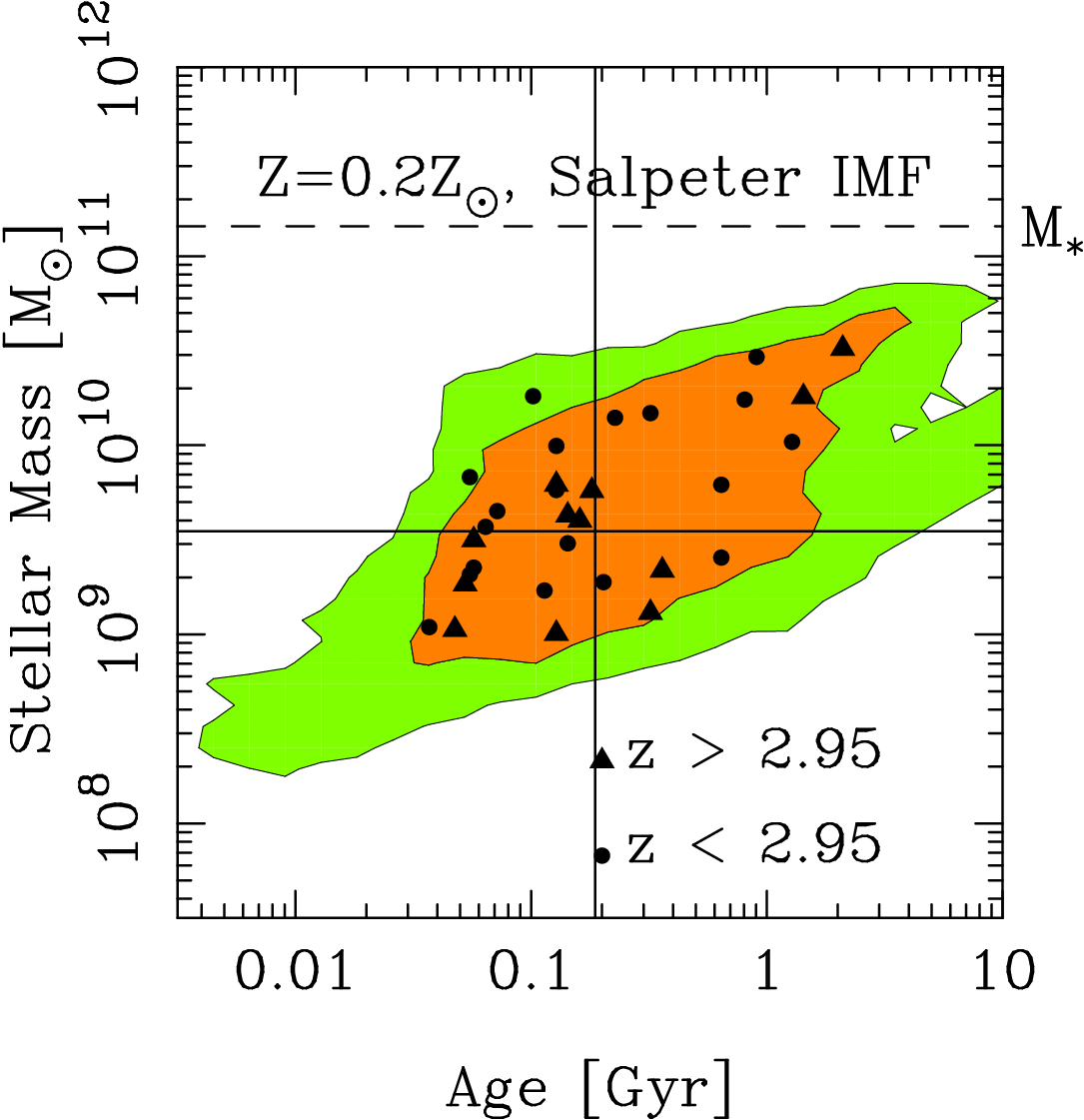}} 
\vspace{0.2in}
\vbox{\plottwo{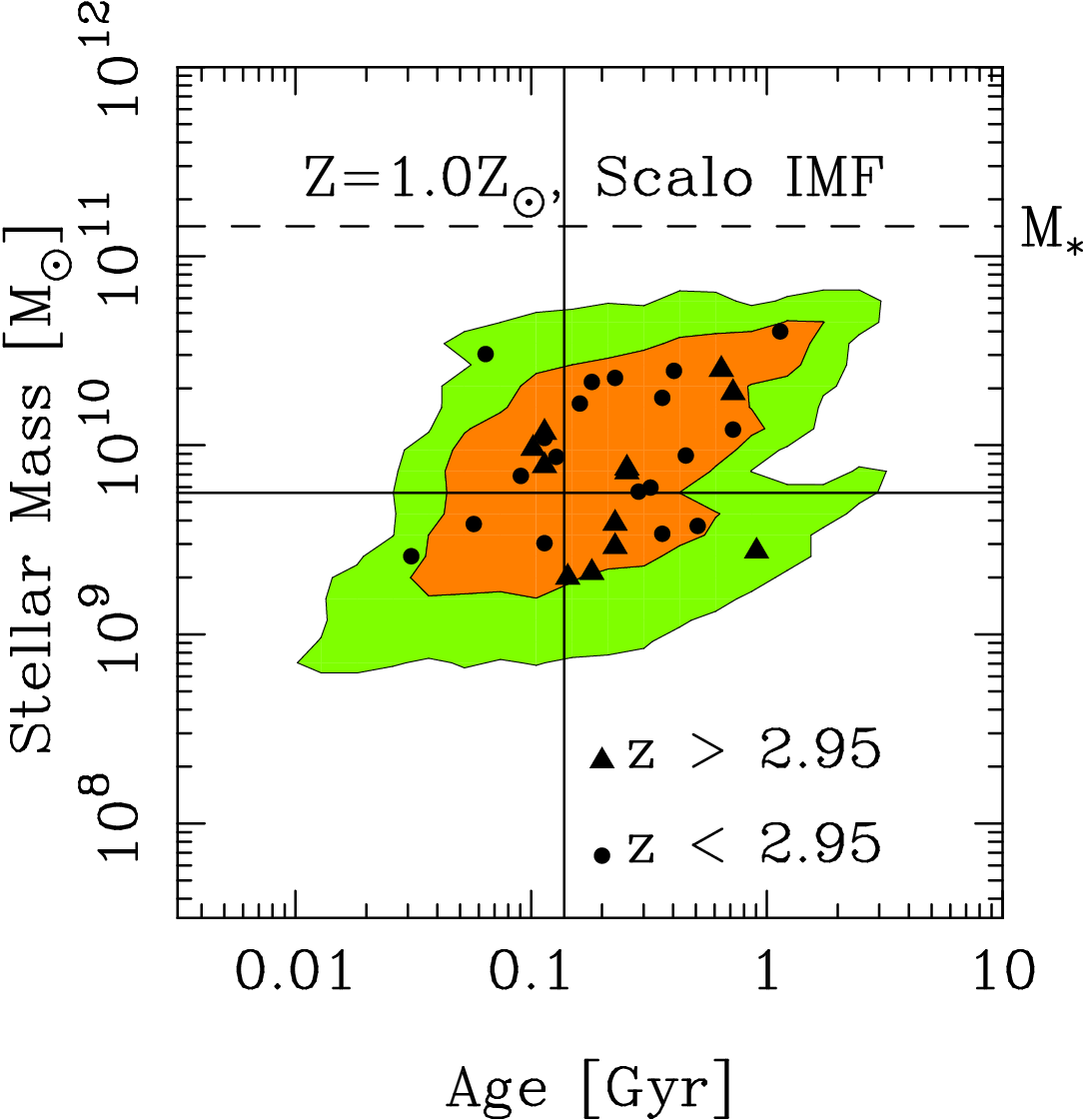}{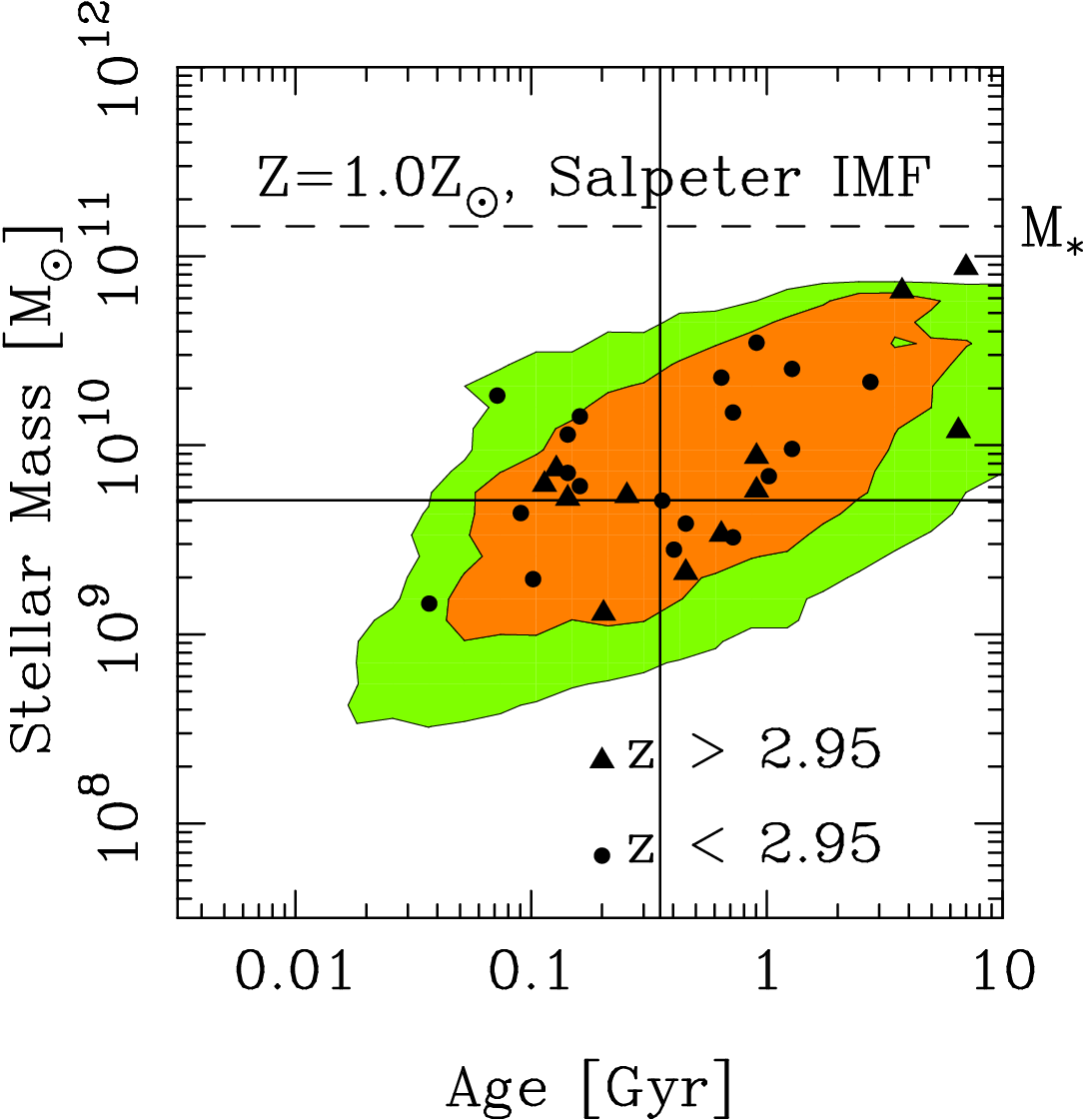}} 
\caption{Same as Figure~\ref{fig:comp_age_mass}, but using the
SMC dust extinction law. \label{fig:comp_age_mass_smc}}
\end{figure*}


\section{Discussion
\label{section:discussion}}

The constraints on the LBG, stellar--population parameters span a wide
proportion of the available multi--dimensional parameter space.   The
intersection of the confidence intervals of all the LBGs in the sample
generally shows that there is no region common to all the objects with
any significance.  There is some small overlap in the age--extinction
confidence region (a compactification of the larger parameter space),
which is common to all the LBGs, albeit with a very small likelihood
(order of a few percent).  Even so, the corresponding star--formation
histories for these models vary for each object.  We also note that
genuine object--to--object variation is suggested from the large range
of observed LBG SED colors, which is beyond what is expected from the
photometric uncertainties and from simply varying only the dust
content in the galaxies.  Thus, this seems to suggest that there is
genuine variation in the objects' stellar--population properties.
However, it is difficult (if not impossible) to prove genuine
variation robustly amongst the LBGs in our sample given the current
data.  In this section, we use the general constraints on stellar
population parameters for the LBG sample, which we use to investigate
the assembly history of LBGs.


\subsection{Properties of the LBG Stellar Populations
\label{section:properties}}

In Figures~\ref{fig:comp_age_extinct}-\ref{fig:comp_age_mass_smc},  we
show combined confidence intervals for the entire sample of LBGs
(excluding objects 989 and 782 for reasons given above).  These plots
were constructed by averaging the likelihood distributions on the
fitting parameters for the individual galaxies, and thus show the
favored (and disfavored) regions of the parameter space among the
entire population.  The best--fit values of the parameters for
individual galaxies are marked by points.  Different symbols code
galaxies at redshifts below and above $z = 2.95$.  For the higher
redshift objects, the 4000~\AA\ break and 3646~\AA\ Balmer break move
through the NICMOS F160W filter bandpass, leaving only the
lower--$S/N$, \ks--band data to sample rest--frame optical light, and
thus making the stellar--population constraints more uncertain.   In
the sections that follow, we discuss the implications of the model
fitting for the stellar population properties of the LBGs.

\subsubsection{Initial Mass Function
\label{section:IMF}}

The best--fitting models using any of the IMFs (Salpeter,  Miller \&
Scalo, and Scalo) are consistent with the data.   The best--fit models
using the Scalo and Miller--Scalo IMFs  tend to favor younger ages and
slightly lower attenuation values because the steeper slope at the
high mass end of these IMFs results in spectra that are inherently
redder.  In a detailed spectral analysis of the gravitationally lensed
galaxy cB--58 ($z=2.7$), \citet{pet00a} show that the  P~Cygni profile
of the \ion{C}{4} $\lambda$1549 line requires  a significant
population of stars with masses $\geq 50$~\Msol, and is inconsistent
with a truncated IMF or one with a very steep  high--mass slope, such
as the Miller--Scalo form.

Varying the IMF lower mass cutoff would also change the derived
stellar masses.  The BC2000 models used here assume upper and lower
mass cutoffs of $\mathM_u = 100$~\msol\ and $\mathM_l = 0.1$~\msol.
For a Salpeter IMF, stars with masses less than 2~\msol\ contribute
typically  less than $\sim 1$\% of the 1700~\AA\ luminosity.  For
fixed 1700~\AA\ luminosity  and fixed upper mass cutoff, the total
stellar mass varies with the IMF lower mass cutoff and slope $x$,
[$N(\mathM) \propto \mathM^{-x}\;d\mathM$] as
\begin{equation}
\mathM_\mathrm{tot} \propto \left\{ \begin{array}{ll}
	(2-x)^{-1}(\mathM_u^{2-x} - \mathM_l^{2-x}) & x \neq 2 \\
	\ln(\mathM_u/\mathM_l) & x = 2.  \end{array} \right.
\end{equation}
So, for example, for a Salpeter IMF ($x=2.35$) with $\mathM_l =
1$~\msol, the total stellar mass would be 39\% the mass derived with
$\mathM_l = 0.1$~\msol.

\subsubsection{Metallicity
\label{section:metallicity}}

Increasing the metallicity reddens the model SEDs.   This can arise
due to the combination of several effects.   Higher metallicity stars
have shorter main sequence lifetimes.   Also, for a fixed age, higher
metallicity populations have lower effective temperatures,  and have
stronger metallic absorption features.  In general,  the broad--band
photometry provides little constraint on the LBG metallicities.
However, the choice of metallicity  can have a significant impact on
the other derived parameters.  As can be seen from
Figures~\ref{fig:comp_age_extinct} and \ref{fig:comp_age_mass},
increasing the model metallicities from $0.2$~\Zsol\ to $1.0$~\Zsol\
has the effect of decreasing the required extinction while increasing
the population age.  This effect is seen across the entire range  of
metallicities we have explored, $Z = 1/200$ to 3~\zsol.   As described
in \S\ref{section:constraints} above, there is  some degeneracy
between age and extinction when fitting the models to the photometry.
In this context, it appears that there is an interplay between the
effects of metallicity, age, and extinction on the galaxy colors.  The
choice of metallicity changes the color of the model SED at fixed age,
and the best--fitting models required  to fit the photometry of a
particular galaxy then tend to slide along the age--extinction
anticorrelation.  We caution again that the theoretical stellar model
atmospheres used by the BC2000 models are relatively uncalibrated by
observational data at low metallicity, and therefore any results that
show a strong metallicity dependence should be regarded with caution.
We also note that this age--extinction degeneracy is  strong for
models using the starburst extinction law, but is much  weaker when
the SMC law is used.

\subsubsection{Star Formation History
\label{section:sfhistory}}

The past star--formation histories for individual galaxies are only
loosely constrained by the current data set, (see the age--$\tau$
panels in Figures~\ref{fig:lbg01b}--\ref{fig:lbg31b},  and also
Figure~\ref{fig:lbgspec}).  Generally, a large region of the parameter
space is consistent with the data, especially for $t/\tau \lsim 1$,
where the SFR is roughly constant and most values provide good fits.
In Figure~\ref{fig:comp_age_tau}, we show the mean age--$\tau$
confidence region for the LBG population.  In most cases, models with
$t \gg \tau$, where the current SFR is  much smaller than the past
average, are strongly disfavored, although a few galaxies can be fit
by ``post--starburst'' models with ages $\sim$100~Myr (see
\S\ref{section:sfrate}).  In other words, the LBGs do not appear to be
UV--bright objects due to a relatively small, residual tail of
on--going star formation in an older galaxy, but rather are seen when
they are still in the process of forming a significant fraction of
their stellar content.  This result may, in part, be due to the choice
of exponential, monotonically decreasing or constant star--formation
histories;  we consider other possible star--formation histories in
\S\ref{section:oldstars} below.  Although for most LBGs the
star--formation timescales are only poorly constrained, we do note
that for all combinations of metallicity and IMF there are a number of
galaxies that are best fit by short timescale ``bursts'' (see
Figure~\ref{fig:comp_age_tau}).

\subsubsection{Dust Extinction
\label{section:extinction}}

For models using the \citet{cal00} starburst extinction law,  there is
some degeneracy in the attenuation--age parameter confidence regions
(see discussion below).  For individual LBGs, the 68\%\ confidence
region  typically spans $\Delta A_{1700} \sim 1.5-2$~mag, which
corresponds  to a range of allowable flux corrections that vary by a
factor  $\sim 6$ at 1700~\AA.  For the population as a whole (see
Figure~\ref{fig:comp_age_extinct}), the distribution of best--fit
values spans $\extinctA \simeq 0-4$~mag ($\simeq 0-3$~mag) for  $Z =
0.2$~\zsol\ (1.0~\zsol), with only a mild dependence  on the chosen
IMF, but with notable sensitivity to metallicity.   For low
metallicity models, a number of galaxies are best fit with  young ages
and heavy reddening, but this is not seen when solar  metallicity
models are used.  The probability--weighted mean UV  attenuation for
the LBG sample is $\langle \extinctA \rangle = 1.6$~mag (1.2~mag) for
$Z = 0.2$~\zsol\ (1.0~\zsol).  The average and range of \extinctA\
values derived here  are similar to those derived by \citet{ste99} and
\citet{ade00} based on optical photometry for a large data set of
several hundred LBGs, using the same dust attenuation law but simpler
stellar population assumptions.   However, the luminosity--weighted
average UV--flux correction  (i.e., the ratio of intrinsic to emitted
luminosity summed over  the whole sample) is considerably larger, a
factor of $\simeq 13.8$  (4.6) for $Z=0.2$~\zsol\ (1.0~\zsol) and a
Salpeter IMF (and similar  for the Scalo IMF).   This reflects the
tendency for LBGs with the  largest fitted dust corrections to have
the highest intrinsic UV  luminosities and SFRs, a point emphasized by
\citet{meu99} and \citet{ade00}.  Those authors argue that this  is a
real trend among LBGs, and not an artifact of the extinction
correction process.

If the SMC extinction law is adopted instead, the anticorrelation
between age and attenuation largely disappears
(Figure~\ref{fig:comp_age_extinct_smc}), and the tail of  young and
high--extinction models found with the ``grey''  starburst dust
attenuation is absent.  The allowed model  extinction at 1700~\AA\ is
$\lsim 1.5$~mag regardless of IMF  and metallicity, and any
correlation with population age is  weak at best.  Most galaxies are
well fit by stellar population  ages $\gsim 40$~Myr.

Regardless of metallicity, IMF, and dust model assumptions, models
with very young ages and low dust content are strongly disfavored.
Very young, unreddened star--forming regions, particularly those with
low metallicity, should have UV spectra that rise (in $f_\nu$ units)
toward shorter wavelengths.   In practice, this is never observed
among the LBGs:  even the youngest galaxies apparently contain
significant  amounts of dust at these redshifts.  This may suggest
that the LBGs have experienced metal enrichment from a previous
stellar generation,  and therefore the LBG population probably are not
``first--generation'' objects (unless the spectra of young, low
metallicity, dust--free objects  are sufficiently different from the
models used here).   Such first--generation objects must be quite rare
at these redshifts, or must pollute themselves with dust on very short
timescales.  Alternatively, perhaps the SFRs (and hence luminosities)
of such objects are too low to allow them to be detected in the the
HDF images.

\begin{figure*}[th]
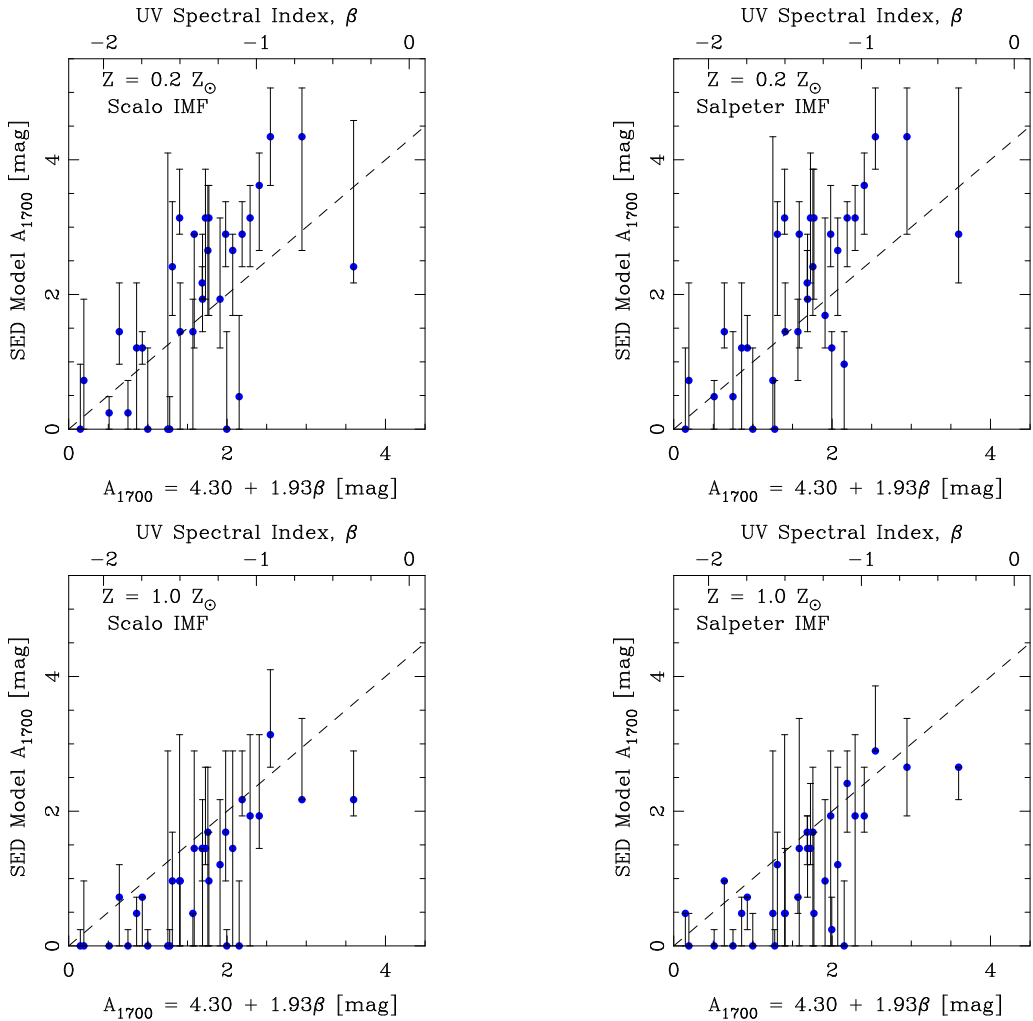

\epsscale{1.4}
\vbox{\plottwo{f14a.eps}{f14b.eps}} \vspace{0.1in}
\vbox{\plottwo{f14c.eps}{f14d.eps}}
\caption{Comparison of the fitted SED model extinction (using  the
starburst attenuation law), in magnitudes at 1700~\AA, with the
attenuation inferred from the U--spectral index, $\beta$, alone
($f_\lambda \propto \lambda^\beta$) using the relation from
\citet{meu99}.  The $\beta$ values measured from the WFPC2 photometry
are indicated  along the top axis.  The points and error bars in the
ordinate show  the best--fit values and 68\% confidence intervals on
dust attenuation  from the population synthesis models fit to the
optical--to--infrared  photometry.
\label{fig:a1700}}
\end{figure*}

The extinction correction factors that we derive from the model
fitting differ somewhat from those obtained by other authors using
various techniques.  Modelling many of the same HDF--N galaxies using
ground--based NIR photometry, \citet{saw98} find SED fits with
somewhat higher median UV--flux--suppression factors,  $\langle
A_{1700} \rangle \simeq 2.7$~mag for $Z = 0.2$~\zsol\ models
(accounting for small differences in the forms of the starburst
attenuation law). The two analyses are in closer agreement at solar
metallicity, where Sawicki \& Yee find $\langle A_{1700} \rangle
\simeq 1.4-1.9$~mag, depending  on the assumed star--formation
history.   The differences between  these analyses may be due to
several factors, including the improved  sensitivity of our NICMOS
photometry, more accurate matching of  optical to infrared photometry,
allowance for a wider range of  possible star--formation histories
(Sawicki \& Yee considered only  $\delta$--function bursts and
continuous star--formation models),  and the use of a larger galaxy
sample.  \citet{meu97} also found large UV--attenuation values ($\sim
15-20$) by fitting the the UV  slopes of a sample of \hdf\ LBGs to the
starburst extinction law  from \citet{cal94}.  However, using a
revised method of a locally  calibrated UV slope to far--infrared
(FIR) distribution, \citet{meu99}  derive more modest mean extinction
of factors $\simeq 5$.  As noted above, the attenuations we derive are
similar to those found by \citet{ade00} using UV rest--frame data for
a larger LBG sample.  Overall, the fact that the derived UV--flux
attenuation depends so strongly on parameters on which we have little
reliable information, such as stellar--population metallicity, leads
us to believe that one must regard the quantitative results of such
analyses with considerable caution, but the necessity for significant
dust attenuation in many or most LBGs seems secure.

In local, UV--bright starburst galaxies, the UV--spectral slope
$\beta$ ($f_\lambda \propto \lambda^{\beta}$) correlates with the flux
emitted at far infrared wavelengths (8-1000~\micron), demonstrating a
close relation between UV extinction and dust reprocessing.
\citet{meu99} calibrate this relationship as $\extinctA = 4.30 +
1.93\beta$ (where we have adjusted their parameters slightly to
account  for a small shift in the reference UV wavelength).    In
Figure~\ref{fig:a1700}, we compare the attenuation derived from the
UV--through--optical rest--frame SED model fitting to that computed
from the UV slope $\beta$ alone, using the prescription of
\citet{meu99} to convert $\wfv - \wfi$ colors to $\beta$.  For a given
set of IMF and metallicity assumptions, there is considerable  scatter
($\simeq 1$~mag) in the comparison of attenuation values.  This is
appreciably larger (by approximately a factor of two) than uncertainty
that results from random and systematic errors in deriving $\beta$
from UV--to--FIR flux ratios (see Meurer \etal\ 1999).  The
uncertainty in \extinctA\ can produce substantially differing UV--flux
correction factors (by an order of magnitude in flux), which directly
affects the inferred SFRs (see discussion below).  In some cases,
galaxies with  $\beta$ values that predict $\extinctA \simeq 0.5 -
2.5$~mag can be fit by models requiring negligible extinction.  For
these objects, the full--SED model fitting would suggest that the
UV--spectral slope is due to an  aging stellar population rather than
dust.  There are also systematic  offsets in the derived attenuations,
depending mainly on the  choice of metallicity for the population
synthesis models.  The lower  metallicity models are bluer, and thus
require more reddening to match  the observed colors.  On average, the
best--fit values of $\extinctA$ for the $Z=0.2$~\zsol\  synthesis
models are greater than those predicted from the UV slope alone by
$\delta(\extinctA) \simeq 0.35$~mag (0.20~mag) for Salpeter (Scalo)
IMF,  while the solar metallicity models nearly always result in
attenuation values lower than the $\beta$ predictions by $\simeq
0.6$~mag for both Salpeter and Scalo IMFs.  Therefore, we emphasize
that while it may be justifiable to apply an average flux correction
to a galaxy ensemble, there are very likely large uncertainties,  both
random and systematic, when making extinction corrections to
individual objects.

\subsubsection{Stellar Population Age
\label{section:age}}

As we have already noted, there is some degeneracy between age and
extinction when fitting models to the broad--band LBG photometry (see,
\eg, \S\ref{section:constraints} and  Figures~\ref{fig:lbg01} and
\ref{fig:lbg31}).  A similar trend is seen for the  composite
distribution in Figure~\ref{fig:comp_age_extinct}.  It is not clear
how much of this anticorrelation can be explained  by the combined
effect of the parameter degeneracies for  individual galaxies, or
whether there is some real tendency for younger galaxies to have more
extinction.  This might not be unexpected, because young star
formation within our own Galaxy tends to be highly dust--obscured, and
at later times, stars may blow away dust or migrate from their dusty
birthplaces.  On the other hand, one might expect very young galaxies
to be relatively free of dust and metals, unless these were produced
during previous episodes of star formation, before the birth of the
current generation of stars.

As can be seen from the composite probability plots
(Figure~\ref{fig:comp_age_extinct}), the best--fit solar metallicity
models span an age range of  $30\;\mathrm{Myr} \lsim t \lsim 1$~Gyr.
The upper range of population ages is similar for both low  ($Z =
0.2$~\zsol) and high (solar) metallicity models, but the low
metallicity models exhibit a tail of solutions with very young ages
(down to $\sim 4$~Myr) and high extinction, $\extinctA \sim 3$ to
4~mag with the Calzetti \etal\ dust model.   As discussed above  in
\S\ref{section:metallicity}, the low metallicity models are
intrinsically bluer, and thus require more extinction to fit the
observed colors at a fixed age.  It is notable that the fitting
solutions with very young ages and high extinction appear only  when
using the starburst attenuation law, which has a rather  ``grey''
wavelength dependence.  Fits with the steeper SMC dust  law avoid this
young and heavily reddened region of parameter space.

Although the population fitting with low metallicity models allows
very young ages ($\lsim 10^7$ yr), these timescales may be somewhat
unphysical based on the kinematics and linear size scales observed for
LBGs.  Nebular--emission--line widths of $\sigma \simeq 60-120$~\kms\
have been measured for several LBGs \citep{pet98,pet01,tep00a,moo00}.
The objects in our sample typically have physical diameters of
$3-10$~kpc.  Attributing the emission--line widths to virialized
motions, the corresponding dynamical timescales for these galaxies are
$t_\mathrm{dyn} \sim 10^{7.5-8}$~yr.  This may arguably set  a lower
bound on the time required for star formation to propagate  across the
galaxies, and thus on the age of their stellar populations.
Therefore, these solutions are possibly disfavored even if formally
allowed by the SED fits, unless one invokes an unusual geometry  in
which the star--formation regions and nebular emission lines  do not
originate in identical regions.  In \S\ref{section:sfrate} below, we
consider constraints on these solutions from limits on sub--mm
emission in the \hdf.

The geometric mean age for the composite distribution is $\simeq
70$~Myr (120~Myr) for models with metallicity 0.2~\zsol\ (1.0~\zsol)
and Salpeter IMF.  For Scalo IMF models, the distributions have
slightly lower mean ages, $\simeq 40$~Myr (70~Myr) for 0.2~\zsol\
(1.0~\zsol) metallicity.  For the solar metallicity, Salpeter IMF
model fits, three galaxies have best--fitting model ages $t \gsim
1$~Gyr, resulting from their moderately red $\nich-K$ colors and model
fits implying little or no dust extinction.  All three galaxies,
however, are at $z > 3.2$, where only the \ks\ data reaches  rest
frame $\lambda > 4000$~\AA, and for one object (NIC 284, by far the
faintest object in the sample), the photometry provides  only a weak
upper limit to the \ks\ flux.  For this reason, the possible older
ages for these galaxies must be regarded  with caution.

\begin{figure*}[th]
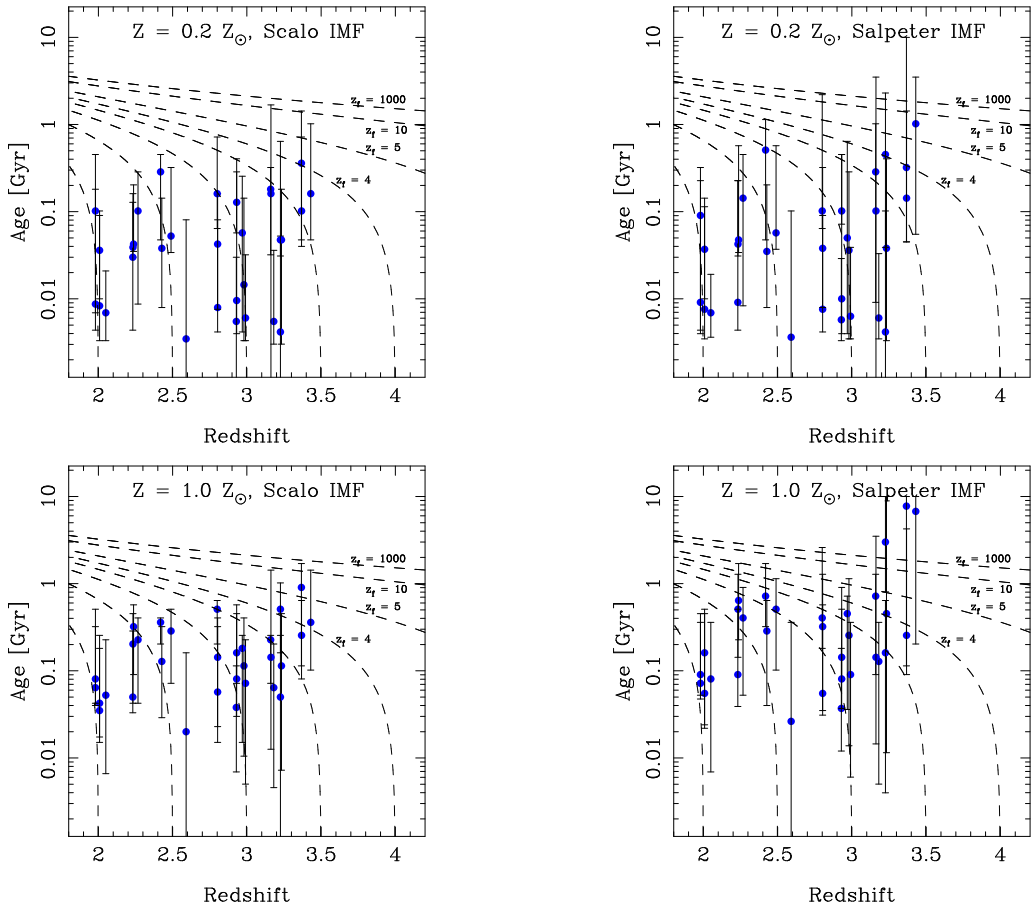

\epsscale{1.4}
\vbox{\plottwo{f15a.eps}{f15b.eps}} \vspace{0.1in}
\vbox{\plottwo{f15c.eps}{f15d.eps}}
\caption{The distribution of LBG stellar population ages as a function
of redshift.  The data points and error bars show the best--fit ages
and their 68\% confidence interval from the stellar population
synthesis models.  The dashed curves show the lookback time from the
observed galaxy redshift to constant ``formation'' redshifts,  $z_f =
2.0$, 2.5, 3.0, 3.5, 4.0, 5.0, 10.0, and 1000 (not all  curves are
labeled).
\label{fig:zedage}}
\end{figure*}

\citet{saw98} derived upper limits of $t < 200$~Myr on the ages of
the dominant stellar populations in most HDF LBGs, and concluded  that
LBGs are dominated by very young populations ($t \sim 25$~Myr).
Their models with $\delta$--function star--formation histories have UV
luminosities that decline rapidly, mandating young ages  for a
UV--bright (by definition) LBG population.  It is therefore more
useful to compare their constant star--formation models to our
results.   As can been seen from the individual galaxy confidence
distributions and the composite age--attenuation plots, many of the
LBGs in our sample have allowable models with ages on the order of
$t\sim 1$~Gyr.  However, it is also true that for the majority of
galaxies, our best--fit ages for low metallicity, Salpeter  IMF models
are also $< 200$~Myr, with a weighted mean  value of $\sim 70$~Myr,
only $\sim 2\times$ the median age  found by Sawicki \& Yee for $z <
3$ galaxies assuming the  same metallicity and constant star
formation.  Adopting more metal rich models or a steeper
UV--extinction law shifts the age distribution to larger values, with
median or weighted mean values in the range of $100-300$~Myr and a
substantial tail extending to $\sim 1$~Gyr or more.   Although we
generally find somewhat older ages than do Sawicki \& Yee, our results
do not {\it exclude} very young  ages of several tens of Myr for many
objects, at least not for  fits using the low metallicity models.

The ages derived from these best--fit stellar population models are
generally much younger than the age of the universe at the LBG
redshifts.  In Figure~\ref{fig:zedage}, we plot the 68\% confidence
interval on the model ages for each LBG versus the observed galaxy
redshift.   We also overplot curves showing the lookback times to
higher redshifts.   The results are somewhat dependent on the
particular choices for the IMF and metallicity, and derived ages only
apply to the youngest stellar populations under the chosen
star--formation histories, which do not necessarily apply to the
galaxies as a whole (\ie, the star--formation history for the entire
galaxy may indeed be more complex than the simple monotonically
decaying models used here).  The derived ages from these models
generally favor a scenario where most of the galaxies have formed
their \textit{observed} stellar populations (\ie, those stars that
dominate the rest--frame UV--to--optical SED) within a rather small
redshift interval prior to the  redshift at which they are observed,
at least for those objects at  $z \lsim 3.0$, where the model age
constraints are best.  If this is the case, then these LBGs assembled
(or, are assembling) their stellar masses relatively rapidly.  The LBG
sample is devoid  of old ($t \gsim 1$~Gyr), quiescent galaxies at $2
\lsim z \lsim 3$,  which would be expected to populate the upper--left
region of the panels  in Figure~\ref{fig:zedage}.  This might be
expected, because the galaxies  are selected to be UV--bright, but as
we have previously noted, there  are few (if any) candidates for old,
quiescent galaxies among the  NICMOS infrared--selected \hdf\ catalog
with \textit{photometric}  redshifts $2.0 \lsim z_\mathrm{ph} \lsim
3.5$ (see Figure~\ref{fig:cmd-vmh}).  I.e., there is little evidence
for a population of ``fading,''  post--burst remnants from evolved
galaxies formed at $z \gsim 3$.   However, the presence of older
stellar components {\it within}  the star--forming LBGs, cannot be
ruled out.  As we will see in  \S\ref{section:oldstars}, such older
stars could, in principle,  provide a significant fraction of the
galaxies' total stellar masses.  Under such a scenario, the fitted
ages for the young stellar components might not represent the ages of
the dominant (by mass) stellar populations.

\begin{figure*}[th]
\epsscale{1.5}
\vbox{\plottwo{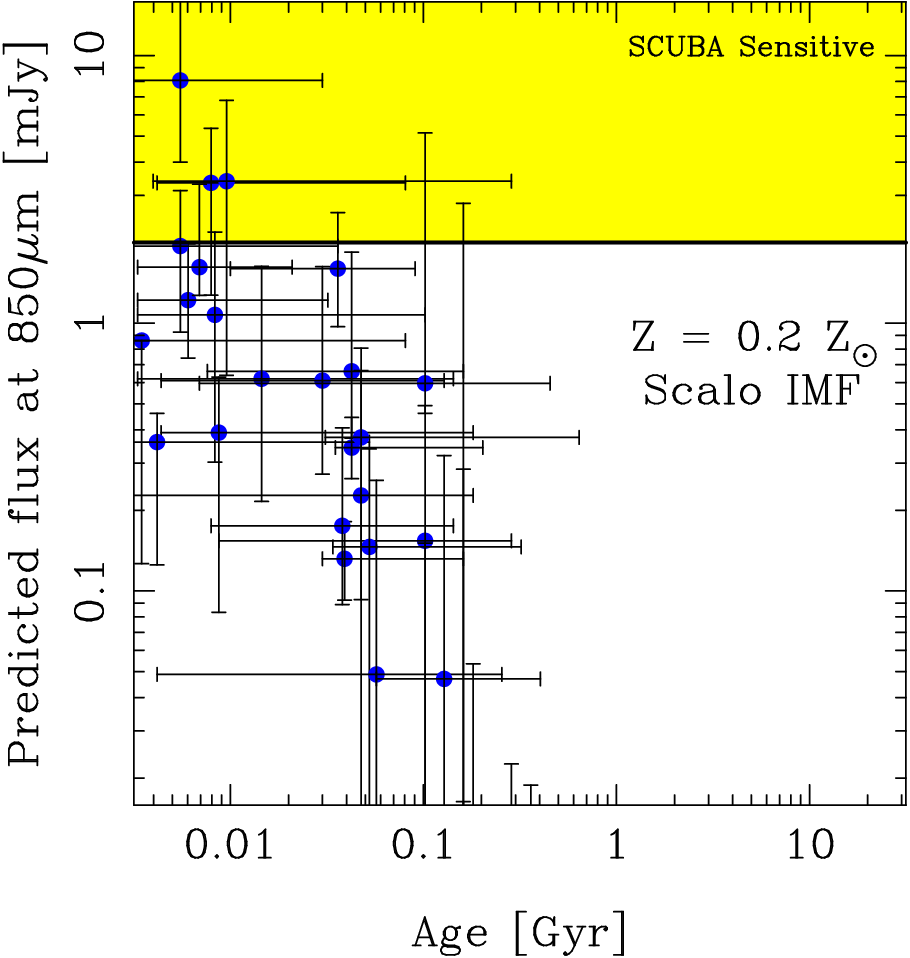}{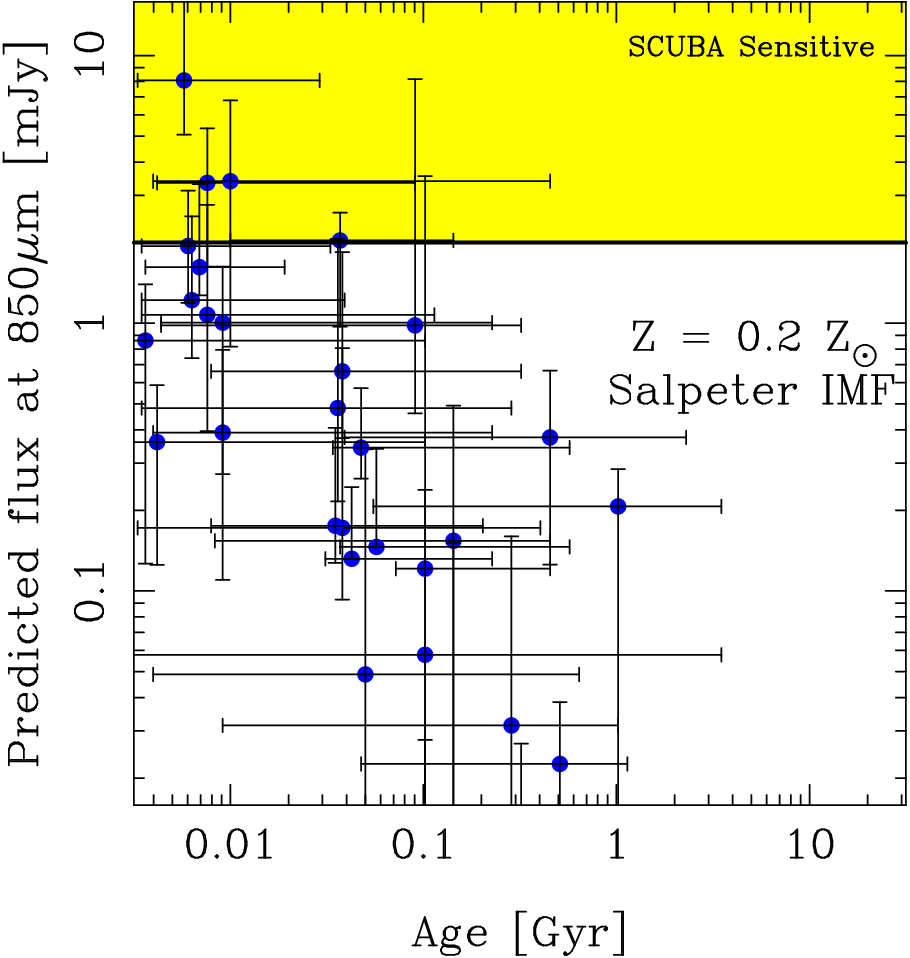}} \vspace{0.1in}
\vbox{\plottwo{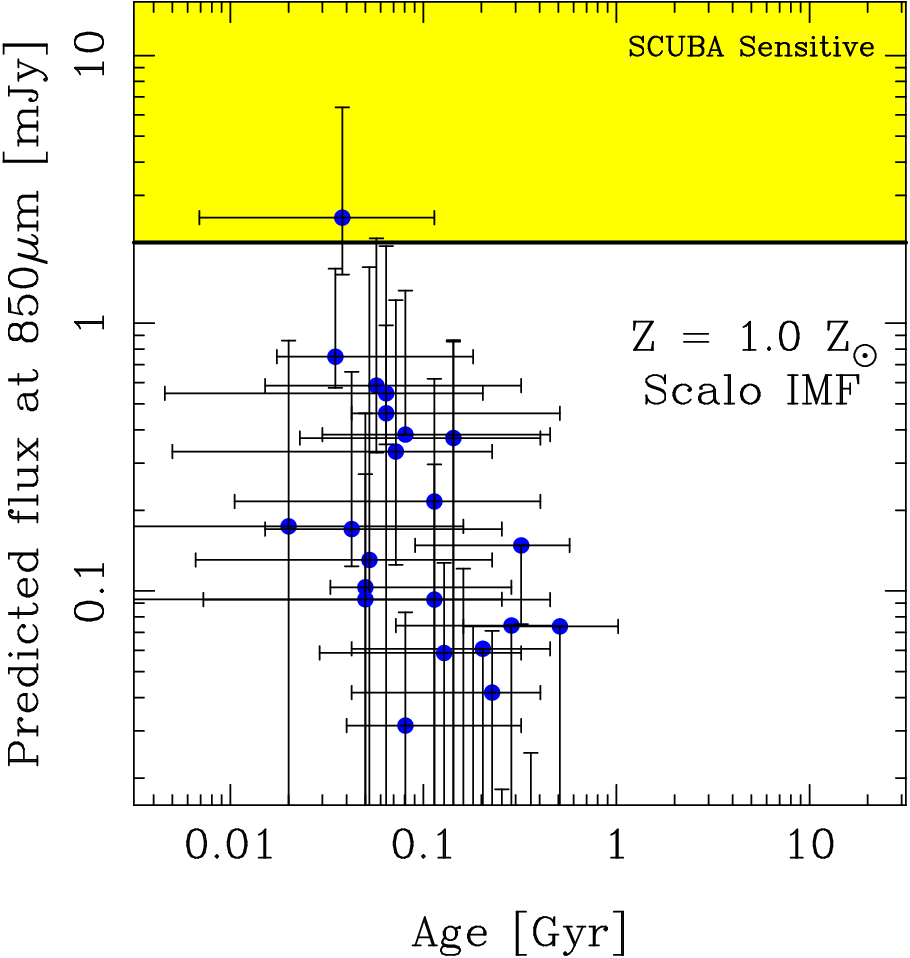}{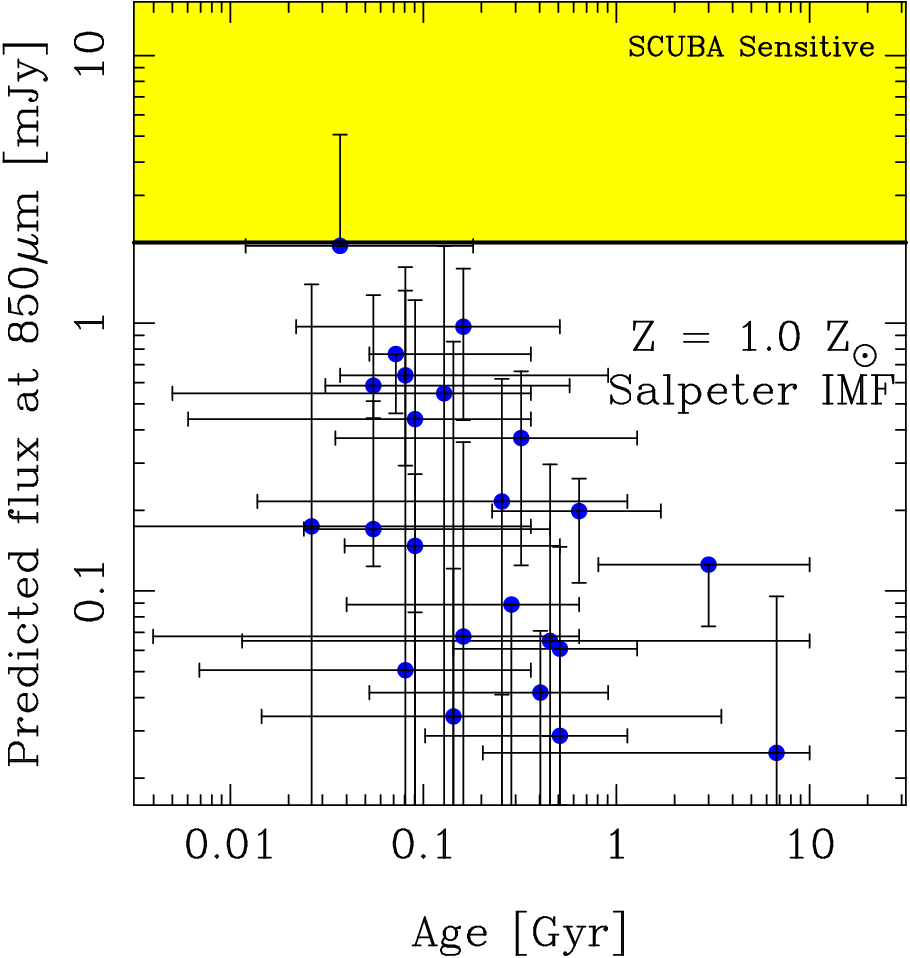}}
\caption{Predicted 850~\micron\ flux densities from the LBGs based on
the stellar population and extinction model fitting.  The  sub--mm
fluxes were predicted using the FIR--UV flux density correlation in
local starburst galaxies \citep{meu99,ade00},  with the dust
attenuations derived from the best--fitting models.  The four panels
correspond to various combinations of IMF and  metallicity, as
labeled.  The yellow shaded region in each panel  indicates
sensitivity range of the SCUBA 850~\micron\ survey of the \hdf,
$f_\nu(850\micron) \gsim 2$~mJy \citep{hug98}.  For the $0.2$~\Zsol\
models, one might expect that a few of the youngest, dustiest, and
most actively star--forming galaxies could have been detected with
SCUBA.
\label{fig:scuba}}
\end{figure*}

As with extinction, the strong dependence of the derived age
distribution on relatively uncertain parameters such as metallicity,
the extinction law, and the assumed prior star--formation histories
means that one must interpret these results with caution.   The
question of LBG ages cannot be completely resolved with the  available
broad--band photometry, and would benefit from a more  detailed
knowledge and independent data on the chemical enrichment  history and
dust content of these galaxies, as well as from  measurements of
starlight at still longer rest--frame wavelengths.

\subsubsection{Star Formation Rate
\label{section:sfrate}}

In Table~\ref{table:bestfits}, we list the instantaneous
star--formation rates for the best--fit models for each LBG.  One
interesting result is the wide range of SFRs [$10^{-2} \lsim \Psi /
(\msol\,\mathrm{yr}^{-1})\, \lsim 10^{3}$] derived for galaxies in the
sample,  and indeed the wide range even among acceptable models for
individual  galaxies.  The highest derived SFRs mostly correspond to
models with  young ages (mostly $< 15$~Myr) and small $t/\tau$.  Most
of these  are the young, low metallicity, high extinction models
discussed above.   For ages $\lsim 10^{7.5}$~yr and small $t/\tau$,
the UV luminosity emitted per unit SFR is lower than its asymptotic
value at later times because the OB stars that produce most of  the UV
light are still building up on the main sequence.  At later times,
their birth and death rates equilibrate, and the UV luminosity per
unit SFR stabilizes.  This, combined with  the large extinction
implied for these models, results in model SFRs that are substantially
larger than those which would be derived from the UV--luminosity
calibrations  generally used in the literature \citep[\eg,][]{mad98}.
The large SFRs are required in order to produce the inferred  stellar
mass on such short timescales.  Conversely, a few galaxies can be fit
with negligible amounts of ongoing star formation.   These are
``post--starburst'' models with $t \gg \tau$ and  ages 90--150~Myr,
where the UV light primarily originates from B stars that have not yet
left the main sequence.

For the young, low metallicity models with high extinction,  the
derived instantaneous SFRs are extremely high,  ($\mathrm{SFR} \gsim
1000$~\msol~yr$^{-1}$).  With this combination  of high SFRs and
extinction, one would expect that  most of the energy from star
formation should be absorbed by dust  and reradiated in the
far--infrared, perhaps with significant  emission at mid--infrared and
radio wavelengths as well.  To date, however, there have been no
robust detections of LBGs in the \hdf\ at mid--IR, far--IR or sub--mm
wavelengths \citep[][~although there are two possible radio
detections, Richards 1998]{gol97,hug98,dow99}.  Therefore we can
potentially use these surveys to constrain the intrinsic SFRs of these
LBGs.

In particular, we considered the SCUBA 850~\micron\ observations of
the \hdf\ \citep{hug98}, because sub--mm observations are, in
principle, quite sensitive to star formation in galaxies from $z\sim
1-10$ due to the strong negative $k$--correction.  Using the relations
given by \citet{ade00}, we predicted LBG flux densities for SCUBA at
850~\micron\ from the observed LBG (rest--frame) UV luminosity and the
dust attenuations  from the best fit models.  Figure~\ref{fig:scuba}
shows the  predicted 850~\micron\ SCUBA flux density as a function of
the LBG stellar population age.  For the solar metallicity models, all
of the LBGs have predicted  850~\micron\ flux densities that are below
(or very close to)  the 2~mJy limit of the \hdf\ SCUBA observations.
For the  $Z=0.2$~\zsol\ models, however, a few of the galaxies would
have  predicted 850~\micron\ flux densities $> 2$~mJy.  In most cases,
the predicted sub--mm emission is only slightly above the SCUBA
detection  threshold.  Given the range in the allowable \extinctA\
values for each  galaxy, the lack of SCUBA detections may not
conclusively rule out any particular model.\footnote{The $z=2.93$
galaxy NIC 522, or HDF 1-54.0, is the only  LBG with predicted
850~\micron\ flux density well above the SCUBA limit for  all model
solutions, with $f_\nu(850\micron) \simeq 3 - 10$~mJy ($\simeq
1-3$~mJy) for $Z=0.2$~\zsol\ (1.0~\zsol).    This object has a clearly
disturbed morphology and a steep UV spectral slope, and we note in
passing that this object lies only $\simeq 7$\arcsec\ from the
reported centroid position of the SCUBA source, 850.3
[$f_\nu(850\micron) = 3.0$~mJy] of \citet{hug98}.}  Moreover, recent
mm and sub--mm observations of gravitationally lensed high redshift
galaxies \citep{wer00,bak01} apparently find less far--IR emission
than predicted from the far--IR/UV correlations derived at $z=0$.
Conversely, \citet{meu00} find  that the UV--FIR correlation for local
starbursts substantially {\it underpredicts} the far--infrared
emission from nearby ultraluminous infrared galaxies.  At this time,
then, it seems difficult to set additional constraints on the SFRs
derived from our SED model fitting.  Ultimately, if the SFR can be
constrained from observations in the FIR, then this could constrain
the viable population synthesis models for each LBG.

\subsubsection{Stellar Mass
\label{section:mass}}

Figures~\ref{fig:comp_age_mass} and \ref{fig:comp_age_mass_smc} show
the distribution of best--fitting stellar--mass estimates  for the
spectroscopic LBG sample, under the assumption of the Calzetti and SMC
extinction laws, respectively.   Recall that due to the simple nature
of the star--formation histories of the models, we treat these stellar
mass derivations strictly as lower limits;  we will consider upper
limits to the stellar masses below (\S\ref{section:oldstars}).  It is
also  important to note that this is not a complete sample of
galaxies, and  certainly not a mass--limited one.  The distribution of
masses reflects  this fact, \eg, in terms of the lower mass limit for
the sample.

The (geometric) mean stellar mass of the composite distribution is
$\simeq 3 (6)\times 10^9$~\msol\ for the $Z=0.2$~\zsol\ (1.0~\zsol)
metallicity  and Salpeter IMF models, assuming the starburst
extinction relation.   Using a Scalo IMF shifts this value to $\simeq
7\times 10^9$~\msol\  (insensitive to metallicity).  The 68\%
confidence range of our composite  model mass distribution spans the
interval $\mathM \sim 10^{9-11}$~\msol,  similar to the range of
best--fit values for individual LBGs.  The inferred stellar mass would
be quite sensitive to the lower mass  cutoff of the IMF, a parameter
that we do not vary here  (see \S\ref{section:IMF}).   Thus, our
conclusions  are based on the assumption that this  cutoff does not
change from galaxy to galaxy.  There is a weak  correlation between
the stellar mass and the population age for  the Salpeter IMF models
(no correlation is apparent for either  the Scalo or Miller \& Scalo
IMF models).  The distribution of stellar masses for the LBG  sample
is nearly independent of the assumed dust--extinction law (cf.\
Figure~\ref{fig:comp_age_mass} and
Figure~\ref{fig:comp_age_mass_smc}).  We attribute this primarily to
the high--quality near--infrared data, which are sensitive to the
later--type stars that constitute the bulk of the stellar mass,  and
where the effects of dust extinction are reduced.   The LBG stellar
masses are nearly all smaller than those of  present--day,  \lstar\
galaxies ($1.4\times10^{11} h_{70}^{-2}$~\Msol\  for a Salpeter IMF,
Cole \etal\ 2001), but are similar to that of the  Galactic bulge,
$\sim 10^{10}$~\Msol\ \citep{ric99}.   \citet{saw98} estimated similar
median masses, $3-8\times 10^9$~\msol,  for $Z=0.2$~\zsol, Salpeter
IMF models (transforming to our assumed  cosmology).  In a new,
infrared photometric study of a sample of more  luminous (mostly $L >
L_{UV}^\ast$) LBGs, \citet{sha01} derive median  stellar masses (for
$Z = \zsol$, Salpeter IMF models) that are  $\sim 4\times$ larger than
those of our fainter HDF objects.

Somerville, Primack, \& Faber (2001) have considered various models
for the nature and evolution of high redshift galaxies.  In
particular, they present distributions of stellar ages and masses for
``mock--HDF'' catalogs derived using their ``collisional starburst''
model, where high redshift star formation is driven primarily by
mergers and tidal interactions.  For bright ($\wfv \leq 25.5$)
galaxies in the redshift interval  $2.0 \leq z \leq 3.5$, Somerville
\etal\ predict a broad range of stellar masses, $10^8 \lsim \mathM /
\msol \lsim 10^{11}$, with  a median value of $\mathM \approx 8\times
10^{9}$~\msol\ (for $h=0.7$).  This is somewhat larger than the mean
value (for Salpeter IMF models, as adopted by Somerville \etal) from
our small \hdf\ sample,  but the comparison is not precise because our
sample is not complete and magnitude limited at $\wfv \leq 25.5$.  The
Somerville \etal\ models include a small tail of objects with masses
greater than those of present--day \lstar\ galaxies,  $\mathM \gsim
10^{11}$~\msol, which are not seen in the \hdf\ sample.  However, this
may simply be due to small number statistics of  the \hdf\ itself.
More notably, the collisional starburst model  also predicts a
significant number of bright ($\wfv \leq 25.5$), low--mass ($10^8
\lsim \mathM / \msol \lsim 10^9$) galaxies, which are  also not found
in our LBG sample.  These are presumably small galaxies that have
recently experienced a massive episode of star formation, which pushes
them above the optical magnitude limit.  They could also be  young,
dust--free objects, which are also not found in our SED  analysis.  A
comparison with \hdf\ LBGs selected from a \wfv--based  catalog shows
that none with $\wfv \leq 25.5$ are missed in the  NICMOS--selected
sample.  Therefore, given that the colors and  fluxes of our
spectroscopically selected LBGs are representative of all galaxies
with photometric redshifts in this interval in  the \hdf\ (see
Figure~\ref{fig:cmd-vmh}), the absence of optically bright but
low--mass LBGs may point to a discrepancy with the collisional
starburst models.

\begin{figure*}[th]
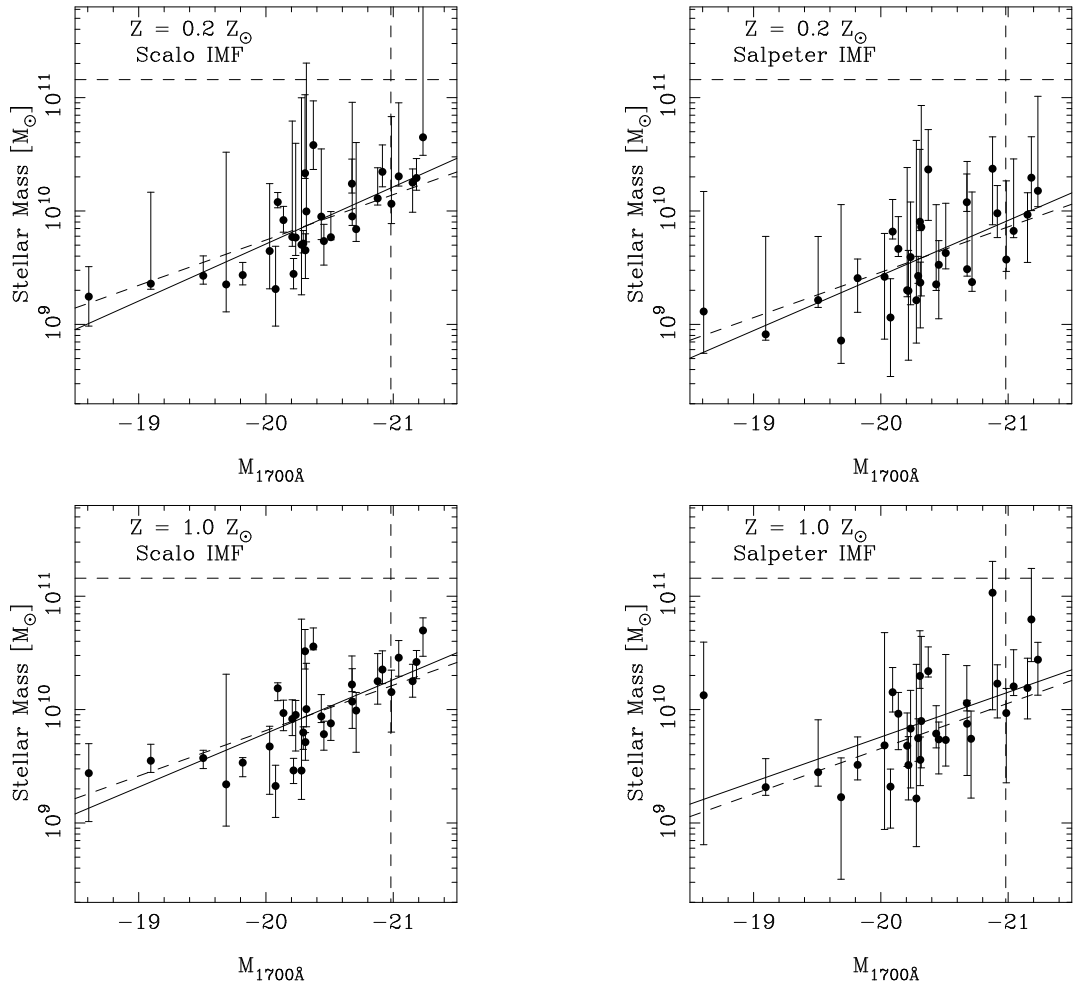

\epsscale{1.5}
\vbox{\plottwo{f17a.eps}{f17b.eps}} 
\vspace{0.1in}
\vbox{\plottwo{f17c.eps}{f17d.eps}}
\caption{ LBG stellar masses versus observed, rest--frame UV
luminosity.  The different panels show results from exponential
star--formation history models with various metallicity and IMF
assumptions,  and all assuming the \citet{cal00} starburst extinction
law.   The points mark the best--fit masses, with error bars
indicating  the 68\% confidence interval.  The UV luminosity is given
as  an absolute magnitude at 1700~\AA\ on the AB scale.  The
characteristic UV luminosity $\lstar_{1700}$ for LBGs at $z \approx 3$
from the luminosity function of Steidel \etal\ (1999) corresponds  to
$M^\ast_{1700} = -21.0$, marked by a vertical dashed line, while  the
stellar mass of a present--day \lstar\ galaxy is indicated by the
horizontal dashed line.  In each panel, the diagonal dashed line
corresponds to a constant mass--to--light ratio
$\mathM/L_\mathrm{UV}$.  The solid lines show a least squares fit  of
a straight line to the the data.
\label{fig:comp_mass}}
\end{figure*}

\begin{figure*}[th]
\epsscale{1.5}
\vbox{\plottwo{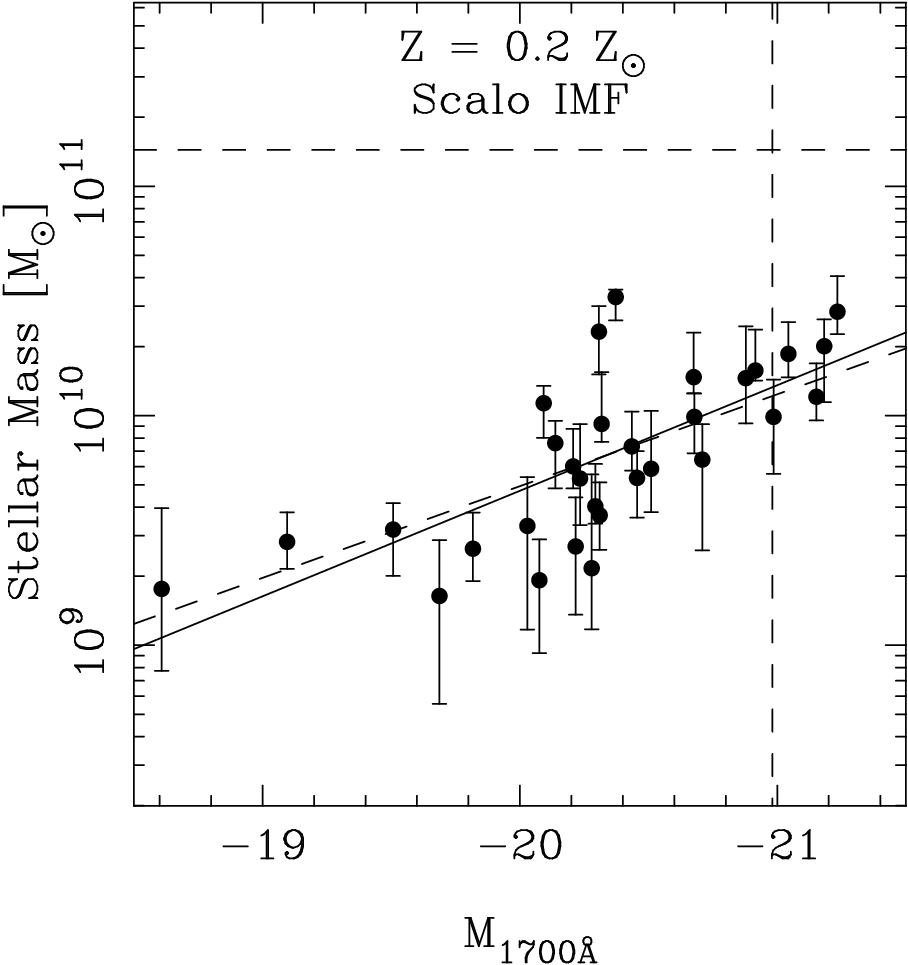}{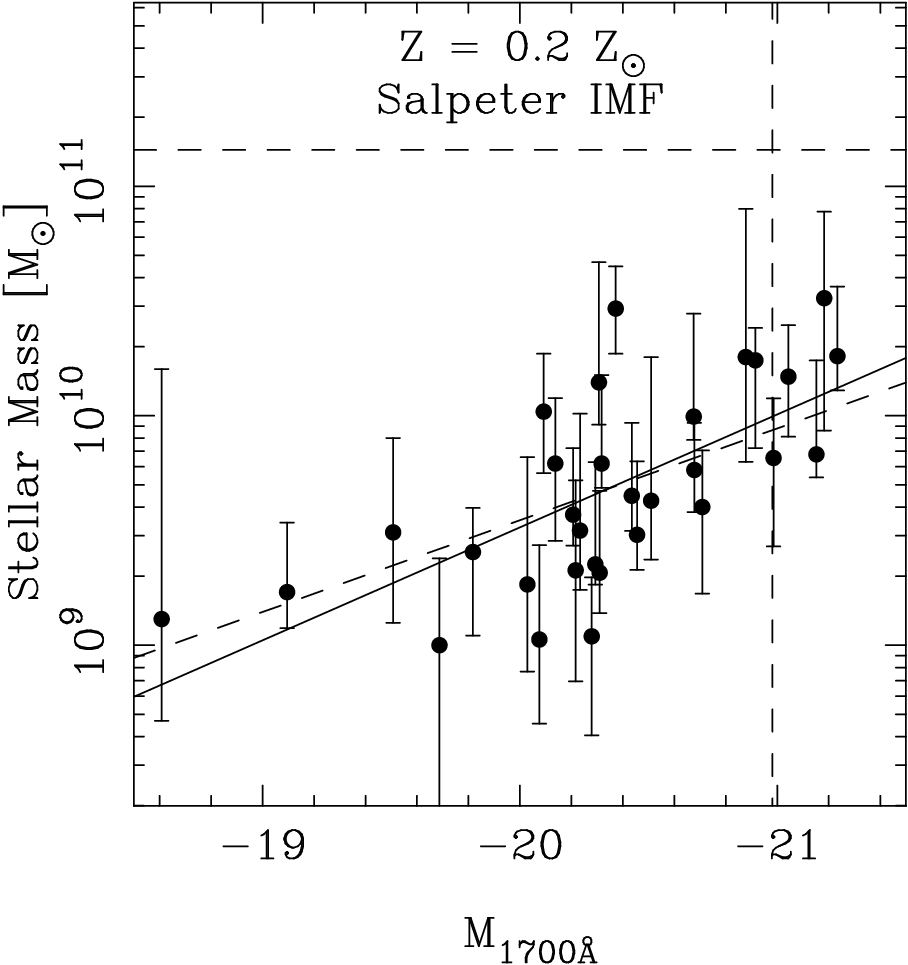}}
\vspace{0.1in}
\vbox{\plottwo{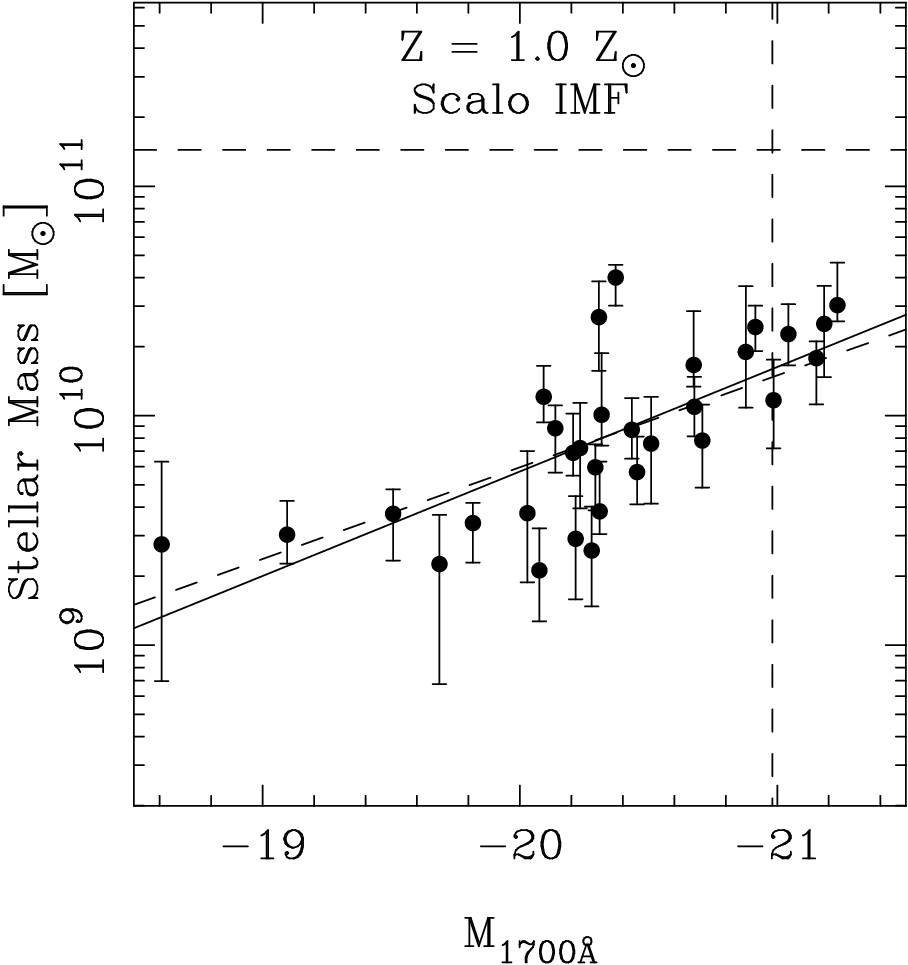}{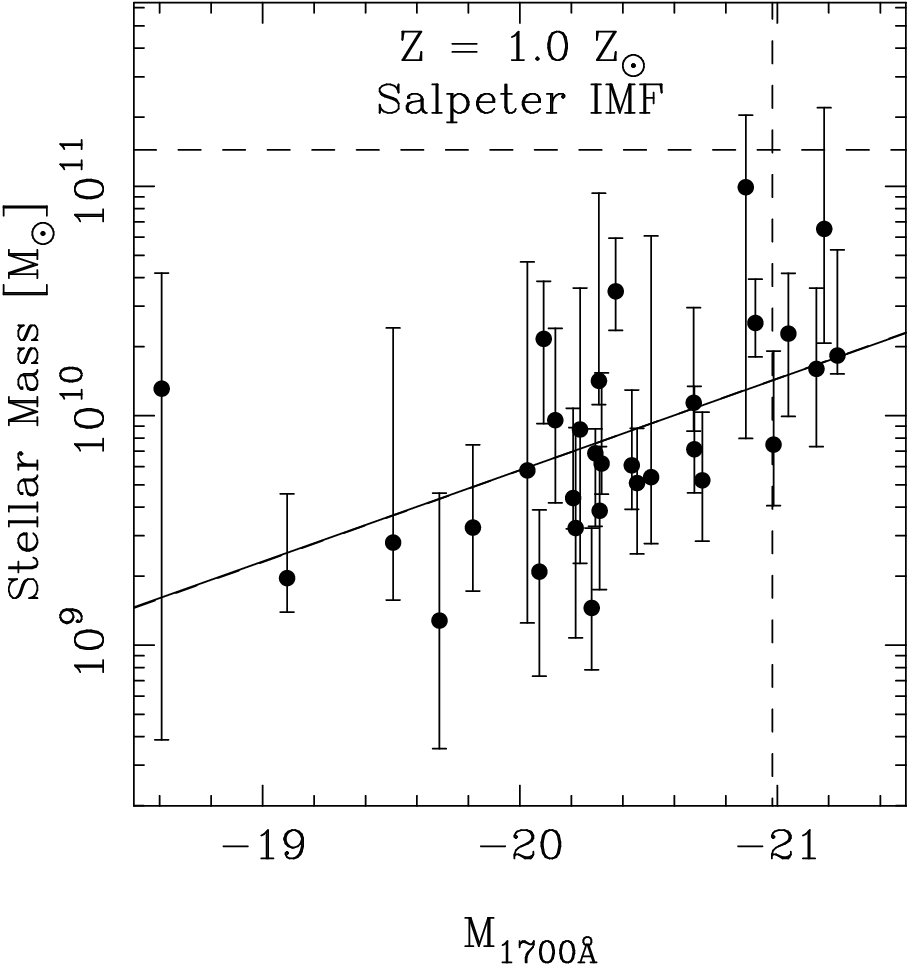}}
\caption{Same as Figure~\ref{fig:comp_mass}, but using the
SMC extinction law.
\label{fig:comp_mass_smc}}
\end{figure*}

We find a fairly good correlation between the derived stellar masses
and the rest--frame UV luminosity in these galaxies (see
Figures~\ref{fig:comp_mass} and \ref{fig:comp_mass_smc}).  This result
is somewhat surprising, because the UV light is generally expected to
trace the instantaneous SFR but not necessarily the stellar mass.  The
scatter of the best--fit values in this correlation is $\sigma(\log
\mathM/L_\mathrm{UV}) \simeq 0.3$~dex,  regardless of IMF and
metallicity.  This correlation could be misleading if the full range
of \hdf\ galaxies with $2.0 \lsim z_\mathrm{ph} \lsim 3.5$ do not
follow this trend.  This would require that the low--mass end of the
photometric redshift galaxy sample contain a large population of old
or highly extincted galaxies, which would produce larger $\wfv-\nich$
colors at fainter magnitudes.  There is no indication of any such
population;  indeed, if anything the  LBG color--magnitude diagram,
Figure~\ref{fig:cmd-vmh}, shows that  most of the redder LBGs are
\textit{brighter}.  In the next section, we will consider limits on
``hidden'' stellar mass from previous generations of star formation,
and it may be that the apparent $L_{UV}$--mass correlation would be
weakened  if such mass were indeed present.

The stellar masses of the brighter LBGs, with UV luminosities near the
\lstar\ value from \citet{ste99}, are of order $10^{10}$~\Msol.  It is
interesting to compare this with kinematically estimated  masses based
on measurements of nebular--emission--line widths for LBGs  (see
\S\ref{section:age}).  A naive application of the virial theorem
combining these line widths  and the typical half--light radii for
LBGs suggests masses that  are also $\sim 10^{10}$~\Msol.  Thus, the
stellar masses that we infer  for $\sim \lstar$ LBGs can account for
most or all of the kinematically  estimated mass.  In fact these
stellar--mass estimates are, if anything,  lower limits, measuring
only the mass of the ``young,'' actively star  forming component of
the galaxy.  Furthermore, as described in \S\ref{section:oldstars}
below, the total stellar mass could be larger if a maximal old stellar
population were present.  As \citet{pet98,pet01} have emphasized, the
nebular emission lines probably originate in  the UV--bright, central
regions of the galaxies, and thus may not  sample the full
gravitational potential well.  It is therefore unclear whether these
virial estimates really measure the total mass of the LBG dark matter
halos.  Unless the ratio of dark to luminous matter in LBGs is much
smaller than that in present--day galaxies, it appears that the simple
virial mass calculations based on the nebular line widths must
significantly underestimate the total masses of the dark--matter halos
in LBGs.

\subsection{Constraints on old stellar populations at 
$2 \lsim z\lsim 3.5$\label{section:oldstars}}

The masses computed from the best--fitting SEDs represent the
integrated stellar mass for the stellar population that dominates the
rest--frame UV--to--optical light of the galaxy.   Strictly speaking,
these estimates constitute only a lower limit  to the total integrated
stellar mass.  We may speculate about the existence of older stars in
these galaxies that do not  substantially contribute to the observed
light, but which might contain a significant fraction of the total
stellar mass.  Observations of star clusters in the Galactic bulge
(\eg, McWilliam \& Rich 1994, Ortolani \etal\ 1995) have shown that
ages of their stellar populations  are consistent with the age of the
Universe, implying very early formation epochs (see also Renzini
1999).  In principle, a significant mass fraction of old stars could
be present in LBGs, but invisible at UV to optical wavelengths, hidden
under the glare of younger stars.

In \S2 we noted that nearly all infrared--selected \hdf\ galaxies with
spectroscopic or photometric redshifts in the range $2 < z < 3.5$ have
blue optical--to--infrared colors, suggesting the presence of
on--going star formation, and that there were few, if any, candidates
for red, non--star--forming, old galaxies at these redshifts.   Could
old objects with $z\sim 3$ have faded past the NICMOS \hdf\ detection
limits?  For a non--star--forming, passively evolving galaxy at $z =
2.7$  with ages of $t= 0.5$, 1.0, 2.0, and 2.4~Gyr (the latter being
the age of  the Universe for the chosen cosmology), and assuming
negligible dust  extinction, the stellar masses corresponding to the
$\nich = 26.5$ limit  of the NICMOS \hdf\ image are $\mathM \simeq
0.2$, 0.5, 1.0, and $1.3\times 10^{10}$~\msol, while the corresponding
$\wfv - \nich$ colors are 3.7, 6.1, 7.3, and 7.4 mag, respectively.
Many of the LBGs in our sample have derived stellar masses that are
greater than these limits.  If old objects with comparable mass exist
at these redshifts, then they should be detected in the deep, NICMOS
data.   Similarly, the luminosities of LBGs from our sample at $3
\lsim z \lsim 3.5$ are large enough that  even if their star formation
were truncated at the observed epoch,  they would still be detectable
above the $\nich < 26.5$ limiting  magnitude $\sim 1$~Gyr later at
$2.0 \lsim z \lsim 2.5$.  While  these hypothetical old objects would
be too red to meet the LBG  selection criteria,  we would expect
photometric redshift techniques to identify them as high redshift,
early--type--galaxy candidates.  The ``$J$--dropout'' HDF-N--JD1 might
be such an object,  but there are few other examples.  In general,
there is little  difference in the color distribution  between the
spectroscopically  confirmed \hdf\ galaxies with  $2.0 \lsim z \lsim
3.5$, and those with  photometrically defined redshifts in the same
range, and thus we conclude  that there are few (if any) massive, old,
non--star--forming objects  in the \hdf\ in this redshift interval.

Although there are few candidates for red, non--star--forming galaxies
in this redshift range, there could still, in principle, be a
significant mass component of old stars underlying the active star
formation in the UV--bright LBGs.  The infrared photometry sets an
upper limit to the allowable total mass of the combined young and old
populations, and the fit to the full UV--to--optical SED sets a
maximum mass ratio of old to young stars, or else the  observed colors
become too red.

\begin{figure*}[th]
\epsscale{1.3}
\plotone{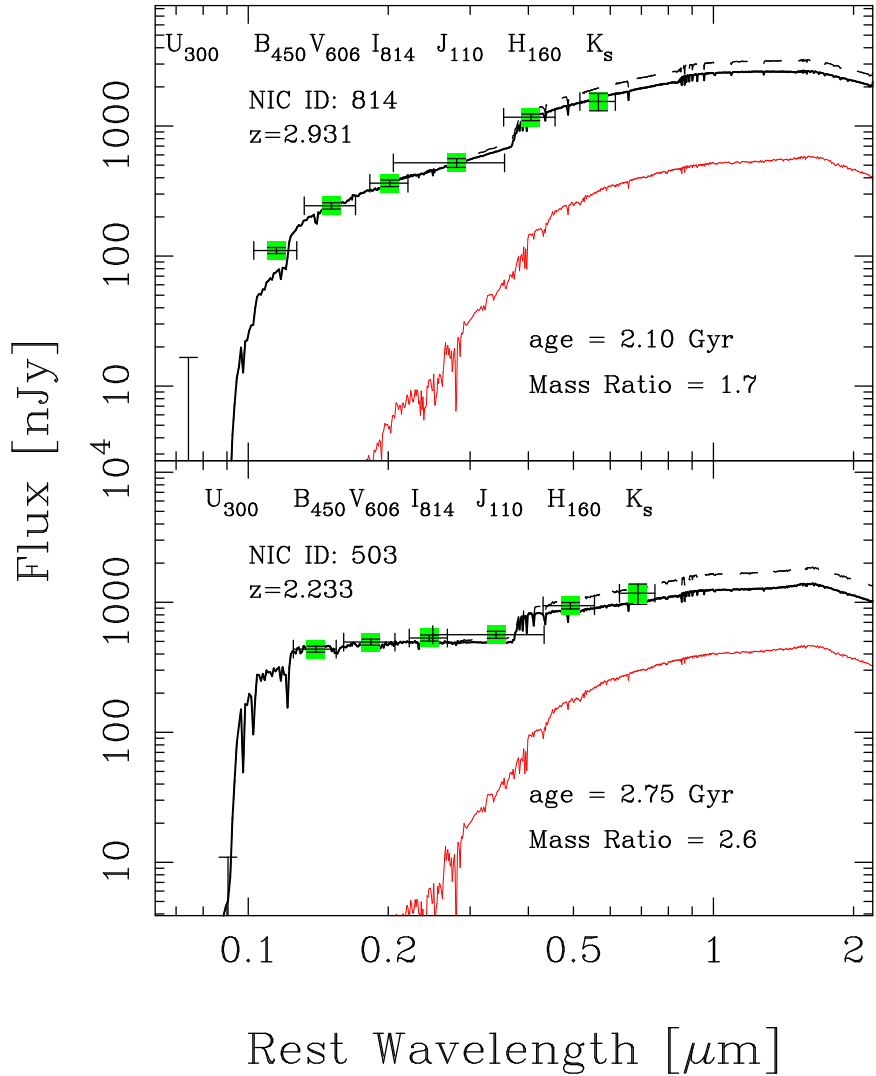}
\caption{Schematic illustration showing the effect of adding a
maximally  old stellar population on the best--fitting SEDs of LBGs
NIC 503 (HDF 2-903.0) and 814 (HDF 3-93.0).  For each object, the data
points show the measured photometry with flux  uncertainties and
bandpass widths indicated by error bars.  The solid black line  shows
the best--fitting ``young'' model SED for the $Z=0.2$~\zsol\ and
Salpeter IMF models from Table~\ref{table:bestfits}.  The red line
shows  the maximum allowable contribution to the SED from an
additional, old stellar population whose age is that of the universe
at the galaxy redshift.  The mass ratios of the old--to--young stellar
populations are given in the panels.  The dashed black line shows the
superposition  of the two components.
\label{fig:oldmassspec}}
\end{figure*}

To set limits on the LBG mass contribution from old stars, we
considered a simple, two--component star--formation--history model.
The ``young'' light from ongoing star formation is represented by the
generic suite of models with monotonically decreasing or constant
star--formation histories from \S4.1 (but restricted here to the
Salpeter IMF and metallicities $Z=0.2$ and 1.0~\zsol).  We then added
a second component to the SED from an old stellar population, formed
in a $\delta$--function burst that evolves passively without further
star formation.  We assumed that this component formed at $z =
\infty$, thus maximizing its age ($1.75-3.15$~Gyr for our adopted
cosmology) and mass--to--light ratio.  The superposition of the
two--component SEDs increases the galaxy luminosity predominantly for
rest--frame wavelengths, $\lambda_0 \gsim 4000$~\AA, as shown for two
fiducial LBGs in Figure~\ref{fig:oldmassspec}.  We then refit each LBG
with the full suite of two--component models using the same LBG
photometry and uncertainties as in \S\ref{section:models}.  This two
component model lets us set an upper limit to the LBG stellar mass,
using a stellar population with maximal $M/L$ while still matching the
observed UV--to--optical SED.

In order to gauge the fit improvement due to adding a model component
from an older stellar population, we applied an $F$--test on the
minimum values of $\chi^2$ per degree of freedom derived for each
galaxy using the two sets of models (the single--component and
two--component star--formation histories).  For most galaxies,  the
$F$--test indicates that both models fit the data equally well.
However, for a small fraction of the LBG sample, the addition of an
old stellar component does improve the goodness--of--fit.  Moreover,
because the additional model constraint does not significantly degrade
the fit to \textit{any} of the LBGs in the sample, this may actually
imply older stellar populations from previous star--formation episodes
are indeed present (or at least for some objects).  However, based on
the limited statistics, it is difficult to assign a significance to
this hypothesis.  One reasonable conclusion from this test is simply
that there is  not clear evidence for an \textit{absence} of old
stellar populations  underlying the active star formation in LBGs in
this sample.

\begin{figure*}[th]
\epsscale{1.6}
\plottwo{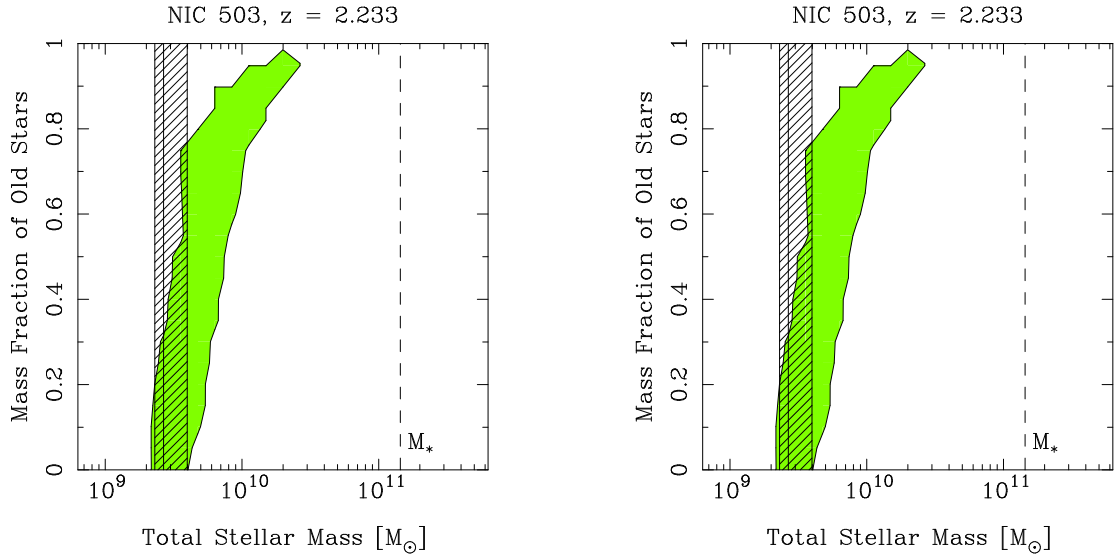}{f20b.eps}
\caption{Constraints on the fraction of stellar mass in a
hypothetical,  old stellar component for two fiducial LBGs.  The solid
line and the hatched  region show the best fitting value and 68\%
confidence region for the  total stellar mass from the single
component, ``young'' models with Salpeter IMF and $Z=0.2$~\zsol.  The
green shaded region shows the  approximate 68\% confidence region for
the additional stellar mass allowable from a maximally old stellar
population, in terms of the mass  fraction $\equiv
\mathM(\mathrm{Old\;Stars})/ [ \mathM(\mathrm{Old}) +
\mathM(\mathrm{Young}) ]$,  as a function of the total stellar mass,
$\mathM(\mathrm{Old}) + \mathM(\mathrm{Young})$,  for best fitting,
two component models.  The vertical dashed line  indicates the mass of
a present--day \lstar\ galaxy.
\label{fig:oldmass}}
\end{figure*}

The two--component models generally favor a slightly different  range
of parameters than those that best fit the single--age models.   The
best--fitting ages of the young stellar component  from the
two--component models are generally somewhat smaller  than those
derived in \S\ref{section:properties}, with correspondingly higher
dust extinction.  The mean values for the LBG sample are $t \approx
15$~Myr (30~Myr), and $\extinctA \approx 2.4$~mag (1.6~mag) for $Z =
0.2$~\Zsol\ (1.0~\Zsol).  This is simply a result of the
``decomposition'' of the observed SEDs into two components, where the
``old'' component influences the SED only past $\sim 4400$~\AA, with
only a small amount of recently formed stars in the ``young''
component, which dominate the UV portion of the SED.

\begin{figure*}[th]
\epsscale{1.6}
\plottwo{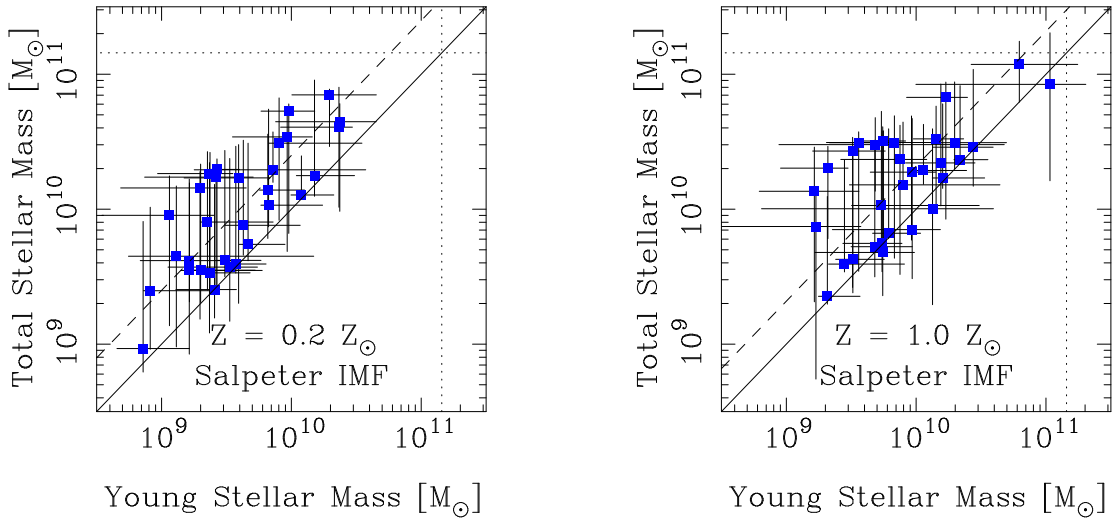}{f21b.eps}
\caption{The best--fitting total stellar mass [$\equiv
\mathM(\mathrm{Old\;Stars}) + \mathM(\mathrm{Young\;Stars})$] for each
LBG, plotted against the mass from the single--component models.  Each
data point shows the range of total stellar mass from the
two--component models as a function of the range of stellar masses
from the single--component model fits.  The error bars correspond to
the 68\% confidence range on each set of models.  The solid line shows
a one--to--one mass ratio, while the dashed line indicates the average
mass ratio, $X$, of the best--fit values for the LBGs, where $X =
(2.9,\; 3.1)$ for the $Z = (0.2,\;1.0)$~\zsol\ models, respectively.
The dotted lines mark the stellar mass of a present--day \lstar\
galaxy.
\label{fig:oldmasscomposite}}
\end{figure*}

The total stellar masses derived for these two--component models are
larger than those from \S\ref{section:properties}.  This is due to
the presence of the older stellar component, which has a large
mass--to--light ratio and does not contribute substantially to the
light, but do contribute significantly more mass than do the younger
stars.  Figure~\ref{fig:oldmass} shows the relative  mass fraction of
old stars as a function of the total stellar mass  (old + young stars)
for the same two fiducial LBGs shown in Figure~\ref{fig:oldmassspec}.
For these galaxies, and indeed for most  of the LBGs, there are viable
models where the old stellar populations contain $\gsim 90$\%\ of the
total stellar mass, while a minority  population of young stars
dominate the observed SED.  However, for  most galaxies, the total
stellar masses (old and young stars)  increase only by factors of a
few over the estimates derived using  the single--components models.
In Figure~\ref{fig:oldmasscomposite}, we compare the best--fitting
stellar masses from one-- and two--component star--formation history
models.  On average, the stellar masses from the two--component models
are $\approx 3\times$ the masses derived from  the single--component
models.  The (68\%) upper mass limits from the  two--component models
can be $\approx 6-8\times$ the best--fit single--component masses,
with these old stars contributing $\lsim 20$\% of  the light at
rest--frame optical wavelengths.  The fraction of mass that  can be
``hidden'' in an older stellar population increases with redshift  as
the observed bandpasses shift blueward, especially for $z \gsim 3$
where the Balmer break shifts through the F160W bandpass.  These
``maximal $\mathM/L$,'' two--component star--formation history models
set an upper bound to the total stellar mass, and the results from the
one-- and two--component models bound the allowable range of LBG
stellar masses for a given choice of IMF.  Tighter constraints on the
total stellar mass will require future observations at still longer
rest--frame wavelengths.  Deep observations with the SIRTF IRAC
instrument will be able to measure the rest--frame $K$--band
luminosities of LBGs, at these redshifts,  and should provide the best
census of LBG stellar mass at these redshifts.

\subsection{Evolution of Lyman Break Galaxies\label{section:evolution}}

The question remains, how do the LBGs fit into the history of galaxy
formation and evolution, and what is their fate at $z=0$?  As
discussed in the introduction, several evolutionary scenarios have
been proposed for LBGs.  Regardless of the physical parameters that
regulate star formation  in these objects, these different scenarios
predict substantially different end products for their evolution.  The
clustering observed for LBGs indicates that these galaxies are
strongly biased relative to the overall mass distribution in the
universe  (Giavalisco \etal\ 1998, Adelberger \etal\ 1998).  This
suggests  that they are hosted by the most massive dark matter halos
at these  redshifts, and that they will ultimately become parts of
fairly massive  galaxies today.  In this scenario, the LBGs may
plausibly represent the formation stages of the bulges and spheroidal
components of massive, present--day galaxies.   In contrast, from
their analyses of \hdf\ LBGs, \citet{saw98}  concluded that the LBGs
will form only  a small fraction ($\sim 5$\%) of the stellar mass
contained in a typical, \lstar\ galaxy today.  That scenario is also
consistent with one of the  models suggested by \citet{low97}, in
which the LBGs could be small,  sub--galactic clumps that will fade by
the present day to represent  present--day dwarf elliptical/spheroidal
galaxies.  This question  hinges on our understanding of the fraction
of stellar mass that  formed has formed at these high redshifts.

By fitting population synthesis models, we have found a range  of
star--formation histories that adequately match the observed
photometry for each galaxy in our sample.  If we assume that models
were to continue forming stars without change or interruption after
the time at which the galaxies are observed, we may predict the
subsequent evolution of these galaxies at lower redshifts.  For each
LBG, we selected all models within the 68\% confidence region of the
parameter space, \ie, those which best fit the photometric data.
Because a range of different models provides acceptable fits to  the
photometry, there is therefore a wide range of possible end--product
masses.  Star--formation models with $t \gsim \tau$  have already
produced the bulk of their stellar populations by the redshift where
the galaxy is observed.  The colors and luminosities of these models
at later times evolve essentially passively, and the stellar masses
remain roughly constant, although some of the mass becomes locked up
in stellar remnants or is returned to the interstellar medium as stars
burn off the main sequence.  Alternatively, models with $t \ll \tau$
continue to build stellar mass over their \efolding\ timescale,
$\tau$, growing by factors  of $\sim \tau/t$ over their lifetimes.
For most of the LBG sample, the best--fitting model remains
significantly less massive than a present--day \lstar\ galaxy.
However,  it is also true that nearly every galaxy has {\it possible}
models (in the sense that they provide acceptable fits to the observed
photometry) that will result in final stellar masses that are  {\it
greater} than those of \lstar\ galaxies today.  Clearly, galaxies
cannot continue to form stars continuously at high rates in this
fashion for long, because it is unlikely that they have sufficiently
large gas reservoirs to fuel such star formation indefinitely.

Although the best--fit stellar masses for nearly all of the \hdf\ LBGs
are significantly smaller than that of a present--day \lstar\ galaxy,
it  is important to remember that most of these galaxies are also
fainter than  the ``typical'' $z \approx 3$ LBGs studied  in the
ground--based survey of Steidel \etal\  This can be seen in
Figures~\ref{fig:comp_mass} and \ref{fig:comp_mass_smc},  where the
masses are plotted against the UV luminosities of the galaxies.  The
characteristic \lstar\ luminosity at 1700~\AA\ for  LBGs from
\citet{ste99} is approximately $M_{1700} = -21.0$, which is brighter
than most of the  \hdf\ LBGs.  \citet{gia01} have found that the
fainter \hdf\ LBGs cluster less strongly than do their brighter
counterparts from the ground--based surveys, suggesting that  they are
generally associated with less massive dark matter halos.  As a
result, it might not be surprising if many of them are not in fact
destined to become $L > \lstar$ galaxies today.  The stellar masses
plotted in Figures~\ref{fig:comp_mass} and \ref{fig:comp_mass_smc}
scale roughly linearly with  the UV luminosity, and the intercept  at
$M_{1700} = -21$ is roughly $10^{10} \Msol$, or $\sim 1/10$th  the
mass of a present--day \lstar\ galaxy.  As discussed in
\S\ref{section:oldstars}, we cannot discount the possibility that LBGs
may have a component of older stars  from previous episodes of star
formation.  This could, in principle,  significantly boost their
stellar masses with little effect  on their observed luminosities, but
unless all LBGs have the maximum  permissible amount of old stellar
mass (i.e., unless they actually formed most of their stars at
extremely high redshift, $z \gg 3$),  then it seems that LBGs with
$L^\ast$ UV luminosities must generally  have smaller stellar masses
than their present--day $L^\ast$ descendents.

\placefigure{fig:comp_mass}
\placefigure{fig:comp_mass2}
\placefigure{fig:comp_mass_smc}

For our adopted cosmology, the co--moving \hdf\ volume  between $z=2$
and 3.5 is approximately 27000$h_{70}^{-3}$~Mpc$^3$.  From the
$K$--band luminosity function of \citet{col01}, we would expect an
average of 23 galaxies with luminosities  $L \geq \lstar_K$ within
such a volume in the local universe.  Modulo variations due to
clustering, their progenitors must be present within this redshift
range in the \hdf, at some earlier stage of evolution -- perhaps in
many smaller pieces, or even as yet uncollapsed.   Of the galaxies
that we have analyzed here, only a few have best--fitting
star--formation histories that, if extrapolated  forward without
change to $z=0$, would produce all the stars found  in an $L^\ast$
galaxy today.  There are many other \hdf\ LBGs without  spectroscopic
redshifts (mostly fainter objects, see  Figure~\ref{fig:cmd-vmh}) that
we have not analyzed here, and some fraction of these may also grow to
comparable masses at $z=0$.  In general, then, we can say that the
best--fitting stellar population models for \hdf\ LBGs have masses
smaller than those  of \lstar\ galaxies today, and that most LBGs are
likely to remain sub--\lstar\ without further episodes  of star
formation (perhaps fueled by gas infall) or substantial merging at
lower redshifts.  These \hdf\ LBGs are less luminous, and thus
arguably less massive, than their counterparts from the larger
ground--based surveys, and it is unclear whether or not there are
enough objects that could ``passively'' evolve to produce the massive
end of the present--day galaxy population without additional star
formation or merging.  However, these histories are not definitive:
for nearly every LBG, there exist allowable models {\it capable} of
producing (or even overproducing) the total stellar mass  of a
present--day, \lstar\ galaxy.

An interesting puzzle results from the fact that the ages of the  $2.0
< z < 2.5$ sub--sample of LBGs are generally much shorter than the
span of cosmic time across the redshift range of our entire sample
[$t(2 < z < 3.5) \simeq 1.5$~Gyr]. Because the masses derived for LBGs
at $z \sim 3 - 3.5$ are comparable to those at $z\sim 2-2.5$,  we
might expect to find a population of old, quiescent galaxies among the
many \hdf\ objects with (photometrically estimated) redshifts $2.0
\lsim z \lsim 2.5$.  Indeed, given the lifetimes we estimate for the
LBG star--formation episodes (with median stellar population  ages
$t\sim 0.1$~Gyr), such ``old'' galaxies should outnumber the observed
LBGs by a factor of $\sim 1.5\;\mathrm{Gyr} / 0.1\;\mathrm{Gyr} = 15$.
We do not see such a population of old, high--redshift objects,  given
the narrow range colors observed for galaxies at these redshifts  (see
Figure~\ref{fig:cmd-vmh}) to the flux limits of the NICMOS data.
Moreover, \citet{ste99} find that the UV luminosity functions of LBGs
are indistinguishable (at least for luminosities around $L^\ast$) at
$z\sim 3$ and $z\sim 4$.  The $z \sim 4$ objects must evolve into {\it
something}, but again there is no evidence  for a population of faded
remnants at $z < 3.5$.  Thus the fate  of the the $z\sim 3$ LBG
population remains somewhat enigmatic.

One possible solution may be found if the simple star--formation
prescription (\ie, constant or declining SFR with time) used in our
models is incorrect for most objects.  Galaxies might maintain  a
youthful appearance, despite long star--formation timescales, if the
SFRs generally {\it increased} with time over this redshift range.
This, however, may be difficult to reconcile with the apparent lack of
evolution in the UV luminosity function between $z = 3$ and 4.
Alternatively, multiple star--forming episodes  with a duty cycle of
$t_d \lsim 1$~Gyr might be the norm.  In this way, the stars from high
redshift LBGs are incorporated into their lower redshift counterparts,
with intermittent episodes of star formation to rejuvenate the
emergent UV--to--optical  spectrum in order to avoid overproducing
galaxies with red colors  or old SED ages.  Interestingly, when we fit
LBGs using the two--component  models (ongoing star formation plus a
maximally--old stellar component, see \S\ref{section:oldstars}), we
frequently derive quite small ages  for the ``young'' component (often
$t \lsim 30$~Myr), which may also be  an indication that short bursts
are the normal mode for star formation at these epochs.

This complex, stochastic evolutionary history is reminiscent of the
collisional starburst models of \citet{som01}, although in detail
their  model appears to predict too broad a range of LBG stellar
masses compared  to the values we derive.  In the general context of
hierarchical models  for galaxy formation, one might naturally expect
that young galaxies are  the sites of vigorous merging activity and
gas accretion, which can trigger  episodic (but frequent) star
formation.  The total stellar masses  resulting from these complex
star--formation histories should be bounded  at the minimum by the
single--component models we have used in this paper,  and at the
maximum by the two--component models with a maximally old stellar
population.


\section{Summary}

Using WFPC2 (\wfu\wfb\wfv\wfi), NICMOS (\nicj\nich), and ground--based
\ks--band data, we have investigated the stellar populations of 33
\hdf\ LBGs  with redshifts $2.0 \lsim z \lsim 3.5$, reaching the
following conclusions:

1. The LBGs have rest--frame UV--to--optical SEDs that are dissimilar
   to present--day, ``Hubble Sequence'' spiral and elliptical
   galaxies, but are comparable to those of local starburst galaxies
   with modest amounts of dust extinction.  Examining complete galaxy
   samples based on photometric redshifts, we find little evidence for
   a substantial population of galaxies with redder colors at similar
   redshifts.

2. By comparing the LBG data in seven broad--bands to stellar
   population synthesis models, we derive good constraints on the
   total stellar masses for the LBGs in our sample.  The minimum value
   is derived using single--component star--formation histories, with
   mean values that range from $3\;(6)\times 10^9$~\msol\ for
   $Z=0.2$~\zsol\ (1.0~\zsol) and a Salpeter IMF to $7\times
   10^9$~\msol\ using a Scalo IMF (for both low and solar metallicity
   models).  For galaxies with $L^\ast$ UV luminosities, the inferred
   masses are $\sim 10^{10}$~\msol, or roughly 1/10th that of  a
   present--day, \lstar\ galaxy.  Furthermore, we constrain the
   maximum  LBG masses by including a second model component,
   corresponding to a maximally--old (and hence maximum  $M/L$)
   stellar population hidden in the glare  of the ongoing
   star--formation.  We find, on average, that a hypothetical old
   stellar population could in principle contain up to $\approx
   3-8\times$ the stellar mass of the young stars  that dominate the
   observed SED.  The stellar masses we derive are  comparable to the
   simple kinematic estimates for LBG masses from  nebular--line
   widths, suggesting that those measurements underestimate  the total
   mass of the LBG dark matter halos.  We find a fairly  tight
   correlation between the derived stellar masses and  UV luminosity,
   with a scatter of $\approx 0.3$~dex.

3. The allowable (68\% confidence) SED models generally place only
   loose constraints on the stellar--population ages, dust
   attenuation, and prior star--formation history.  For solar
   metallicity models, the best--fitting ages for individual galaxies
   span a range of  30 Myr to $\sim 1$~Gyr, and have dust attenuation
   factors of $0-3$~mag  at 1700~\AA.  For models with sub--solar
   metallicity, however,  a number of objects are well fit with very
   young  ages ($\lsim 10$~Myr) and high extinction ($A_{1700} =
   2.5-4$~mag).  The mean ages and extinctions for our LBG sample thus
   depend  somewhat on the particular choice of IMF and metallicity,
   with ages  of 70~Myr (120~Myr) for 0.2~\zsol\ (1.0~\zsol)
   metallicity and Salpeter IMF, and 40~Myr (70~Myr) for 0.2~\zsol\
   (1.0~\zsol) metallicity and Scalo IMF.  The typical dust
   attenuation  correction at 1700~\AA\ is a factor of 4.4 (3.0) for
   $Z = 0.2$~\zsol\ (1.0~\zsol), although the total, luminosity
   weighted mean flux correction for the sample as a whole is larger,
   13.8 (4.6), due to larger fitted extinctions for some galaxies with
   the largest inferred SFRs.  These values are consistent with those
   from \citet{ste99}  based on rest--frame UV observations alone of
   nearly 1000 LBGs.  These average correction factors, while possibly
   applicable to the  LBG population as a whole, are probably not very
   useful for  individual galaxies  due to the wide allowable range of
   parameter  space for each LBG.  We also  find that no galaxy in our
   LBG sample  is well fit by young, dust--free,
   ``first--generation'' stellar  populations.  Such objects must be
   quite  rare at these redshifts,  or too faint to meet the LBG color
   selection  criteria, or must  have SEDs that are significantly
   different from those produced  by the stellar population models we
   use here.

4. The extrapolated evolution of the best--fit SED models generally
   indicates that LBGs will remain sub--\lstar--sized galaxies today
   unless they undergo subsequent merging or gas accretion.  However,
   for nearly all the LBGs, there are allowable (68\% confidence)
   models  capable of producing the mass of a present--day \lstar
   galaxy, and in  some cases many times that of an \lstar--galaxy by
   $z=0$.  Based on  the derived mass and age distributions, we
   suggest that star formation  at high redshifts may be driven by
   discrete, recurrent events,  which could result from mergers, tidal
   interactions, etc.   These starbursts must occur with a relatively
   short duty cycle ($\sim 10^8\,\mathrm{Gyr}/10^9$~Gyr), and a
   timescale between star--formation events of $t \lsim 1.5$~Gyr,  in
   order to keep rejuvenating the galaxy SEDs and explain the
   apparent absence of non--star--forming, evolved, massive galaxies
   at $z \sim 2$.   The LBGs may indeed represent the building sites
   of  massive present--day galaxies, but the dominant stellar--mass
   assembly  may occur via repeating starburst events rather than
   quiescent, continuous star formation.


\acknowledgements

We would like to thank the other members of our HDF--N/NICMOS GO team
who have contributed to many aspects of this program, and the STScI
staff who helped to ensure that the observations were carried out in
an optimal manner.  We are grateful to the referee, Kurt Adelberger,
for a thorough report, and suggestions that greatly improved the
quality and clarity of this paper.  We thank Matthew Bershady, Peter
Eisenhardt, Richard Elston, and Adam Stanford who helped collect the
ground--based $K$--band data, and Tamas Budav\'ari for providing his
photometric redshift catalog.  Adam Riess offered valuable guidance
with statistical issues related to the model fitting.  We also thank
Stephane Charlot and Gustavo Bruzual for providing the most recent
version of their population  synthesis code, and Peter Eisenhardt,
Jennifer Lotz, Gerhardt Meurer, and Max Pettini for useful discussions
and suggestions.   Support for this work was provided by NASA through
grant number GO-07817.01-96A from the Space Telescope Science
Institute, which is operated by the Association of Universities for
Research in Astronomy, Inc., under NASA contract NAS5-26555.



\begin{deluxetable}{lccccc}
\tablewidth{0pt}
\tablecaption{Observations of the \hdf.\label{table:hdfdata}}
\tablehead{
	\colhead{} & \colhead{} & 
	\colhead{$\lambda_\mathrm{eff}$} & \colhead{Exposure Time} &
	\colhead{$m_\mathrm{lim}$} & \colhead{PSF} \\
	\colhead{Camera} & \colhead{Bandpass} & \colhead{$(\mu\mathrm{m})$} &
	\colhead{(ksec)} & \colhead{(AB)\tablenotemark{a}} &
	\colhead{FWHM}
}
\startdata
WFPC2  &  F300W  &  0.30  &  153.7  &  27.0 & 0\farcs14 \\
       &  F450W  &  0.46  &  120.6  &  27.9 & \\
       &  F606W  &  0.60  &  109.1  &  28.2 & \\
       &  F814W  &  0.80  &  123.6  &  27.6 & \\[5pt]
NIC3   &  F110W  &  1.10  &  \phn 12.6   &  26.5 & 0\farcs22 \\
       &  F160W  &  1.60  &  \phn 12.6   &  26.5 & \\[5pt]
KPNO IRIM & \ks\  &  2.20  &  \phn 82.4   &  23.8 & 1\arcsec\phn\phn
\enddata
\tablenotetext{a}{The WFPC2 and NICMOS limiting magnitudes are $10\sigma$ in a 0.2 arcsec$^2$ aperture.  The IRIM \ks-band limiting magnitude is $5\sigma$ in a 2 arcsec$^2$ aperture.}
\end{deluxetable}


\begin{deluxetable}{llccccccc}
\tablewidth{0pt}
\tablecaption{\hdf\ Lyman Break Galaxy Sample\label{table:lbgdata}}
\tablehead{
	\colhead{} & \colhead{} & \colhead{} & 
	\colhead{RA\tablenotemark{c}} & 
	\colhead{DEC\tablenotemark{c}}& \colhead{} & 
	\colhead{} & \colhead{} &
	\colhead{} \\
	\colhead{NIC ID\tablenotemark{a}} &
	\colhead{WFPC2 ID\tablenotemark{b}} &
	 \colhead{$z$} & 
	\colhead{(s)} & 
	\colhead{(mm:ss.s)}& \colhead{\nich} & 
	\colhead{$V_{606}-H_{160}$} & \colhead{$H_{160}-K_s$} &
	\colhead{Ref.}
}
\startdata
162\tablenotemark{d}  & 2-585.2    & 1.980 & 49.74 & 14:15.6 & 24.46 & 1.10 & $< 1.02$ & 5,8 \\
163\tablenotemark{d}  & 2-585.1    & 1.980 & 49.82 & 14:14.8 & 22.44 & 1.82 & \phs0.32 & 5,8 \\
110  & 2-449.0    & 2.005 & 48.30 & 14:16.6 & 22.52 & 1.21 & \phs0.28 & 1 \\
109  & 2-454.0    & 2.009 & 48.23 & 14:18.5 & 23.49 & 0.97 & \phs0.08 & 7  \\
1513 & 3-875.0    & 2.050 & 37.07 & 12.25.3 & 23.06 & 0.94 & $-0.09$ & 8 \\
843  & 3-118.0    & 2.232 & 54.71 & 13:14.8 & 23.56 & 0.92 & \phs0.19 & 6 \\
503  & 2-903.0    & 2.233 & 55.06 & 13:47.1 & 23.94 & 0.70 & \phs0.34 & 2 \\
274  & 2-525.0    & 2.237 & 50.09 & 14:01.1 & 23.58 & 1.24 & \phs0.51 & 6 \\
67   & 2-82.1     & 2.267 & 44.07 & 14:10.1 & 23.86 & 0.67 & $-0.12$ & 2 \\
561  & 2-824.0    & 2.419 & 54.61 & 13:41.4 & 24.67 & 0.63 & $< -0.73$ & 2 \\
97\tablenotemark{e} & 2-239.0 & 2.427 & 45.89 & 14:12.1 & 23.99 & 0.75 & \phs0.69 & 6 \\
741  & 2-591.2   & 2.489 & 53.18 & 13:22.7 & 23.99 & 0.90 & \phs0.29 & 6 \\
989  & 4-445.0    & 2.500 & 44.64 & 12:27.4 & 22.61 & 1.55 & \phs0.26 & 3,8 \\
804  & 4-639.0    & 2.591 & 41.72 & 12:38.8 & 24.53 & 0.47 & \phs0.52 & 1 \\
1352 & 4-497.0    & 2.800 & 45.74 & 11:57.3 & 25.11 & 0.82 & \phs0.32 & 8 \\
782  & 4-316.0    & 2.801 & 45.02 & 12:51.1 & 23.20 & 1.17 & \phs0.34 & 8 \\
1357\tablenotemark{d} & 4-555.1    & 2.803 & 45.41 & 11:53.1 & 23.21 & 1.26 & \phs0.28 & 1 \\
1358\tablenotemark{d} & 4-555.2    & 2.803 & 45.30 & 11:52.2 & 23.49 & 0.85 & \phs0.45 & 1 \\
522  & 1-54.0     & 2.929 & 44.10 & 13:10.8 & 22.96 & 1.44 & \phs0.36 & 6 \\
813\tablenotemark{d}  & 4-52.0     & 2.931 & 47.72 & 12:55.8 & 23.64 & 1.29 & \phs0.13 & 2 \\
814\tablenotemark{d}  & 3-93.0     & 2.931 & 47.86 & 12:55.4 & 23.65 & 1.70 & \phs0.40 & 2 \\
1063 & 4-289.0    & 2.969 & 46.94 & 12:26.1 & 25.26 & 0.18 & \phs0.76 & 6 \\
1541 & 4-363.0    & 2.980 & 48.31 & 11:45.8 & 24.51 & 0.88 & \phs0.56 & 2 \\
661  & 2-643.0    & 2.991 & 53.42 & 13:29.4 & 24.15 & 0.79 & \phs0.02 & 2 \\
273  & 2-76.11    & 3.160 & 45.34 & 13:47.0 & 25.24 & 0.24 & \phs1.23 & 2 \\
367  & 2-565.0    & 3.162 & 51.19 & 13:48.8 & 24.15 & 1.11 & $-0.99$ & 6 \\
282  & 2-901.0    & 3.181 & 53.58 & 14:10.2 & 23.97 & 0.91 & $-0.27$ & 2 \\
1114\tablenotemark{d} & 4-858.11   & 3.220 & 41.17 & 12:02.9 & 25.68 & 0.53 & \phs 0.98 & 1,2,4,8 \\
1115\tablenotemark{d} & 4-858.0    & 3.220 & 41.26 & 12:03.0 & 23.89 & 0.77 & \phs1.00 & 1,2,4,8 \\
947  & 3-243.0    & 3.233 & 49.80 & 12:48.8 & 25.51 & 0.55 & \phs1.64 & 2 \\
284  & 2-834.2    & 3.367 & 52.98 & 14:08.5 & 26.53 & 0.65 & $< 1.41$ & 8 \\ 
516  & 2-637.0    & 3.368 & 52.74 & 13:39.1 & 24.88 & 0.54 & \phs0.38 & 2 \\
553  & 2-604.0    & 3.430 & 52.40 & 13:37.8 & 24.43 & 0.80 & \phs1.16 & 2 \\
\enddata
\tablenotetext{a}{ID numbers refer to the catalog of Dickinson \etal\ 2001.}
\tablenotetext{b}{ID numbers refer to the catalog of Williams \etal\ 1996.}
\tablenotetext{c}{RA add 12h36m; Declination add +62d (J2000)}
\tablenotetext{d}{Part of a close pair (separation $\lsim 1^{\prime\prime}$) split in our catalog and assumed (or confirmed) to have the same redshift.}
\tablenotetext{e}{Multiple faint components in our catalog have been remerged (NICMOS IDs 96, 97, and 98).}
\tablerefs{(1) Steidel \etal\ 1996; (2) Lowenthal \etal\ 1997; (3) Cohen \etal\ 1996; (4) Zepf \etal\ 1997; (5) Elston \etal, private communication; (6) Dickinson 1998; (7) Lowenthal, Simard, \& Koo 1998; (8) Cohen \etal\ 2000}
\end{deluxetable}


\begin{deluxetable}{lccccccccc}
\tablewidth{0pt}
\tablecaption{Best Fit Stellar Population Synthesis Model Parameters\label{table:bestfits}}
\tablehead{
        \colhead{} & \colhead{} & \colhead{$Z$} & \colhead{} &
        \colhead{Age} & \colhead{$\tau$} & 
        \colhead{\extinctA\tablenotemark{\dag}} & 
        \colhead{$\log \mathcal{M}$} & \colhead{SFR} & \colhead{} \\
        \colhead{NIC ID} & \colhead{$z$} & \colhead{(\Zsol)} & 
        \colhead{IMF} & \colhead{(Myr)} & \colhead{(Myr)} & \colhead{(mag)} &
        \colhead{(\Msol)} &     \colhead{(\Msol~yr$^{-1}$)} &
        \colhead{$\chi^2_0$} \\
        \colhead{(1)} & \colhead{(2)} & \colhead{(3)} & 
        \colhead{(4)} & \colhead{(5)} & \colhead{(6)} & \colhead{(7)} &
        \colhead{(8)} & \colhead{(9)} & \colhead{(10)}
}
\startdata
162     &  1.98 & 0.2 & Salpeter &  \phn\phn\phn   9.1 & \phn\phn\phn   10 & 3.1 & \phn  9.0 & \phn\phn\phn 68.9\phn\phn &  \phn  9.2 \\* 
        &       & 1.0 & Salpeter &      \phn\phn  90.5 & \phn\phn\phn   10 & 0.5 & \phn  9.4 & \phn\phn\phn\phn 0.035 &  \phn  9.2 \\* 
        &       & 0.2 & Scalo    &  \phn\phn\phn   8.7 & \phn\phn\phn   10 & 3.1 & \phn  9.4 & \phn\phn 201\phd\phn\phn\phn & \phn  9.1 \\* 
        &       & 1.0 & Scalo    &      \phn\phn  80.6 & \phn\phn\phn   20 & 1.0 & \phn  9.6 & \phn\phn\phn\phn 4.12\phn &  \phn  9.1 \\
\tableline
163     &  1.98 & 0.2 & Salpeter &      \phn\phn  90.5 & \phn\phn\phn   10 & 2.2 &      10.5 & \phn\phn\phn\phn 0.392 &  \phn  9.5 \\* 
        &       & 1.0 & Salpeter &      \phn\phn  71.9 & \phn\phn\phn   10 & 2.2 &      10.4 & \phn\phn\phn\phn 2.33\phn &  \phn  9.7 \\* 
        &       & 0.2 & Scalo    &          \phn 102.0 & \phn\phn\phn   30 & 2.4 &      10.7 & \phn\phn\phn 57.0\phn\phn &  \phn  9.1 \\* 
        &       & 1.0 & Scalo    &      \phn\phn  64.1 & \phn\phn\phn   10 & 2.2 &      10.6 & \phn\phn\phn\phn 7.56\phn &  \phn  9.4 \\
\tableline
110     & 2.008 & 0.2 & Salpeter &      \phn\phn  37.0 & \phn 2000 & 3.1 &      10.1 & \phn\phn 345\phd\phn\phn\phn & \phn  3.3 \\* 
        &       & 1.0 & Salpeter &          \phn 161.0 & \phn 2000 & 2.4 &      10.3 & \phn\phn 148\phd\phn\phn\phn & \phn  4.2 \\* 
        &       & 0.2 & Scalo    &      \phn\phn  34.0 & \phn\phn\phn   50 & 2.9 &      10.4 & \phn\phn 585\phd\phn\phn\phn & \phn  3.5 \\* 
        &       & 1.0 & Scalo    &      \phn\phn  35.0 & \phn\phn\phn   10 & 2.2 &      10.4 & \phn\phn\phn 90.0\phn\phn &  \phn  4.3 \\
\tableline
109     & 2.008 & 0.2 & Salpeter &  \phn\phn\phn   7.6 & \phn\phn\phn   10 & 3.1 & \phn  9.4 & \phn\phn 217\phd\phn\phn\phn & \phn  1.2 \\* 
        &       & 1.0 & Salpeter &      \phn\phn  55.0 & \phn\phn\phn   10 & 1.2 & \phn  9.8 & \phn\phn\phn\phn 2.75\phn &  \phn  2.5 \\* 
        &       & 0.2 & Scalo    &  \phn\phn\phn   7.9 & \phn\phn\phn   30 & 3.1 & \phn  9.9 & \phn\phn 810\phd\phn\phn\phn & \phn  1.2 \\* 
        &       & 1.0 & Scalo    &      \phn\phn  42.5 & \phn\phn\phn   10 & 1.4 &      10.0 & \phn\phn\phn 15.1\phn\phn &  \phn  2.3 \\
\tableline
1513    &  2.05 & 0.2 & Salpeter &  \phn\phn\phn   6.9 & \phn\phn\phn   10 & 3.1 & \phn  9.6 & \phn\phn 372\phd\phn\phn\phn & \phn  1.8 \\* 
        &       & 1.0 & Salpeter &      \phn\phn  80.6 & \phn\phn\phn   10 & 0.5 & \phn 10.0 & \phn\phn\phn\phn 0.333 &  \phn  5.8 \\* 
        &       & 0.2 & Scalo    &  \phn\phn\phn   6.9 & \phn\phn\phn   10 & 3.1 &      10.0 & \phn 1067\phd\phn\phn\phn &  \phn  1.7 \\* 
        &       & 1.0 & Scalo    &      \phn\phn  52.5 & \phn\phn\phn   10 & 1.0 &      10.2 & \phn\phn\phn\phn 7.71\phn &  \phn  5.5 \\
\tableline
843     & 2.232 & 0.2 & Salpeter &  \phn\phn\phn   9.1 & \phn\phn\phn   10 & 2.9 & \phn  9.4 & \phn\phn 190\phd\phn\phn\phn & \phn  1.1 \\* 
        &       & 1.0 & Salpeter &      \phn\phn  90.5 & \phn\phn\phn   30 & 1.2 & \phn  9.9 & \phn\phn\phn 15.2\phn\phn &  \phn  0.6 \\* 
        &       & 0.2 & Scalo    &      \phn\phn  30.0 & \phn\phn\phn   30 & 2.4 &      10.0 & \phn\phn 217\phd\phn\phn\phn & \phn  0.9 \\* 
        &       & 1.0 & Scalo    &      \phn\phn  50.0 & \phn\phn\phn   10 & 1.0 &      10.0 & \phn\phn\phn\phn 7.55\phn &  \phn  0.5 \\
\tableline
503     & 2.233 & 0.2 & Salpeter &      \phn\phn  42.5 & \phn\phn\phn   10 & 1.2 & \phn  9.5 & \phn\phn\phn\phn 5.31\phn &  \phn  3.2 \\* 
        &       & 1.0 & Salpeter &          \phn 509.0 & \multicolumn{1}{c}{$\infty$} & 0.7 & \phn  9.8 & \phn\phn\phn 16.4\phn\phn &  \phn  4.6 \\* 
        &       & 0.2 & Scalo    &      \phn\phn  39.0 & \phn\phn\phn   10 & 1.2 & \phn  9.8 & \phn\phn\phn 13.4\phn\phn &  \phn  3.9 \\* 
        &       & 1.0 & Scalo    &          \phn 203.0 & \multicolumn{1}{c}{$\infty$} & 0.7 & \phn  9.9 & \phn\phn\phn 39.9\phn\phn &  \phn  6.4 \\
\tableline
274     & 2.237 & 0.2 & Salpeter &      \phn\phn  47.5 & \phn\phn\phn   10 & 2.2 & \phn  9.9 & \phn\phn\phn\phn 7.60\phn &  \phn  0.9 \\* 
        &       & 1.0 & Salpeter &          \phn 641.0 & \multicolumn{1}{c}{$\infty$} & 1.7 &      10.2 & \phn\phn\phn 32.1\phn\phn &  \phn  2.1 \\* 
        &       & 0.2 & Scalo    &      \phn\phn  42.5 & \phn\phn\phn   10 & 2.2 &      10.2 & \phn\phn\phn 20.8\phn\phn &  \phn  1.6 \\* 
        &       & 1.0 & Scalo    &          \phn 321.0 & \multicolumn{1}{c}{$\infty$} & 1.4 &      10.3 & \phn\phn\phn 59.9\phn\phn &  \phn  4.0 \\
\tableline
67      & 2.267 & 0.2 & Salpeter &          \phn 143.0 & \phn\phn  300 & 1.2 & \phn  9.6 & \phn\phn\phn 25.3\phn\phn &  \phn  0.9 \\* 
        &       & 1.0 & Salpeter &          \phn 404.0 & \phn 5000 & 0.7 & \phn  9.8 & \phn\phn\phn 18.9\phn\phn &  \phn  1.0 \\* 
        &       & 0.2 & Scalo    &          \phn 102.0 & \phn 2000 & 1.2 & \phn  9.8 & \phn\phn\phn 64.6\phn\phn &  \phn  0.7 \\* 
        &       & 1.0 & Scalo    &          \phn 227.0 & \phn 3000 & 0.5 & \phn  9.9 & \phn\phn\phn 32.4\phn\phn &  \phn  0.8 \\
\tableline
561     & 2.419 & 0.2 & Salpeter &          \phn 509.0 & \multicolumn{1}{c}{$\infty$} & 0.5 & \phn  9.5 & \phn\phn\phn\phn 7.31\phn &  \phn  4.6 \\* 
        &       & 1.0 & Salpeter &          \phn 719.0 & \phn 1000 & 0.0 & \phn  9.6 & \phn\phn\phn\phn 4.59\phn &  \phn  4.2 \\* 
        &       & 0.2 & Scalo    &          \phn 286.0 & \phn 3000 & 0.2 & \phn  9.5 & \phn\phn\phn 11.4\phn\phn &  \phn  4.7 \\* 
        &       & 1.0 & Scalo    &          \phn 360.0 & \multicolumn{1}{c}{$\infty$} & 0.0 & \phn  9.6 & \phn\phn\phn 11.9\phn\phn &  \phn  5.2 \\
\tableline
97      & 2.427 & 0.2 & Salpeter &      \phn\phn  35.0 & \phn\phn\phn   10 & 1.4 & \phn  9.5 & \phn\phn\phn\phn 9.05\phn &  \phn  6.8 \\* 
        &       & 1.0 & Salpeter &          \phn 286.0 & \multicolumn{1}{c}{$\infty$} & 1.0 & \phn  9.6 & \phn\phn\phn 16.9\phn\phn &  \phn  7.4 \\* 
        &       & 0.2 & Scalo    &      \phn\phn  38.0 & \phn\phn\phn   20 & 1.4 & \phn  9.7 & \phn\phn\phn 45.6\phn\phn &  \phn  7.2 \\* 
        &       & 1.0 & Scalo    &          \phn 128.0 & \phn\phn  200 & 0.7 & \phn  9.8 & \phn\phn\phn 34.0\phn\phn &  \phn  7.7 \\
\tableline
741     & 2.489 & 0.2 & Salpeter &      \phn\phn  57.1 & \phn\phn\phn   10 & 1.4 & \phn  9.8 & \phn\phn\phn\phn 2.16\phn &  \phn  2.6 \\* 
        &       & 1.0 & Salpeter &          \phn 509.0 & \phn 7000 & 1.4 &      10.1 & \phn\phn\phn 26.3\phn\phn &  \phn  2.6 \\* 
        &       & 0.2 & Scalo    &      \phn\phn  52.5 & \phn\phn\phn   10 & 1.4 &      10.0 & \phn\phn\phn\phn 5.54\phn &  \phn  2.7 \\* 
        &       & 1.0 & Scalo    &          \phn 286.0 & \phn\phn  700 & 1.0 &      10.1 & \phn\phn\phn 34.4\phn\phn &  \phn  3.0 \\
\tableline
989     &   2.5 & 0.2 & Salpeter &  \phn\phn\phn   2.5 & \phn\phn\phn   10 & 5.8 &      10.8 & 23043\phd\phn\phn\phn &  \phn  7.9 \\* 
        &       & 1.0 & Salpeter &  \phn\phn\phn   4.8 & \phn\phn\phn   20 & 5.1 &      10.3 & \phn 3918\phd\phn\phn\phn &  \phn  8.1 \\* 
        &       & 0.2 & Scalo    &  \phn\phn\phn   2.6 & \phn\phn\phn   10 & 5.3 &      11.1 & 42544\phd\phn\phn\phn &  \phn  9.6 \\* 
        &       & 1.0 & Scalo    &  \phn\phn\phn   4.2 & \phn\phn\phn   10 & 5.1 &      10.8 & 12250\phd\phn\phn\phn &  \phn  8.6 \\
\tableline
804     & 2.591 & 0.2 & Salpeter &  \phn\phn\phn   3.6 & \phn\phn\phn   20 & 2.9 & \phn  9.3 & \phn\phn 505\phd\phn\phn\phn & \phn  2.5 \\* 
        &       & 1.0 & Salpeter &      \phn\phn  26.3 & \phn\phn\phn   10 & 1.4 & \phn  9.3 & \phn\phn\phn 17.5\phn\phn &  \phn  1.4 \\* 
        &       & 0.2 & Scalo    &  \phn\phn\phn   3.5 & \phn 2000 & 2.9 & \phn  9.8 & \phn 1795\phd\phn\phn\phn &  \phn  2.6 \\* 
        &       & 1.0 & Scalo    &      \phn\phn  20.0 & \phn\phn\phn   10 & 1.4 & \phn  9.6 & \phn\phn\phn 57.8\phn\phn &  \phn  1.5 \\
\tableline
1352    &   2.8 & 0.2 & Salpeter &          \phn 102.0 & \phn\phn\phn   10 & 0.0 & \phn  9.3 & \phn\phn\phn\phn 0.009 &  \phn  6.0 \\* 
        &       & 1.0 & Salpeter &          \phn 404.0 & \phn\phn  200 & 0.0 & \phn  9.5 & \phn\phn\phn\phn 3.08\phn &  \phn  5.2 \\* 
        &       & 0.2 & Scalo    &          \phn 161.0 & \phn\phn\phn   50 & 0.0 & \phn  9.5 & \phn\phn\phn\phn 2.77\phn &  \phn  6.1 \\* 
        &       & 1.0 & Scalo    &          \phn 509.0 & \phn 3000 & 0.0 & \phn  9.7 & \phn\phn\phn\phn 8.46\phn &  \phn  5.5 \\
\tableline
782     & 2.801 & 0.2 & Salpeter &  \phn\phn\phn   1.0 & \phn\phn\phn   10 & 5.1 &      10.6 & 32518\phd\phn\phn\phn &       13.8 \\* 
        &       & 1.0 & Salpeter &  \phn\phn\phn   1.0 & \phn\phn\phn   10 & 5.1 &      10.4 & 23486\phd\phn\phn\phn &       16.8 \\* 
        &       & 0.2 & Scalo    &  \phn\phn\phn   2.5 & \phn\phn\phn   10 & 4.8 &      10.9 & 29417\phd\phn\phn\phn &       14.3 \\* 
        &       & 1.0 & Scalo    &  \phn\phn\phn   1.0 & \phn\phn\phn   10 & 4.8 &      10.8 & 56860\phd\phn\phn\phn &       16.8 \\
\tableline
1357    & 2.803 & 0.2 & Salpeter &  \phn\phn\phn   7.6 & \phn\phn\phn   10 & 3.6 & \phn  9.9 & \phn\phn 745\phd\phn\phn\phn & \phn  0.4 \\* 
        &       & 1.0 & Salpeter &      \phn\phn  55.0 & \phn\phn\phn   10 & 1.7 &      10.3 & \phn\phn\phn\phn 9.48\phn &  \phn  0.6 \\* 
        &       & 0.2 & Scalo    &  \phn\phn\phn   7.9 & \phn\phn\phn   20 & 3.6 &      10.4 & \phn 2588\phd\phn\phn\phn &  \phn  0.5 \\* 
        &       & 1.0 & Scalo    &      \phn\phn  57.1 & \phn\phn\phn   20 & 1.9 &      10.6 & \phn\phn 114\phd\phn\phn\phn & \phn  0.5 \\
\tableline
1358    & 2.803 & 0.2 & Salpeter &      \phn\phn  38.0 & \phn\phn\phn   10 & 1.9 &      10.1 & \phn\phn\phn 29.7\phn\phn &  \phn  1.5 \\* 
        &       & 1.0 & Salpeter &          \phn 321.0 & \phn 7000 & 1.4 &      10.3 & \phn\phn\phn 72.1\phn\phn &  \phn  2.1 \\* 
        &       & 0.2 & Scalo    &      \phn\phn  42.5 & \phn\phn\phn   20 & 1.9 &      10.3 & \phn\phn 156\phd\phn\phn\phn & \phn  2.1 \\* 
        &       & 1.0 & Scalo    &          \phn 143.0 & \phn 7000 & 1.4 &      10.3 & \phn\phn 162\phd\phn\phn\phn & \phn  2.9 \\
\tableline
522     & 2.929 & 0.2 & Salpeter &  \phn\phn\phn   5.8 & \phn\phn  200 & 4.3 &      10.3 & \phn 3095\phd\phn\phn\phn &  \phn  1.5 \\* 
        &       & 1.0 & Salpeter &      \phn\phn  37.0 & \phn\phn\phn   10 & 2.9 &      10.5 & \phn\phn\phn 94.4\phn\phn &  \phn  2.1 \\* 
        &       & 0.2 & Scalo    &  \phn\phn\phn   5.5 & \phn\phn  500 & 4.3 &      10.7 & \phn 9661\phd\phn\phn\phn &  \phn  1.5 \\* 
        &       & 1.0 & Scalo    &      \phn\phn  38.0 & \phn\phn\phn   30 & 3.1 &      10.8 & \phn\phn 806\phd\phn\phn\phn & \phn  2.0 \\
\tableline
813     & 2.931 & 0.2 & Salpeter &          \phn 102.0 & \phn\phn\phn   10 & 1.0 &      10.2 & \phn\phn\phn\phn 0.062 &  \phn  3.0 \\* 
        &       & 1.0 & Salpeter &          \phn 143.0 & \phn\phn\phn   10 & 0.0 &      10.2 & \phn\phn\phn\phn 0.001 &  \phn  1.1 \\* 
        &       & 0.2 & Scalo    &          \phn 128.0 & \phn\phn\phn   10 & 0.5 &      10.3 & \phn\phn\phn\phn 0.006 &  \phn  2.5 \\* 
        &       & 1.0 & Scalo    &          \phn 161.0 & \phn\phn\phn   30 & 0.0 &      10.3 & \phn\phn\phn\phn 3.11\phn &  \phn  1.3 \\
\tableline
814     & 2.931 & 0.2 & Salpeter &      \phn\phn  10.0 & \phn\phn\phn   10 & 4.3 &      10.0 & \phn\phn 569\phd\phn\phn\phn & \phn  1.0 \\* 
        &       & 1.0 & Salpeter &      \phn\phn  80.6 & \phn\phn\phn   10 & 1.9
 &      10.4 & \phn\phn\phn\phn 0.882 &       \phn 0.2 \\* 
        &       & 0.2 & Scalo    &  \phn\phn\phn   9.5 & \phn\phn\phn   10 & 4.3 &      10.4 & \phn 1565\phd\phn\phn\phn &  \phn  0.9 \\* 
        &       & 1.0 & Scalo    &      \phn\phn  80.6 & \phn\phn\phn   20 & 2.2 &      10.6 & \phn\phn\phn 36.1\phn\phn &  \phn  0.2 \\
\tableline
1063    & 2.969 & 0.2 & Salpeter &      \phn\phn  50.0 & \phn\phn\phn   20 & 0.7 & \phn  9.2 & \phn\phn\phn\phn 7.26\phn &  \phn  1.1 \\* 
        &       & 1.0 & Salpeter &          \phn 453.0 &      10000 & 0.0 & \phn  9.4 & \phn\phn\phn\phn 6.98\phn &  \phn  1.1 \\* 
        &       & 0.2 & Scalo    &      \phn\phn  40.0 & \phn\phn\phn   20 & 0.7 & \phn  9.4 & \phn\phn\phn 20.5\phn\phn &  \phn  1.1 \\* 
        &       & 1.0 & Scalo    &          \phn 181.0 & \multicolumn{1}{c}{$\infty$} & 0.0 & \phn  9.4 & \phn\phn\phn 15.5\phn\phn &  \phn  1.2 \\
\tableline
1541    &  2.98 & 0.2 & Salpeter &      \phn\phn  36.0 & \phn\phn\phn   20 & 2.4 & \phn  9.7 & \phn\phn\phn 44.9\phn\phn &  \phn  0.3 \\* 
        &       & 1.0 & Salpeter &          \phn 255.0 &      10000 & 1.7 & \phn  9.9 & \phn\phn\phn 33.2\phn\phn &  \phn  0.3 \\* 
        &       & 0.2 & Scalo    &      \phn\phn  14.5 & \phn\phn\phn   10 & 2.7 & \phn  9.9 & \phn\phn 190\phd\phn\phn\phn & \phn  0.4 \\* 
        &       & 1.0 & Scalo    &          \phn 114.0 & \phn 7000 & 1.7 &      10.0 & \phn\phn\phn 85.9\phn\phn &  \phn  0.4 \\
\tableline
661     & 2.991 & 0.2 & Salpeter &  \phn\phn\phn   5.8 & \phn\phn\phn   10 & 2.9 & \phn  9.5 & \phn\phn 367\phd\phn\phn\phn & \phn  1.2 \\* 
        &       & 1.0 & Salpeter &      \phn\phn  80.6 & \phn\phn\phn   70 & 1.7 & \phn  9.8 & \phn\phn\phn 49.6\phn\phn &  \phn  1.6 \\* 
        &       & 0.2 & Scalo    &  \phn\phn\phn   6.0 & \phn\phn\phn   30 & 2.9 & \phn  9.9 & \phn 1247\phd\phn\phn\phn &  \phn  1.2 \\* 
        &       & 1.0 & Scalo    &      \phn\phn  71.9 & \phn\phn  700 & 1.7 &      10.1 & \phn\phn 163\phd\phn\phn\phn & \phn  1.6 \\
\tableline
273     &  3.16 & 0.2 & Salpeter &          \phn 286.0 &      10000 & 0.5 & \phn  9.4 & \phn\phn\phn\phn 8.67\phn &  \phn  6.6 \\* 
        &       & 1.0 & Salpeter &          \phn 641.0 & \phn 7000 & 0.0 & \phn  9.6 & \phn\phn\phn\phn 6.56\phn &  \phn  6.1 \\* 
        &       & 0.2 & Scalo    &          \phn 143.0 & \phn\phn  200 & 0.2 & \phn  9.5 & \phn\phn\phn 14.7\phn\phn &  \phn  6.8 \\* 
        &       & 1.0 & Scalo    &          \phn 227.0 & \multicolumn{1}{c}{$\infty$} & 0.0 & \phn  9.6 & \phn\phn\phn 14.2\phn\phn &  \phn  6.8 \\
\tableline
367     & 3.162 & 0.2 & Salpeter &      \phn\phn  90.5 & \phn\phn\phn   10 & 1.0 & \phn 10.0 & \phn\phn\phn\phn 0.118 &       10.5 \\* 
        &       & 1.0 & Salpeter &          \phn 114.0 & \phn\phn\phn   20 & 0.5 & \phn  9.9 & \phn\phn\phn\phn 1.58\phn &       10.4 \\* 
        &       & 0.2 & Scalo    &          \phn 128.0 & \phn\phn\phn   10 & 0.2 &      10.1 & \phn\phn\phn\phn 0.004 &       10.0 \\* 
        &       & 1.0 & Scalo    &      \phn\phn  90.5 & \phn\phn\phn   10 & 0.2 &      10.1 & \phn\phn\phn\phn 0.141 &       10.1 \\
\tableline
282     & 3.181 & 0.2 & Salpeter &  \phn\phn\phn   5.8 & \phn\phn\phn   30 & 3.1 & \phn  9.7 & \phn\phn 705\phd\phn\phn\phn & \phn  4.7 \\* 
        &       & 1.0 & Salpeter &          \phn 128.0 & \phn 1000 & 1.9 &      10.1 & \phn\phn\phn 94.1\phn\phn &  \phn  5.2 \\* 
        &       & 0.2 & Scalo    &  \phn\phn\phn   5.5 & \phn\phn  100 & 3.1 &      10.2 & \phn 2340\phd\phn\phn\phn &  \phn  4.7 \\* 
        &       & 1.0 & Scalo    &      \phn\phn  64.1 & \phn\phn  500 & 1.9 &      10.3 & \phn\phn 255\phd\phn\phn\phn & \phn  5.2 \\
\tableline
1114    & 3.226 & 0.2 & Salpeter &  \phn\phn\phn   4.2 & \phn\phn\phn   10 & 2.7 & \phn  9.0 & \phn\phn 174\phd\phn\phn\phn & \phn  3.9 \\* 
        &       & 1.0 & Salpeter &          \phn 161.0 & \phn 5000 & 1.2 & \phn  9.3 & \phn\phn\phn 15.3\phn\phn &  \phn  4.1 \\* 
        &       & 0.2 & Scalo    &  \phn\phn\phn   4.0 & \phn\phn\phn   10 & 2.7 & \phn  9.4 & \phn\phn 542\phd\phn\phn\phn & \phn  4.0 \\* 
        &       & 1.0 & Scalo    &      \phn\phn  50.0 & \multicolumn{1}{c}{$\infty$} & 1.4 & \phn  9.4 & \phn\phn\phn 56.3\phn\phn &  \phn  4.2 \\
\tableline
1115    & 3.226 & 0.2 & Salpeter &          \phn 453.0 & \multicolumn{1}{c}{$\infty$} & 1.4 &      10.4 & \phn\phn\phn 63.4\phn\phn &  \phn  8.2 \\* 
        &       & 1.0 & Salpeter &              3000.0 & \multicolumn{1}{c}{$\infty$} & 0.7 &      10.9 & \phn\phn\phn 32.5\phn\phn &  \phn  5.5 \\* 
        &       & 0.2 & Scalo    &      \phn\phn  47.5 & \phn\phn\phn   10 & 1.4 &      10.4 & \phn\phn\phn 21.1\phn\phn &  \phn  8.7 \\* 
        &       & 1.0 & Scalo    &          \phn 509.0 & \phn 7000 & 0.5 &      10.5 & \phn\phn\phn 64.5\phn\phn &  \phn  9.1 \\
\tableline
947     & 3.233 & 0.2 & Salpeter &      \phn\phn  36.0 & \phn\phn\phn   10 & 1.7 & \phn  9.5 & \phn\phn\phn\phn 9.44\phn &       20.8 \\* 
        &       & 1.0 & Salpeter &          \phn 453.0 & \phn 7000 & 1.0 & \phn  9.8 & \phn\phn\phn 16.2\phn\phn &       20.9 \\* 
        &       & 0.2 & Scalo    &      \phn\phn  45.0 & \phn\phn  300 & 1.9 & \phn  9.7 & \phn\phn 111\phd\phn\phn\phn &      21.2 \\* 
        &       & 1.0 & Scalo    &      \phn\phn  90.5 & \phn\phn  200 & 1.2 & \phn  9.7 & \phn\phn\phn 51.3\phn\phn &       21.3 \\
\tableline
284     & 3.367 & 0.2 & Scalo    &          \phn 360.0 & \phn\phn  200 & 0.0 & \phn  9.3 & \phn\phn\phn\phn 1.90\phn &  \phn  1.3 \\* 
        &       & 1.0 & Salpeter &              6000.0 & \multicolumn{1}{c}{$\infty$} & 0.5 &      10.1 & \phn\phn\phn\phn 2.24\phn &  \phn  2.0 \\* 
        &       & 0.2 & Salpeter &          \phn 321.0 & \phn\phn  100 & 0.0 & \phn  9.2 & \phn\phn\phn\phn 0.685 &  \phn  1.2 \\* 
        &       & 1.0 & Scalo    &          \phn 905.0 & \multicolumn{1}{c}{$\infty$} & 0.0 & \phn  9.5 & \phn\phn\phn\phn 3.35\phn &  \phn  2.2 \\
\tableline
516     & 3.368 & 0.2 & Salpeter &          \phn 143.0 & \phn\phn\phn   30 & 0.0 & \phn  9.7 & \phn\phn\phn\phn 1.82\phn &  \phn  0.7 \\* 
        &       & 1.0 & Salpeter &          \phn 255.0 & \phn\phn  100 & 0.0 & \phn  9.8 & \phn\phn\phn\phn 6.99\phn &  \phn  0.6 \\* 
        &       & 0.2 & Scalo    &          \phn 102.0 & \phn\phn\phn   20 & 0.0 & \phn  9.9 & \phn\phn\phn\phn 2.45\phn &  \phn  0.7 \\* 
        &       & 1.0 & Scalo    &          \phn 255.0 & \phn\phn  200 & 0.0 &      10.0 & \phn\phn\phn 19.6\phn\phn &  \phn  0.6 \\
\tableline
553     &  3.43 & 0.2 & Salpeter &              1020.0 & \multicolumn{1}{c}{$\infty$} & 1.2 &      10.5 & \phn\phn\phn 31.2\phn\phn &  \phn  6.5 \\* 
        &       & 1.0 & Salpeter &              6500.0 & \multicolumn{1}{c}{$\infty$} & 0.7 &      11.1 & \phn\phn\phn 22.3\phn\phn &  \phn  5.4 \\* 
        &       & 0.2 & Scalo    &          \phn 161.0 & \phn\phn\phn   30 & 0.0 &      10.2 & \phn\phn\phn\phn 2.29\phn &  \phn  6.8 \\* 
        &       & 1.0 & Scalo    &          \phn 360.0 & \phn\phn  200 & 0.0 &      10.3 & \phn\phn\phn 20.0\phn\phn &  \phn  6.4 \\
\tableline
\enddata
\tablenotetext{\dag}{For starburst extinction law \citep{cal00}.}
\end{deluxetable}


\end{document}